%% file: main.tex
\documentclass[manuscript,screen,acmsmall]{acmart}

\renewcommand\footnotetextcopyrightpermission[1]{} 

\AtBeginDocument{%
  }

\setcopyright{acmlicensed}
\copyrightyear{2025}
\acmYear{2025}
\acmDOI{XXX.XXX}

\usepackage{booktabs}
\usepackage{array}
\usepackage{multirow}
\usepackage{colortbl}
\usepackage{subfigure}
\usepackage{tikz}
\usepackage{mdframed}
\usepackage{tcolorbox}
\usepackage{colortbl}
\usepackage{makecell}
\usepackage{lipsum}
\usetikzlibrary{backgrounds, mindmap, trees, arrows.meta}
\usepackage[normalem]{ulem}
\useunder{\uline}{\ul}{}

\newcommand{\ie}{\emph{i.e.,}~}
\newcommand{\eg}{\emph{e.g.,}~}
\newcommand{\aka}{\emph{a.k.a.,}~}
\newcommand{\etal}{\emph{et al.}~}
\newcommand{\paratitle}[1]{$\bullet$~\textbf{#1}}
\newcommand{\wrt}{\emph{w.r.t.}~}
\newcommand{\ignore}[1]{}

\newcommand{\revision}[1]{{#1}}

\newcommand{\our}{LLM-RS}

\begin{document}

\title{Tapping the Potential of Large Language Models as Recommender Systems: A Comprehensive Framework and Empirical Analysis}

\author{Lanling Xu}
\email{xulanling@ruc.edu.cn}
\orcid{0000-0002-7464-3776}
\author{Junjie Zhang}
\email{junjie.zhang@ruc.edu.cn}
\orcid{0009-0008-8864-915X}
\affiliation{%
  \institution{Gaoling School of Artificial Intelligence, Renmin University of China}
  \city{Beijing}
  \country{China}
}

\author{Bingqian Li}
\orcid{0009-0000-5103-0750}
\affiliation{
  \institution{Gaoling School of Artificial Intelligence, Renmin University of China}
  \city{Beijing}
  \country{China}
}


\author{Jinpeng Wang}
\orcid{0000-0002-0080-8988}
\author{Sheng Chen}
\orcid{0009-0001-7186-4960}
\affiliation{%
  \institution{Meituan Group}
  \city{Beijing}
  \country{China}
}

\author{Wayne Xin Zhao}
\orcid{0000-0002-8333-6196}
\authornote{Wayne Xin Zhao~(batmanfly@gmail.com) is the corresponding author.}
\author{Ji-Rong Wen}
\orcid{0000-0002-9777-9676}
\affiliation{%
  \institution{Gaoling School of Artificial Intelligence, Renmin University of China}
  \city{Beijing}
  \country{China}
}



\renewcommand{\shortauthors}{Lanling Xu et al.}
\renewcommand{\shorttitle}{Leveraging Large Language Models as Recommender Systems}

\begin{abstract}
Recently, Large Language Models~(LLMs) such as ChatGPT have showcased remarkable abilities in solving general tasks, demonstrating the potential for applications in recommender systems. To assess how effectively LLMs can be used in recommendation tasks, our study primarily focuses on employing LLMs as recommender systems through prompting engineering. We propose a general framework for utilizing LLMs in recommendation tasks, focusing on the capabilities of LLMs as recommenders. To conduct our analysis, we formalize the input of LLMs for recommendation into natural language prompts with two key aspects, and explain how our framework can be generalized to various recommendation scenarios. As for the use of LLMs as recommenders, we analyze the impact of public availability, tuning strategies, model architecture, parameter scale, and context length on recommendation results based on the classification of LLMs. As for prompt engineering, we further analyze the impact of four important components of prompts, \ie task descriptions, user interest modeling, candidate items construction and prompting strategies. In each section, we first define and categorize concepts in line with the existing literature. 
Then, we propose inspiring research questions followed by detailed experiments on two public datasets, in order to systematically analyze the impact of different factors on performance. Based on our empirical analysis, we finally summarize promising directions to shed lights on future research.
\end{abstract}

\ccsdesc[500]{Information systems~Recommender systems}
\ccsdesc[500]{Information systems~Language models}

\keywords{Large Language Models, Recommender Systems, Empirical Study}

\received{6 June 2024}
\received[revised]{16 Jan 2025}

\maketitle

\input{sec-intro}
\input{sec-related}
\input{sec-overall}
\input{sec-llm}
\input{sec-prompt}

\input{sec-con}

\bibliographystyle{ACM-Reference-Format}
\bibliography{reference}

\appendix

\end{document}

%% file: sec-intro.tex
\section{Introduction}
\label{sec:intro}

In order to alleviate the problem of information overload~\cite{ni2019amazon2018,hou2022towards}, recommender systems explore user needs and provide them with recommendations based on their historical interactions, which are widely studied both in industry and academia~\cite{rendle2009bpr,guo2017deepfm,he2017neural,he2020lightgcn}. 
During the past decade, various recommendation algorithms have been proposed to solve recommendation tasks by capturing personalized interaction patterns from user behaviors~\cite{kang2018sasrec,zhou2020s3}.
Despite the progress of conventional recommenders, the performance is highly dependent on the limited training data from a few datasets and domains, and there are two major drawbacks. 
On the one hand, traditional models lack the general knowledge of the world beyond interaction sequences. For complex scenarios that need to think or plan, existing methods do not have commonsense knowledge to solve such tasks~\cite{wei2023llmrec,harte2023leveraging4seqrec,xi2023towards-kar,liu2023genre-news}. On the other hand, traditional models cannot truly understand intentions and preferences of users. The results of the recommendation do not have explainability, and the requirements expressed by users in explicit forms such as natural languages are difficult to consider~\cite{li2023recformer,yuan2023morec,li2023tcf}. 

Recently, Large Language Models~(LLMs) such as ChatGPT have demonstrated impressive abilities in solving general tasks~\cite{haleem2022ChatGPTera,wu2023ChatGPTbrief}, showing their potential in the development of next-generation recommender systems. The advantages of incorporating LLMs into recommendation tasks are two-fold. Firstly, the excellent performance of LLMs in complex reasoning tasks indicates the rich world knowledge and superior inference ability, which can effectively compensate for the local knowledge of traditional recommenders~\cite{sanner2023cold-start,mysore2023largeNBR,agrawal2023beyond-metadata}. Secondly, the language modeling abilities of LLMs can seamlessly integrate massive textual data, enabling them to extract features beyond IDs and even explicitly understand user preferences~\cite{he2023zero-shot-CRS,li2023prompt-distillation}. 
Therefore, researchers have attempted to leverage LLMs for recommendation tasks. 
Typically, there are three ways to employ LLMs to make recommendations: (1) LLMs can serve as the recommender to make recommendation decisions, encompassing both discriminative and generative recommendations~\cite{hou2023large,zhang2023recommendation,dai2023uncovering,gao2023chat,bao2023tallrec}. (2) LLMs can be used to improve traditional recommendation models by extracting semantic representations of users and items from text corpora. The extensive semantic information and robust planning capabilities of LLMs are integrated into traditional models~\cite{hou2022towards,du2023enhancing,wang2023enhancing,harte2023leveraging4seqrec,agrawal2023beyond-metadata,wei2023llmrec,xi2023towards-kar,liu2023genre-news}. (3) LLMs are used as the recommendation simulator to execute external generative agents in the recommendation process, where LLMs can empower users and items to stimulate the virtual environment~\cite{wang2023recagent,wang2023survey,zhang2023generative,zhang2023agentcf,friedman2023recllm}.
We mainly focus on the first scenario in this paper.

Considering the gap between general knowledge from LLMs and domain knowledge from recommendation models~\cite{zhang2023collm,bao2023tallrec}, there are two key factors to leverage LLMs as recommenders, \ie how to select an LLM as the foundation model and how to construct task-specific prompts. As for LLMs, a growing number of open-source and closed-source models have emerged, and the same model also has different variants due to settings such as parameter scales and context lengths~\cite{zhao2023survey}. It is worth discussing how to select corresponding LLMs for specific scenarios and develop corresponding training strategies. 
As for prompts, it is an important medium for interactions between humans and language models, and a well-designed prompt can better stimulate the powerful capabilities of LLMs~\cite{liu2023prompting-survey, le2021many}. 
\ignore{Improving the performance of artificial intelligence by designing and improving prompts is known as prompt engineering, which is also applicable in the field of recommender systems.} 
To stimulate the recommendation ability of LLMs, prompt engineering should involve not only task description and prompting strategies for general tasks, but also the incorporation of user interest and candidate items in recommender systems~\cite{yao2023DOKE,fan2023survey,liu2023prompting-survey}.

\begin{table}[t]
\caption{An overview of the primary discoveries presented in our work. We summarize our findings in the second column as ``our findings'', and verify findings in existing literature as ``re-validated findings''.}
\small
\begin{tabular}{@{}p{0.03\textwidth}p{0.42\textwidth}p{0.42\textwidth}@{}}
\toprule
\multicolumn{1}{c}{\textbf{Aspect}}                       & \multicolumn{1}{c}{\textbf{Our Findings}}         & \multicolumn{1}{c}{\textbf{Re-validated Findings}} \\ \midrule
\multicolumn{1}{c}{\textbf{LLMs}}        & \begin{tabular}[c]{@{}p{0.42\textwidth}@{}}  
$\bullet$ \revision{The long-chain reasoning capabilities and slow-thinking process of LLMs are optimized for Mathematics and coding tasks, which may not be suitable for the recommendation task.} \\
$\bullet$ The larger the parameter scale, the better the ability, while a longer maximum context length leads to worse recommendations. \\ $\bullet$ Fine-tuning all parameters of LLMs is more effective than parameter-efficient fine-tuning such as LoRA, but more training time is required. \revision{Similarly, a larger rank in LoRA means that more parameters can be fine-tuned, generally achieving better recommendations.} 
\end{tabular} & \begin{tabular}[c]{@{}p{0.42\textwidth}@{}}
$\bullet$ LLMs have zero-shot recommendation capabilities, but are inferior to fully-trained recommenders. Fine-tuning LLMs can surpass traditional recommendation models~\cite{dai2023uncovering,hou2023large}. \\
$\bullet$ Instruction tuning can enhance fine-tuning results of LLMs on recommendations~\cite{bao2023tallrec,touvron2023llama2}. \\ $\bullet$ In few-shot training scenarios, LLMs are more capable of adapting to recommendation tasks compared to traditional models~\cite{liao2023llara,fu2023uni-ctr}. \\ $\bullet$ There are limitations leveraging LLMs as recommenders, such as position bias~\cite{hou2023large,ma2023STELLA} and lack of domain knowledge~\cite{yao2023DOKE,zheng2023LC-Rec}. \end{tabular}                                                  \\ \midrule
\multirow{25}{*}{\textbf{Prompts}} & \multicolumn{2}{c}{\emph{User Interest Modeling}}       \\ \cmidrule(l){2-3} 
                                             & \begin{tabular}[c]{@{}p{0.42\textwidth}@{}} $\bullet$ \revision{The recommendation performance of retrieval-based and generation-based methods for user interest modeling is generally better than that of memory-based methods, indicating the importance of the interest memory.} \\ $\bullet$ \revision{Long-term interest and short-term interest can complement each other, and it is preferable to combine the two with personalized descriptions for personalized query and memory.}    \end{tabular}    & \begin{tabular}[c]{@{}p{0.42\textwidth}@{}} $\bullet$ For short-term interest, only truncating the most recent items is not optimal~\cite{lin2023rella}. \\ $\bullet$ Increasing the number of historical items to represent users brings insignificant gains~\cite{hou2023large}. \\ $\bullet$ Personalized user profiles and customized item descriptions assist in user modeling for LLM-based recommendations~\cite{yao2023DOKE,shu2023rah}. \\ $\bullet$ \revision{Retrieval-enhanced items can improve sequential comprehension~\cite{lin2023rella}}.
                                             \end{tabular}    \\ \cmidrule(l){2-3} 
                                             & \multicolumn{2}{c}{\emph{Candidate Items Construction}} \\ \cmidrule(l){2-3} 
                                             & \begin{tabular}[c]{@{}p{0.42\textwidth}@{}} $\bullet$ \revision{LLMs exhibit limitations in retaining critical information regarding the candidate list from prompts, potentially leading to hallucination and responses that deviate from the task.} \\ $\bullet$ \revision{Description-based identifiers outperform token-based when grounding items.}  \end{tabular}    & \begin{tabular}[c]{@{}p{0.42\textwidth}@{}} $\bullet$ Retrieving candidate items by traditional models and then re-ranking them by LLMs can further improve the results~\cite{hou2023large,yang2023palr}. \\ $\bullet$ Indexing methods and grounding strategies of items are important factors affecting the effectiveness of recommendations~\cite{lin2023transrec,liao2023llara,li2023e4srec}. \end{tabular}    \\ \cmidrule(l){2-3} 
                                             & \multicolumn{2}{c}{\emph{Prompting Strategies}}         \\ \cmidrule(l){2-3} 
                                             & \begin{tabular}[c]{@{}p{0.42\textwidth}@{}} $\bullet$ Specific problem decomposition is required for recommendation in CoT prompting. \\ $\bullet$ The few-shot prompting strategy has insignificant advantage in recommendations. \\ $\bullet$ \revision{Explicit feedback is better than implicit feedback to fine-tune LLMs for recommendation.} \end{tabular}
& \begin{tabular}[c]{@{}p{0.42\textwidth}@{}} $\bullet$ Although provided with historical items in chronological order, LLMs still need explicit guidance to understand recent items~\cite{hou2023large,ma2023STELLA}. \\ $\bullet$ Role-playing and expert-like prompts can leverage the capabilities of LLMs~\cite{jin2023core,wang2023rethinkingCRS}.  \end{tabular}  \\
\bottomrule
\end{tabular}
\label{tab:intro-summary}
\end{table}

Although existing studies have made initial attempts to explore the recommendation capabilities of LLMs like ChatGPT~\cite{shu2023rah,dai2023uncovering,gao2023chat,hou2023large,liu2023chatgpt}, and some studies have used paradigms such as fine-tuning and instruction tuning to train LLMs in the field of recommender systems~\cite{zhang2023recommendation,zheng2023LC-Rec,bao2023bi,bao2023tallrec}, they focus on exploring the performance of a certain task instead of constructing a comprehensive framework to formalize the potential applications of LLM-based recommender systems.
There are also systematic reviews concentrating on the progress of LLMs~\cite{zhao2023survey} and surveys of recommender systems empowered by LLMs~\cite{li2023survey, lin2023survey, wu2023survey}. However, previous surveys generally use specific criteria to classify existing work and introduce them separately. They mainly focus on showcasing related work and summarizing advantages and limitations, rather than conducting additional experiments to validate existing results and explore new discoveries. Our work focuses on the capabilities of LLMs to serve as recommender systems, aiming to establish a general framework of \revision{leveraging {\ul \textbf{L}}arge {\ul \textbf{L}}anguage {\ul \textbf{M}}odels as {\ul \textbf{R}}ecommender {\ul \textbf{S}}ystems, namely~\our}. 

In order to conduct our analysis for~\our, we formalize the input of LLMs for recommendation into natural language prompts with two key aspects: LLMs and prompts, and explain how our framework can be generalized to various recommendation scenarios and tasks. As for the use of LLMs as recommenders, we analyze the impact of the public availability, tuning strategies, model architecture, parameter scale, and context length on recommendation results based on the classification of LLMs. As for prompt engineering, we further analyze the impact of four important components of prompts, \ie task description, user interest modeling, candidate items construction, and prompting strategies. Given personalized prompts that include task description and user interest, the LLM selects, generates, or explains candidate items based on general world knowledge and personalized user profiles. For each module, we first define and categorize concepts in line with the existing literature. Then, we propose inspiring research questions, followed by detailed experiments to analyze the impact of different conditions on the recommendation performance. Based on the empirical analysis, we finally summarize empirical findings for future research.

\ignore{We conduct a more detailed empirical analysis to explore the recommendation capabilities of LLMs with different conditions. Our proposed framework~\our~is a high-level summary of recommendation tasks leveraging LLMs. It not only outlines the core components and key conclusions of relevant work, but also sheds light on future research such as the combination of long-term and short-term interest for LLM-based recommendation. Finally, we conduct empirical analysis in detail and summarize future directions to draw inspiration for LLM-based recommender systems.}

In general, the contributions of our work can be summarized as follows:
\begin{itemize}
    \item We derive a general framework~\our~to sum up existing work of utilizing LLMs as foundation models for recommendation, which can be generalized to multiple scenarios and tasks by different LLMs and prompts.
    \item We provide a systematic analysis on leveraging LLMs as recommender systems, focusing on two aspects: LLMs and prompt engineering. The use of LLMs includes analysis of public availability, tuning strategies, model architecture, parameter scale, and context length. Moreover, prompt engineering consists of discussions on task description, user interest modeling, candidate items construction and prompting strategies. For each aspect, we define and describe with concepts first, and then provide reference solutions with experiments. 
    \item Extensive experiments on two public datasets conclude key findings for recommendation with LLMs. As listed in Table~\ref{tab:intro-summary}, our findings include experimental settings on each aspect of our proposed framework, and obtain empirical experience on evaluating the performance of LLMs on recommendation tasks for future research.
\end{itemize}

In what follows, we first review the related work in Section~\ref{sec:related}. In Section~\ref{sec:overall}, we present our proposed general framework and its instantiation, and introduce overall settings of the following experiments. As the core components of this paper, we discuss two main aspects of~\our, \ie LLMs and prompts in Section~\ref{sec:llm} and Section~\ref{sec:prompt}, respectively. For each aspect, we generalize key factors that affect recommendation results, and conduct corresponding experiments to summarize empirical findings. At last, Section~\ref{sec:con} concludes this paper and sheds lights on future directions.

%% file: sec-related.tex
\section{RELATED WORK}
\label{sec:related}

\subsection{Recommender Systems}
\label{sec:related-recommender-systems}

For tackling the challenge of information overload~\cite{ni2019amazon2018,yuan2023go}, recommender systems have become pivotal tools for delivering personalized contents for users across various domains. 
Based on previous studies, research has delineated two primary categories for common recommendation methods: interaction-based recommendation and content-based recommendation~\cite{thorat2015survey}. 
In line with previous studies, recommendation algorithms aim to derive user preferences and behavioral patterns from their historical interactions. The most common technique for the interaction-based recommendation is Collaborative Filtering~(CF)~\cite{sarwar2001item, 
su2009survey}, which recommends items based on preferences of similar users. Matrix Factorization~(MF)~\cite{koren2009matrix} is a prevalent approach in collaborative filtering, and it constructs embedding representations for users and items from the interaction matrix, facilitating the algorithm to calculate similarity scores efficiently. Furthermore, Neural Collaborative Filtering~(NCF)~\cite{he2017neural}, integrating deep neural networks, replaces the inner product used in MF with a neural architecture, thereby demonstrating better performance than previous methods. Contemporary advancements in deep neural network architectures have enhanced the integration of user and item embeddings~\cite{liu2023SAGCN}. For example, since recommendation data can be represented as graph-structured data, Graph Neural Network~(GNN)~\cite{wu2022graph} can be utilized to encode the information of the interaction graph~(nodes consist of users and items), and generate meaningful representations via message propagation and contrastive learning strategies~\cite{wang2019ngcf,he2020lightgcn,wu2021sgl,lin2022ncl}. As Pre-trained Language Models~(PLM) gain prominence, there is a growing interest in pre-trained large-scale recommendation models powered by PLMs~\cite{zhou2020s3,zhang2021LLMRecSys,hou2022towards}. Besides user-item pairs and IDs, content-based recommendation algorithms leverage auxiliary modalities such as textual and visual information to augment user and item representations in recommendation tasks~\cite{qiu2021u,yuan2023go,wang2023EnhancedRepre}.


\subsection{Large Language Models for Recommender Systems}
\label{sec:related-LLMs-for-recommender-systems}

Large Language Models~(LLMs) are a cutting-edge advancement in artificial intelligence that excel in understanding and generating human-like texts~\cite{touvron2023llama,zeng2022glm,raffel2020T5}. LLMs are usually transformer-based models and trained on vast amounts of textual data with billions of parameters, allowing them to comprehend contexts, generate coherent sentences, and even mimic human conversations~\cite{haleem2022ChatGPTera,wu2023ChatGPTbrief}. Through this process, LLMs have shown prominent potentials in the field of Natural Language Processing~(NLP), and have demonstrated various incredible capabilities in dealing with complex NLP tasks, including but not limited to In-Context Learning~(ICL)~\cite{brown2020language}, instruction following~\cite{touvron2023llama2} and step-by-step reasoning abilities~\cite{zhao2023survey}. Recently, LLMs have been increasingly integrated into recommender systems to provide personalized recommendation~\cite{li2023survey,wu2023survey,fan2023survey}. Recent studies have explored the fusion of LLMs with recommender systems, which can be divided into the three paradigms, \ie \emph{LLMs as recommender systems}~(Section~\ref{sec:related-llm-as-rs}), \emph{LLMs improve recommender systems}~(Section~\ref{sec:related-llm-improves-rs}) and \emph{LLMs as recommendation simulator}~(Section~\ref{sec:related-llm-as-simulator}) as follows.

\subsubsection{LLMs as Recommender Systems}
\label{sec:related-llm-as-rs}

This paradigm takes LLMs as recommender systems. Employing diverse strategies like pre-training, fine-tuning, or prompting, LLMs can combine general knowledge with input data to yield personalized recommendations for users~\cite{hou2023large,dai2023uncovering,ji2023genrec}.
Due to the variety of recommendation tasks, LLMs as recommender systems can be categorized into two types: \emph{discriminative recommendation} and \emph{generative recommendation}.

\paratitle{Discriminative recommendation} instructs LLMs to make recommendation decisions on the given candidate items, usually focusing on item scoring~\cite{fu2023uni-ctr} and re-ranking tasks~\cite{dai2023uncovering}. 
For Click-Through Rate~(CTR) prediction tasks, \revision{TALLRec~\cite{bao2023tallrec} used the recommendation data to fine-tune LLaMA~\cite{touvron2023llama} as the CTR prediction model, showcasing the powerful capabilities of LLMs in few-shot learning.}
Liu \etal~\cite{liu2023chatgpt} designed specific zero-shot and few-shot prompts to evaluate abilities of LLMs on rating predictions. LLMs were required to assign a score for the item according to the previous rating history of users and the score range given in prompts, while the result indicated that LLMs can outperform classical rating methods in few-shot conditions~\cite{liu2023chatgpt}. Kang \etal~\cite{kang2023llms-ctr} further formulated the rating prediction task as multi-class classification and regression task, investigating the influence of model size on recommendation performance. Different from these methods, Hou \etal~\cite{hou2023large} structured a re-ranking task, employing ICL approaches for LLMs to rank items in the candidate pool. Previous studies highlighted the sensitivity of LLMs to the sequence of interaction histories provided in prompts~\cite{ma2023STELLA}, which can be alleviated by strategies such as recency-focused prompting~\cite{hou2023large}. 

\paratitle{Generative recommendation} requires LLMs to generate items recommended to users, either from candidate item lists within prompts or from LLMs with general knowledge~\cite{li2023survey}. 
GenRec~\cite{ji2023genrec} leveraged the contextual comprehension ability of LLMs to
transform interaction histories into formulated prompts for next-item predictions. To address instances where GenRec might propose items absent in candidate lists, GPT4Rec~\cite{li2023gpt4rec} came up with the method that used Best Matching~(BM25)~\cite{robertson2009bm25} algorithm to retrieve the most similar item in candidate item list with the item generated by LLMs. 
\revision{Faced with the challenge of item identifiers for compatibility with LLMs in recommender systems, LC-Rec~\cite{zheng2023LC-Rec} employed vector quantization to convert natural languages into trainable tokens, enabling the alignment with input formats of LLMs.
To further address the limitations of previous methods reliant on historical data, 
To make the recommendation results generated by LLMs break through the limitations of existing items,
SpecGR~\cite{ding2024SpecGR} introduced a draft-then-verify framework to recommend new items in inductive settings. }
In addition to top-n recommendations, LLMs can be leveraged for generative tasks such as explainable recommendations~\cite{lei2023recexplainer,colas2023knowrec,li2023PEPLER,yang2023RecInterpreter,luo2023LLMXRec} and review summarization~\cite{geng2022p5,liu2023llmrec,wang2023recmind}.
Moreover,
with the incredible abilities in dialogue comprehension and communication, LLMs are naturally considered as the backbone of conversational and interactive recommender systems. ChatRec~\cite{gao2023chat} designed an interactive recommendation framework based on ChatGPT, which can comprehend requirements of users through multi-turn dialogues and traditional recommendation models. Moreover, RecLLM~\cite{friedman2023recllm} combined the dialogue management module with a ranker module and a controllable LLM-based user simulator to generate synthetic conversations for tuning system modules. 
Apart from these methods, InteRecAgent~\cite{huang2023recommender} employed LLMs as the brain and recommender models as tools, combining their respective strengths to create an interactive recommender system~\cite{huang2023recommender}. As a Conversational Recommender System~(CRS), InteRecAgent enabled traditional recommender systems to become interactive systems with a natural language interface through the integration of LLMs. 

The methods for adapting LLMs as recommender systems mainly have two paradigms, \ie \emph{non-tuning paradigm} and \emph{tuning paradigm} as follows. 

\paratitle{Non-tuning paradigm} keeps parameters of LLMs fixed and extracts the general knowledge of LLMs with prompting strategies. Existing work of non-tuning paradigm focuses on designing appropriate prompts to stimulate recommendation abilities of LLMs~\cite{yao2023DOKE,wang2023recmind,li2023ChatGPT-news}. Liu \etal~\cite{liu2023chatgpt} proposed a prompt construction framework to evaluate abilities of ChatGPT on five common recommendation tasks, each type of prompts contained zero-shot and few-shot versions. Hou \etal~\cite{hou2023large} not only used prompts to evaluate abilities of LLMs on sequential recommendation, but also introduced recency-focused prompting and ICL strategies to alleviate order perception and position bias issues of LLMs. ChatRec~\cite{gao2023chat} and InteRecAgent~\cite{huang2023recommender} mentioned above are also within the classic non-tuning paradigm. 

\paratitle{Tuning paradigm} aims to update parameters of LLMs to inject recommendation capabilities into LLM itself. \revision{The tuning strategies include fine-tuning~\cite{zheng2023LC-Rec,hu2021lora, bao2023bi,zhang2024causality} and instruction tuning~\cite{luo2023recranker,qiu2023controlrec}.} P5~\cite{geng2022p5} proposed five types of instructions targeting at different recommendation tasks to fine-tune a T5~\cite{raffel2020T5} model. The instructions were formulated based on conventional recommendation datasets with designed templates, which equipped LLMs with generation abilities for unseen prompts or items~\cite{geng2022p5}. 
InstructRec~\cite{zhang2023recommendation} further designed abundant instructions for tuning, including 39 manually designed templates with preference, intention, task form and context of a user. Compared with these methods, TallRec~\cite{bao2023tallrec} used Low-Rank Adaptation of LLMs~(LoRA)~\cite{hu2021lora}, a parameter-efficient tuning method, to handle the two-stage tuning for LLMs.
\revision{To further align LLMs with personalized human preferences, recent research~\cite{chen2024softmax,liao2024rosepo,gao2024sprec} has applied post-training techniques such as Direct Preference Optimization~(DPO)~\cite{rafailov2024direct} to fine-tune LLMs for recommendation, thereby integrating negative samples and user preference information within the tuning process.}


Although LLMs as recommender systems present a way of utilizing the common knowledge of LLMs, it still encounters problems to be coped with. Due to the high computational cost~\cite{touvron2023llama2,liao2023llara} and slow inference time~\cite{li2023e4srec}, LLMs are struggled to be efficient enough compared to traditional recommendation methods~\cite{hou2023large,gao2023chat}. Additionally, constraints on the input sequence length will limit the amount of external information~(\eg candidate item lists)~\cite{lin2023rella}, leading to degrading performance of LLMs in scenarios such as sequential recommendation. Furthermore, since information in recommendation tasks is challenging to be expressed in natural language~\cite{lin2023transrec,yuan2023morec}, it is hard to formulate appropriate prompts that make LLMs truly understand what they are required to do.

\subsubsection{LLMs Improve Recommender Systems}
\label{sec:related-llm-improves-rs}

This method mainly utilizes LLMs to generate auxiliary information to enhance the performance of recommendation models~\cite{du2023enhancing,wang2023enhancing,wei2023llmrec}, based on the reasoning abilities and common knowledge. 
The research on how to improve recommendation models with LLMs can be divided into three categories, \ie \emph{LLMs as feature encoder}, \emph{LLMs for data augmentation} and \emph{LLMs co-optimized with domain-specific models}.

\paratitle{LLMs as feature encoder}. The representation embeddings of users and items are important factors in classical recommender systems~\cite{rendle2009bpr,he2020lightgcn}. LLMs serving as feature encoders can generate related textual data of users and items, and enrich their representations with semantic information. U-BERT~\cite{qiu2021u} injected user representations with user review texts, item review texts and domain IDs, augmenting the contextual semantic information in user vectors.  Wu \etal~\cite{wu2021empowering}, on the other hand, employed language models to generate item representations for news recommendation. With the development of LLMs and prompting strategies, BDLM~\cite{zhang2023bridging} constructed the prompt consisting of interaction and contextual information into LLMs, and obtained top-layer feature embeddings as user and item representations, injecting user-item interaction information into embeddings. 

\paratitle{LLMs for data augmentation}. For this paradigm, LLMs are required to generate auxiliary textual information for data augmentation~\cite{agrawal2023beyond-metadata,wei2023llmrec,liu2023SAGCN,liu2023genre-news}. By using prompting or ICL strategies, the related knowledge will be extracted out in different text forms to facilitate recommendation tasks~\cite{xi2023towards-kar,du2023enhancing,wang2023enhancing}. 
One form of auxiliary textual information is summarization or text generation, enabling LLMs to enrich representations of users or items~\cite{wang2023EnhancedRepre}. For example, Du \etal~\cite{du2023enhancing} proposed a job recommendation model which utilized the capability of LLMs for summarization to extract user information and job requirements. 
\revision{Mysore \etal~\cite{mysore2023largeNBR}, on the other hand, used the ICL strategy to instruct LLMs for narrative queries according to the interaction history of users, which was then utilized as the source data for training narrative driver recommendation models.}
Considering item descriptions and user reviews, KAR~\cite{xi2023towards-kar} extracted the reasoning knowledge on user preferences and the factual knowledge on items through specifically designed prompts, while SAGCN~\cite{liu2023SAGCN} utilized a chain-based prompting strategy to generate semantic information. 
Another form of using the textual features generated from LLMs is for graph augmentation in the recommendation field. LLMRG~\cite{wang2023enhancing} leveraged LLMs to extend nodes in recommendation graphs. The resulting reasoning graph was encoded using GNN, which served as additional input to enhance sequential models. LLMRec~\cite{wei2023llmrec} adopted three types of prompts to generate information for graph augmentation, including implicit feedback, user profile and item attributes. 

\paratitle{LLMs co-optimized with domain-specific models}. The categories mentioned above mainly focus on the impact of common knowledge for domain-specific models~\cite{wang2023EnhancedRepre}. However, LLM itself often struggles to handle domain-specific tasks due to the lack of task-related information~\cite{yao2023DOKE,kang2023llms-ctr}. Therefore, some studies conducted experiments to bridge the gap between LLMs and domain-specific models. BDLM~\cite{zhang2023bridging} proposed an information sharing module serving as an information storage mechanism between LLMs and domain-specific models. The user embeddings and item embeddings stored in the module were updated in turn by the LLM and the domain-specific model, enhancing the performance of both sides. CoLLM~\cite{zhang2023collm} combined LLMs with a collaborative model, which formed collaborative embeddings for LLM usage. By tuning LLM and collaborative module, CoLLM showed great improvements in both warm and cold-start scenarios. In CRS, approaches such as ChatRec~\cite{gao2023chat} and InteRecAgent~\cite{huang2023recommender} considered LLMs as the backbone, and leveraged traditional recommendation models for candidate item retrieval.

\ignore{With the incredible abilities in dialogue comprehension and communication, LLMs are naturally considered as the backbone of conversational and interactive recommender systems. ChatRec~\cite{gao2023chat} designed an interactive recommendation framework based on ChatGPT, which can comprehend requirements of users through multi-turn dialogues and call existing recommender systems to provide results. Moreover, RecLLM~\cite{friedman2023recllm} combines the dialogue management module with a ranker module and a controllable LLM-based user simulator to generate much synthetic conversations for tuning system modules. 
Apart from these methods, InteRecAgent~\cite{huang2023recommender} employs LLMs as the brain and recommender models as tools, combining their respective strengths to create an interactive recommender system~\cite{huang2023recommender}. As a conversational recommender system, InteRecAgent enables traditional recommender systems to become interactive systems with a natural language interface through the integration of LLMs. }

In addition to the context limitation and computational cost of LLMs~\cite{touvron2023llama2}, the paradigm that LLM improves recommendation models also encounters other problems. (1) Although LLMs can enhance offline recommender systems to avoid online latency, this paradigm also limits the ability of LLMs to model real-time collaborative filtering information, neglecting the key factor for recommendation~\cite{wang2023EnhancedRepre,xi2023towards-kar,wei2023llmrec}. (2) Feature encoding, data augmentation, and collaborative training inevitably expose the user data to LLMs, which may bring privacy, security and ethical issues~\cite{weidinger2021ethical,shen2023chatgpt-reliability,carranza2023privacy}.


\subsubsection{LLM as Recommendation Simulator}
\label{sec:related-llm-as-simulator}

Due to the gap between offline metrics and online performance of recommendation methods~\cite{pan2022multimodal,he2017neural}, it is necessary for the designed approach to get intents of users by simulating real-world elements. In this way, LLM as the recommendation simulator is introduced by taking LLMs as the foundational architecture of generative agents, and agents simulate the virtual users in the recommendation environment~\cite{wang2023recagent,zhang2023generative,zhang2023agentcf}. 

Recently, there emerged a lot of work studying the performance of LLMs as the recommendation simulator. Agent4rec~\cite{zhang2023generative} was a movie simulator consisting of two core fractions: LLM-empowered generative agents and recommendation environment. The work equipped each agent with user profile, memory and actions modules, mapping basic behaviors of real-world users. \ignore{Specifically, the user profile module of each agent was initialized by the MovieLens dataset.} AgentCF~\cite{zhang2023agentcf}, on the other hand, considered not only users but also items as agents. It captured the two-sided relations between users and items, and optimized these agents by prompting them to reflect on and adjust the misleading simulations collaboratively~\cite{zhang2023agentcf}. Moreover, in addition to behaviors within the recommender system, RecAgent~\cite{wang2023recagent,wang2023survey} took external influential factors of user agent simulation into account, such as friend chatting and social advertisement. In order to describe users accurately, RecAgent applied five features for users, and implemented two global functions including real-human playing and system intervention to operate agents flexibly. 

Although LLM as recommendation simulator aims to imitate real-world recommendation behaviors to enhance the recommendation performance, it still has deficiencies in some aspects. Firstly, since current work is mainly demo systems that operate a few agents~\cite{wang2023recagent,wang2023survey}, there still exists a gap between virtual agent environment and real-world practical recommendation applications, which requires further research and development. Additionally, LLMs may arise privacy and safety concerns. Many studies take ChatGPT as the architecture of agents, presenting security risks to the recommended information for users~\cite{zhang2023agentcf}. Moreover, Zhang \etal~\cite{zhang2023generative} have explored that hallucination in LLMs can exert huge impact on recommendation simulations. The LLM sometimes fails to accurately simulate human users, such as providing inconsistent score for an item and fabricating non-existent items for rating.

\subsection{Differences with Existing Work}
\label{sec:related-differences-with-existing}

Previous surveys of LLMs for recommender system usually categorized existing work with a classification standard and introduced studies. Lin \etal ~\cite{lin2023survey} categorized existing work into the targets and the methods of adapting LLMs to recommendation tasks. Wu \etal ~\cite{wu2023survey} mainly focused on the form of information between LLMs and recommender systems. Zhao \etal ~\cite{fan2023survey} summarized the framework into four sections, involving deep representation learning, pre-training, fine-tuning and prompting of LLMs. 
\revision{Li \etal~\cite{li2023survey} concentrated on LLMs for generative recommendation, and summarized the different recommendation tasks that LLMs can serve as in detail. Chen \etal~\cite{chen2024LLMPersonal} analyzed challenges and opportunities of LLMs in the field of personalization, and Li \etal~\cite{li2024SurveyGenSearchRec} systematically classified existing work on generative retrieval and generative recommendation. }
These surveys mainly concentrated on demonstrating related work and summarizing the advantages and limitations, \revision{while they lack performance analysis supported by experimental results}. 
\revision{Although existing work like LLMRec~\cite{liu2023llmrec} and LLM-REC~\cite{tan2024IDGenRec} has evaluated the recommendation performance of LLMs on multiple tasks, their analysis of LLMs selection and prompt design is not deep enough.}
Compared to previous work, we concentrate on the performance of LLMs leveraged as recommender systems, and provide a systematic empirical analysis on LLM-based recommendations by devising a general framework~\our. We mainly focus on two aspects, \ie LLMs and task-specific prompts, providing definitions and solutions from both conceptual and methodological perspectives. 
Furthermore, we conduct experiments to discover new findings and validate results previously discussed in existing research, serving as an inspiration for future research efforts.


%% file: sec-overall.tex
\section{GENERAL FRAMEWORK AND OVERALL SETTINGS}
\label{sec:overall}

\begin{figure}[t]
    \centering
    \includegraphics[width=0.95\linewidth]{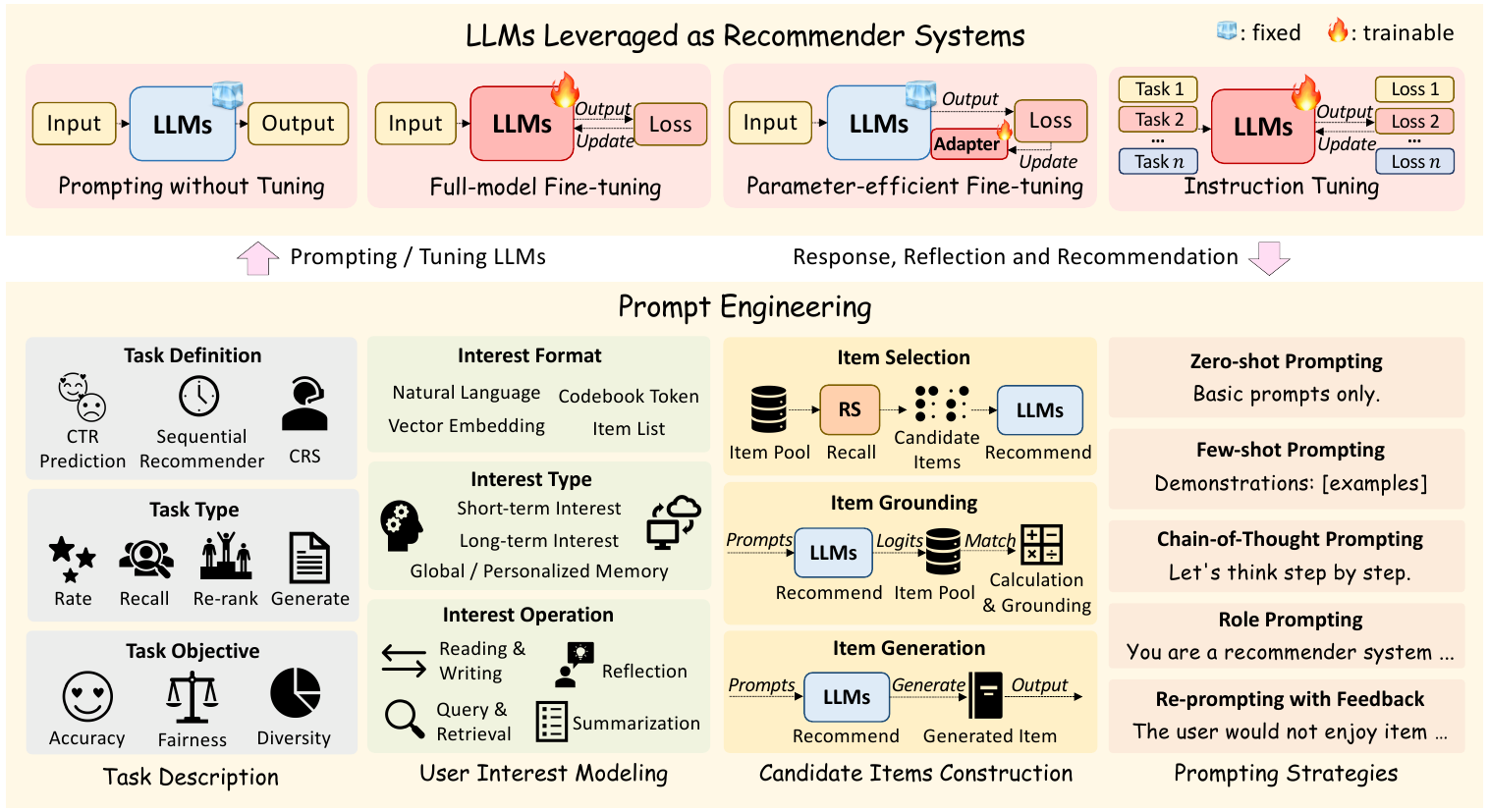}
    \caption{\revision{The overall framework of our proposed~\our. In our framework, LLMs are leveraged as recommender systems in four ways: prompting without tuning, full-model fine-tuning, parameter-efficient fine-tuning and instruction tuning; Prompt engineering consists of four components: task description, user interest modeling, candidate items construction and prompting strategies. LLMs act as recommender systems through task-specific prompts, while LLMs provide response and feedback to optimize prompts.}}
    \Description{
        The framework consists of five components: (1) LLMs Leveraged as Recommender Systems; (2) Task Description; (3) User Interest; (4) Candidate Items and (5) Prompting Strategies. (2) - (5) constitute Task-specific Prompts.
    }
    \label{fig:fig-framework}
\end{figure}

In this section, we present our proposed~\our~with two important components, namely \emph{LLMs} and \emph{prompts}. Generally speaking, the rich world knowledge and general capabilities of LLMs demonstrate the potential to develop LLM-based recommender systems. Nevertheless, it is essential to introduce appropriate prompting engineering to provide domain-specific knowledge on recommendation tasks for LLMs to serve as personalized recommenders. In the following, we first describe our general framework by defining several key elements~(Section~\ref{sec:overall-approach}). Then, we explain how our framework can be generalized to various tasks by framework instantiation~(Section~\ref{sec:overall-instant}). Finally, we introduce our overall experimental settings for further analysis~(Section~\ref{sec:overall-exp-settings}).

\subsection{Overview of the Approach}
\label{sec:overall-approach}
\ignore{
When it comes to employing large language models (LLMs) in recommendation services, the first step involves choosing suitable \textbf{LLMs} tailored to specific recommendation scenarios, with an emphasis on their distinct capabilities. Following this, it is imperative to engage in meticulous \textbf{prompt engineering} to instruct LLMs to \textbf{model user interest} and provide recommendations from a pool of \textbf{candidate items}.  Although recent work on LLM4Rec has proposed various algorithms and architectures, most of them focus on are directed towards enhancing these four key aspects~\cite{}.
Therefore, as illustrated in Eq.~\eqref{eq:eq-framework}, we propose a general framework to formulate the recommendation task in the era of large language models. In what follows, we will thoroughly examine their impact on recommendation performance.}

When it comes to leveraging LLMs as recommender systems, the first step involves choosing suitable \emph{LLMs} tailored to specific scenarios, with an emphasis on their distinct capabilities. \revision{In our framework, LLMs are leveraged as recommender systems in four ways: \emph{prompting without tuning}, \emph{full-model fine-tuning}, \emph{parameter-efficient fine-tuning} and \emph{instruction tuning}}. Subsequently, it is important to conduct \emph{prompt engineering} to instruct LLMs to perform effective recommendations. Typically, in the realm of recommender systems using LLMs, despite the task settings in existing studies vary a lot, the overall prompting format remains relatively consistent with minor variations~\cite{hou2023large,gao2023chat,dai2023uncovering,wang2023recmind}. Therefore, to unify existing prompting approaches for different recommendation purposes, as illustrated in Fig.~\ref{fig:fig-framework}, we establish a general framework for prompt engineering in~\our, consisting of four key factors in the prompts. Firstly, \emph{task description} is required to clearly express the specific goal of recommendation tasks. Secondly, prompts need to be carefully designed, to express \emph{user interest} and meanwhile enable LLMs to provide personalized recommendations with both world and domain knowledge. Thirdly, the purpose of a recommender system is to provide users with candidate items, and potential \emph{candidate items} should be constructed for LLMs to facilitate the recommendation with the understanding of domain-specific item information. Furthermore, special \emph{prompting strategies} can be further employed to enhance the specialized recommendation capabilities of LLMs.
\ignore{Given specialized prompts that include task description and
interest of users, the LLM selects, generates, or explains candidate items based on general world knowledge and
personalized user profiles. }
In what follows, we will thoroughly examine the impact of each factor on the recommendation performance.

\ignore{
When it comes to leveraging large language models for downstream tasks, task-oriented prompts are tailored to the model for desired outputs. Especially for sequential recommendation, the modeling of user interest and the construction of candidate items are key to the recommender system. During this process, large language models~(LLMs), prompt engineering, user interest modeling, and candidate items construction are the four main aspects to be considered. Therefore, we concentrate on these four aspects to discuss their influence on recommendation results as illustrated in Eq.~\eqref{eq:eq-framework}. In general, we propose a general framework to formulate the sequential recommendation task in the era of large language models and name it~\our. }


\subsubsection{Key Elements in Our Framework} To carry out the experiments, we first describe the key elements in~\our, to clarify their definitions and scope in this work. 
Specially, we introduce the following five elements for our framework: 

\paratitle{Large language models~(LLMs)}. \ignore{Recently, the advent of LLMs has brought a technological revolution in artificial intelligence~(AI). LLMs exhibit excellent capacities in language generation and task solving, which also shed light on the development of more powerful recommendation models. However, a}
As proposed in previous research~\cite{hou2023large,zhang2023collm,li2023ctrl,bao2023bi}, there exists a large gap between general language modeling and personalized user behavioral modeling, making it non-trivial to utilize LLMs in recommender systems. In this work, we investigate the efficacy of LLMs from perspectives of \emph{public availability}, \emph{tuning strategies}, \emph{model architecture}, \emph{parameter scale}, and \emph{context length}, aiming to gain insights into the selection of appropriate LLMs for performing recommendation tasks. Observations and discussions on the use of LLMs are presented in Section~\ref{sec:llm}.

\paratitle{Task description}. To adapt LLMs to the scenario of recommendation, it is necessary to clearly express the context and target of recommender systems for LLMs in the prompt, \ie task description. With different prompt descriptions of tasks, LLMs can be leveraged to various recommendation scenarios and tasks such as click-through rate predictions~\cite{kang2023llms-ctr,bao2023tallrec,bao2023bi}, sequential recommendation~\cite{hou2023large,dai2023uncovering,zheng2023LC-Rec} and conversational recommender systems~\cite{he2023zero-shot-CRS,gao2023chat,wang2023rethinkingCRS}. \revision{In addition, the well-designed task description also helps to improve the fairness~\cite{wang2023fair,zhang2023chatgptfair} and diversity~\cite{carraro2024DiverseReranking,deldjoo2024bias-llmrs} of recommendation results.}

\paratitle{User interest modeling}. The modeling of user interest is the key to recommendation tasks~\cite{rendle2009bpr,he2017neural}.   
\revision{From a temporal perspective, user interest can be categorized into short-term instant interest and long-term stable preferences, which collectively constitute the user profile~\cite{tan2024IDGenRec,liu2023llmrec}.}
When leveraging LLMs as recommender systems, users are generally expressed in natural language text~\cite{hou2023large,bao2023tallrec}, which is different from traditional approaches capturing user preference from ID-based behavior sequences~\cite{kang2018sasrec,wu2021sgl}. 
\revision{Moreover, learnable tokens~\cite{wang2024LETTER,kim2024SC-Rec,shao2024ULMRec} and 
distilled soft prompts~\cite{ramos2024PeaPOD,li2023prompt-distillation} for user interest modeling in LLMs have attracted attention of researchers.}
In this paper, we mainly consider the reflected user interest based on his or her interaction behaviors with interacted items. Especially, as detailed in Section~\ref{sec:interest}, we employ item description texts, user profiles, and historical interactions between users and items to reveal the underlying user interest using natural languages~\cite{wang2023recmind,lin2023rella,shu2023rah,yao2023DOKE}. 

\paratitle{Candidate items construction}. The purpose of recommender systems is to provide users with items to choose from, so candidate items construction is a crucial step in our framework~\cite{hou2023large,zhang2023recommendation,dai2023uncovering}. A simple approach is to provide several candidate items in prompts, \eg the items recalled by traditional recommendation models~\cite{lin2023rella,yue2023llamarec}. Due to the input length limitation of LLMs, it is not possible to include all items in the prompts. Besides selecting candidate sets, there are also methods that directly generate candidate items by LLMs, utilizing strategies such as output probability distribution~\cite{yue2023llamarec} and vector quantization~\cite{zheng2023LC-Rec} for item indexing and grounding. Section~\ref{sec:candidate} will focus on the construction strategies of candidate items, including selection, grounding and generation.
    
\paratitle{Prompting strategies}. Despite the impressive capabilities of LLMs, they tend to exhibit unsatisfactory performance in providing personalized recommendations~\cite{hou2023large,dai2023uncovering,liu2023chatgpt,kang2023llms-ctr}. The reason may stem from the significant semantic gap between the general knowledge encoded in LLMs and the domain-specific behavioral pattern and item catalogs of recommender systems~\cite{zhang2023collm,li2023ctrl}. To specialize LLMs to recommender systems, we summarize and propose several prompting strategies specialized for recommendation tasks. Details will be discussed in Section~\ref{sec:prompting-strategies}.

\begin{table}[t]
\centering
\small
\caption{\revision{Instantiation of existing work for~\our. We summarize three paradigms of leveraging LLMs as recommender systems within the framework of our proposed~\our: (1) prompting without tuning, (2) (parameter-efficient) fine-tuning, and (3) instruction tuning. For each paradigm, we provide a comprehensive overview of related work and its instantiation in our framework.}}
\resizebox{\textwidth}{!}{
\begin{tabular}{@{}p{0.09\textwidth}p{0.18\textwidth}p{0.17\textwidth}p{0.15\textwidth}p{0.18\textwidth}p{0.15\textwidth}@{}}
\toprule
\textbf{LLMs} &
  \textbf{Task Descriptions} &
  \textbf{User Interest} &
  \textbf{Candidates} &
  \textbf{Prompt Strategies} &
  \textbf{Related Work} \\ \midrule
\multicolumn{6}{c}{\cellcolor[HTML]{ECF4FF}\textit{Prompting without Tuning}} \\ \midrule
 &
  CTR predictions, rating, re-ranking &
  recent and relevant items~(with attributes), user profile &
  pointwise, pairwise, listwise item(s) &
  chain-of-thoughts, in-context learning, self-inspiring, role prompting &
  \cite{di2023evaluating,kang2023llms-ctr,dai2023uncovering,shu2023rah,hou2023large,wang2023zero,wang2023recmind,liu2023chatgpt,yao2023DOKE,zhiyuli2023bookgpt,liu2023llmrec,li2023ChatGPT-news,wang2023DRDT,liu2023recprompt, ma2023STELLA, carraro2024DiverseReranking, spurlock2024Reprompting, liu2024Rec-GPT4V, sanner2023NearCold-start}  \\ \cmidrule(l){2-6} 
\multirow{-3}{*}{\begin{tabular}[c]{@{}c@{}}\\ ChatGPT, \\ GPT-4, \\ \revision{GPT-4o}\end{tabular}} &
  conversational recommender systems &
  user explicit interest, interactive feedback &
  recalled from traditional models &
  role prompting &
  \cite{gao2023chat,friedman2023recllm,he2023zero-shot-CRS,jin2023core,wang2023rethinkingCRS,huang2023recommender} \\ \cmidrule(l){2-6} 
 &
  generative recommendation &
  recent items~(with attributes), user profile &
  (not provided, generation methods) &
  basic prompts &
  \cite{liu2023genre-news,wang2023GeneRec}  \\ \midrule
\multicolumn{6}{c}{\cellcolor[HTML]{ECF4FF}\textit{(Parameter-efficient) Fine-tuning}} \\ \midrule
\multirow{-2}{*}{\begin{tabular}[c]{@{}c@{}}\\ \\ \\ \\ LLaMA, \\ Vicuna, \\ Flan-T5, \\ BART, \\ GPT, \\ PaLM\end{tabular}} &
  CTR predictions, rating, re-ranking &
  recent and relevant items~(with attributes), user profile, collaborative embedding &
  pointwise, listwise item(s) &
  chain of thoughts, role prompting, soft prompting &
  \cite{bao2023tallrec,zhang2023collm,kang2023llms-ctr,yang2023palr,lin2023rella,yue2023llamarec,liao2023llara,wang2023key-value,di2023evaluating,li2023ChatGPT-news-finetune,fu2023uni-ctr,shi2023LSAT,qin2024PRP, wu2024GLRec, mao2023UniTRec} \\ \cmidrule(l){2-6} 
 &
  recall, retrieving tasks, conversational recommender systems &
  recent and relevant items~(with attributes) &
  (not provided, item grounding methods) &
  basic prompts &
  \cite{lin2023transrec,bao2023bi,zheng2023LC-Rec,li2023e4srec,petrov2023gptrec,li2023gpt4rec,ji2023genrec,zhang2021LLMRecSys,zhu2023cllm4rec, feng2023LLMCRS, yang2022MESE, lin2024DEALRec, rajput2023tiger, chen2024SIIT, wang2024LETTER, yang2024LEGER, li2024AlignToken, paischer2024Mender, yin2024TTDS, penha2024BrideSandR, li2024CALRec, ding2024SpecGR, kim2024SC-Rec, shao2024ULMRec, liu2024P2Rec} \\ \midrule
\multicolumn{6}{c}{\cellcolor[HTML]{ECF4FF}\textit{Instruction Tuning}} \\ \midrule
\multirow{-1}{*}{\begin{tabular}[c]{@{}c@{}}\\ (Flan-)T5, \\ LLaMA\end{tabular}} &
  rating, ranking, explanation generation, review summarization &
  recent items, user profile, short-term intentions &
  pointwise, pairwise, listwise item(s) &
  basic prompts &
  \cite{geng2022p5,zhang2023recommendation,li2023pbnr,chu2023RecSysLLM,qiu2023controlrec,luo2023recranker,li2024PAP-REC, li2023prompt-distillation, hua2023index, geng2023VIP5, wei2024UniMP, ramos2024PeaPOD, tan2024IDGenRec, wang2024RDRec} \\ \bottomrule
\end{tabular}}
\label{tab:overall-instant}
\end{table}

\subsection{Instantiation of~\our} 
\label{sec:overall-instant}

By combining the key elements mentioned above, we can instantiate various types of recommender systems in our framework with the following five steps. Specifically, (1) we can employ LLMs with varying levels of public availability, different tuning strategies, model architectures, parameter scales, and context lengths. (2) We can define a range of task description, such as retrieving, rating, recalling, and ranking. (3) Regarding user interest modeling, we can employ different types of interest, representation forms, and modeling methods. (4) When collecting candidate items, we take into account their different representation types, sources, and grounding methods. (5) Additionally, we can introduce several well-designed prompting strategies to effectively guide the recommendation capabilities of LLMs. To demonstrate the compatibility and versatility of our framework, we summarize previous work on LLM-based recommender systems in Table~\ref{tab:overall-instant} based on various settings in our framework~\our.

\subsection{Experimental Settings}
\label{sec:overall-exp-settings}

In this section, we introduce the overall settings of the following experiments. We first describe the basic information of datasets, and then present the configurations and implementations. 

\subsubsection{Datasets} The domain characteristics of movies and books are closer to the general knowledge of LLMs, which facilitates the further analysis. Considering the scale, popularity and side information of public datasets, we select two representative datasets to conduct our study, \ie MovieLens-1M~\cite{harper2015movielens} and Amazon Books~(2018)~\cite{ni2019amazon2018}. Next, we present the details of the two selected datasets:

\paratitle{MovieLens-1M}~\cite{harper2015movielens} is one of the most widely used benchmark datasets in the field of recommender systems, covering movie ratings and attributes on the website~\url{movielens.org}. We use the one million version from the MovieLens datasets, and it contains 1,000,209 ratings from 6,040 users on 3,706 movies. It is collected by the GroupLens research group.

\paratitle{Amazon Books~(2018)}~\cite{ni2019amazon2018} is an updated version of the Amazon review dataset. 
Amazon review datasets include reviews and product metadata collected from Amazon, one of the largest online e-commerce company in the United States. 
At first, Amazon only operated online book sales business, so the data in the book field is the most abundant. To improve the data quality, we filter out inactive users and unpopular products, and remove data without necessary attributes.

In our research, we are concerned about how LLMs can fully utilize the domain knowledge to make recommendations, and use the title of items as the input for prompts. However, titles are not enough to describe items, and there are deviations between the text of the title and the content of the item itself~(\eg the movie Twelve Monkeys). Therefore, we further investigate the benefits of detailed item descriptions on the recommendation effect.
As shown in Table~\ref{tab:overall-datasets}, there are no item descriptions in the original dataset of MovieLens, only the release year, title, and genre. 
If LLMs do not include knowledge about a movie, only the release year, title and genre may lead to ambiguity and hallucination. 
To enrich the movie dataset, we use the general knowledge of ChatGPT~\footnote{The URL of ChatGPT Application Programming Interface~(API): \url{https://chat.openai.com/}. Note that there are multiple versions of the ChatGPT API. In the absence of clear annotations, the ChatGPT used in this article is ``gpt3.5-turbo-4k-0613''.} to generate text descriptions for movies. We mainly use the interaction information between users and items in the recommendation dataset, as well as the title and description of items. 

\begin{table}[t]
\caption{Statistics of two public datasets for~\our. Since the original MovieLens-1M dataset does not provide description information, we used ChatGPT to generate text descriptions based on item attributes.}
\centering
\small
\begin{tabular}{@{}crrrrc@{}}
\toprule
\textbf{Dataset}      & \textbf{\#User} & \textbf{\#Item} & \textbf{\#Interaction} & \textbf{Sparsity} & \textbf{Item Attributes}                     \\ \midrule
\textbf{MovieLens-1M}        & 6,040           & 3,706           & 1,000,209              & 4.4642\%          & release year, title, genre                   \\
\textbf{Amazon Books} & 13,469          & 12,984          & 1,142,940              & 0.6536\%          & title, categories, brand, price, description \\ \bottomrule
\end{tabular}
\label{tab:overall-datasets}
\end{table}

\subsubsection{Configuration and Implementation}
As for~\our, we can evaluate the cold-start recommendation ability of LLMs in the zero-shot setting, as well as evaluate the fine-tuned performance with a few or full recommendation samples in the fine-tuning setting. Considering the typical scenarios of LLMs as recommender systems, we conduct experiments on two representative task settings, \ie (1) the zero-shot ranking task without modifying parameters of LLMs \revision{(generative recommendation)}~\cite{dai2023uncovering, hou2023large, gao2023chat, ma2023STELLA}; and (2) the Click-Through Rate~(CTR) prediction task with LLMs tuned \revision{(discriminative recommendation)}~\cite{bao2023tallrec,kang2023llms-ctr,fu2023uni-ctr,shi2023LSAT}. 

\begin{figure}[t]
\centering  
\subfigure[\revision{Zero-shot ranking task~\cite{hou2023large}.}]{   
\begin{minipage}{0.48\textwidth}
\centering    
\includegraphics[scale=0.75]{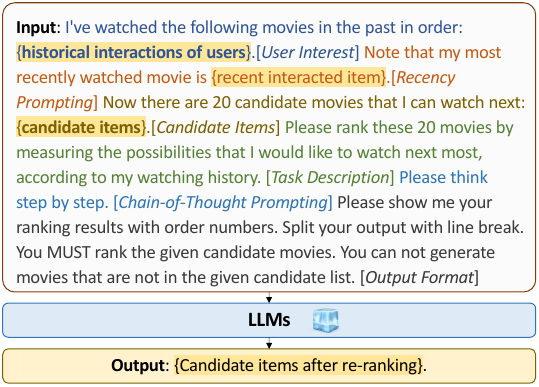}  
\end{minipage}\label{fig:basic-prompt-ranking}
}
\subfigure[\revision{Fine-tuned CTR prediction task~\cite{bao2023tallrec}.}]{ 
\begin{minipage}{0.48\textwidth}
\centering    
\includegraphics[scale=0.75]{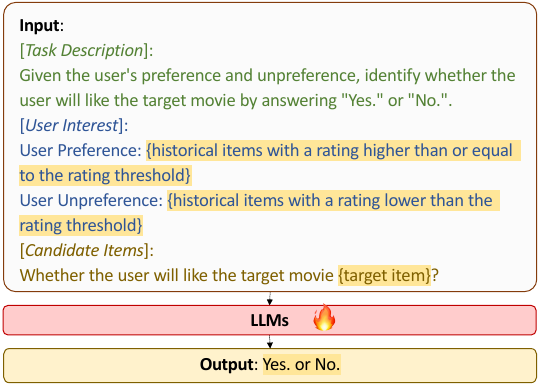}
\end{minipage}\label{fig:basic-prompt-ctr}
}
\caption{\revision{Basic prompts for experiments on the MovieLens-1M dataset in~\our. We explore two application scenarios: (1) prompting LLMs without tuning, and (2) fine-tuning LLMs as recommender systems. The experiments focus on sequential re-ranking and Click-Through Rate (CTR) prediction tasks, respectively.}}    
\label{fig:basic-prompt}    
\end{figure}


\paratitle{Zero-shot ranking task}. On the one hand, we evaluate the zero-shot recommendation performance of LLMs for cold-start scenarios to study the effect of LLMs and the design of prompts.
In this paper, our approach mainly concentrates on ranking tasks that better reflect the capabilities of LLMs~\cite{dai2023uncovering, hou2023large, gao2023chat, ma2023STELLA}.
As shown in Fig.~\ref{fig:fig-framework}, information of users and items is encoded into the prompt as inputs for LLMs. In this setting, we do not modify the parameters of LLMs, so the evaluated models are closed-source LLMs or open-source models without fine-tuning. 
To conduct experiments on the impact of each factor on recommendation results with LLMs, we implement the overall architecture based on the open-source recommendation library \textsc{RecBole}~\cite{recbole[1.0], recbole[2.0], recbole[1.1.1]} and the zero-shot re-ranker LLMRank~\cite{hou2023large}. Our basic prompt used in the MovieLens-1M dataset for the zero-shot ranking task is shown in Fig.~\ref{fig:basic-prompt-ranking}.

\paratitle{CTR prediction task with LLMs tuned}. 
On the other hand, we evaluate the fine-tuned recommendation performance of LLMs to explore how LLMs adapt to recommendation scenarios with data provided~\cite{bao2023tallrec,kang2023llms-ctr,fu2023uni-ctr,shi2023LSAT}. Although our framework can be generalized to various recommendation tasks, we concentrate on exploring the fine-tuning performance of LLMs with point-wise CTR prediction tasks to reduce selection bias. In this setting, we not only consider fine-tuning LLMs using recommendation data from scratch, but also devise a two-stage approach of using instruction data to fine-tune LLMs first, and then implement recommendation fine-tuning for further adaptation. Specifically, we compare the fine-tuned recommendation performance of the original LLM and the LLM after instruction tuning, respectively. \revision{As for the selection of LLMs, LLaMA-7B~\cite{touvron2023llama}, LLaMA2-7B~\cite{touvron2023llama2} and LLaMA3-8B~\cite{dubey2024llama3} are the original models, while Alpaca-lora-7B~\cite{alpaca} and LLaMA2-chat-7B~\cite{touvron2023llama2} are LLMs after instruction tuning.} In terms of the tuning strategies of LLMs, we report results with both \revision{Parameter-Efficient Fine-Tuning~(PEFT) and full-model} fine-tuning. We implement the fine-tuning framework based on the open-source library \texttt{transformers} and the instruction tuning code of LLaMA with Stanford Alpaca data~\footnote{The repository of Alpaca-LoRA: \url{https://github.com/tloen/alpaca-lora}.}. Our basic prompts to fine-tune LLMs for the CTR prediction task on the MovieLens-1M dataset is illustrated in Fig.~\ref{fig:basic-prompt-ctr}.

\subsubsection{Evaluation Metrics}
As for the zero-shot ranking task, considering economic and efficiency factors, we refer to existing literature~\cite{hou2023large, shu2023rah} to randomly sample 200 users instead of evaluating results on the whole dataset. 
The average result of 200 users is the final indicator for performance comparison.
For each user from the sample set, we sort all items that the user interacts with in the chronological order. Then, we evaluate the results based on the leave-one-out strategy and treat the last interacted item as the ground truth. For performance comparison, we fix the length of candidate items to 20 as in~\cite{hou2023large}, and mix the other 19 items from the ground truth in random positions by default. As for evaluation, we utilize two widely used ranking metrics in recommender systems, \ie Recall~\cite{buckland1994recall} and Normalized Discounted Cumulated Gain~(NDCG)~\cite{NDCG}. Since 20 items are selected for candidate generation, we set $k=20$ for Recall@$k$ to measure whether all ground-truth items have been recalled, \revision{and the metric Recall@$20$ can also be expressed as Coverage}. 
In existing literature, the re-generation method of multiple results until the format requirements are met can be employed to obtain the final response~\cite{li2023ChatGPT-news-finetune}, while we consider the recall capability within one inference for fair evaluation.
Furthermore, we set $k=1,10,20$ for NDCG to explore the detailed recommendation performance in terms of ranking abilities.
For a scientific research, we repeat each experiment three times and take the average value as the results.

As for the CTR prediction task, we first sort the original dataset by timestamp, use the latest 10,000 records for training and evaluation, and regard other previous data as the interaction history of users. Then, we split the interactions into the training, validation and test sets in ratio of 8:1:1. For each interaction, we retain 10 historical interacted items as user representations, and set thresholds based on rating data to obtain user preferences~\cite{bao2023tallrec}. Interactions with a rating higher than or equal to the threshold will be considered items that the user likes, while the opposite indicates dislikes. For the MovieLens-1M dataset, the rating threshold is 4, and we set 5 for the Amazon Books dataset. We employ the training data to fine-tune LLMs and evaluate the recommendation performance on the test set. The validation set is used for selecting best checkpoints during the training process. As for evaluation of prediction, we utilize the widely used metric for CTR predictions, \ie accuracy. In a random state, the accuracy is around 0.5.


\subsubsection{Discussion on Variable Factors}

To conduct our analysis for~\our, we mainly focus on using LLMs as recommender systems with two key aspects, \ie LLMs and prompts. 
As for the effects of LLMs, we analyze the impact of 
public availability, tuning strategies, model architecture, parameter scale, and context length on recommendation results based on the classification of LLMs. As for prompt engineering, we further analyze the impact of four important components of prompts, \ie task description, user interest modeling, candidate items construction and prompting strategies. 
Due to the fact that all factors have an impact on the final result, it is also crucial to select the other aspects when focusing on one aspect. 
Limited by resources and efficiency, it is neither necessary nor feasible to exhaust all possibilities. 
When not explicitly specified, the LLM uses ChatGPT released in June 2023 to ensure the consistent recommendation quality. For the design of prompts, we refer to the template in~\cite{hou2023large} and emphasize the most recently interacted items to re-rank 20 candidate items. The default method for modeling user interest is to concatenate the title sequence of recently interacted items using natural languages. 

%% file: sec-llm.tex
\section{THE IMPACT OF LARGE LANGUAGE MODELS AS RECOMMENDER SYSTEMS}
\label{sec:llm}

LLMs are the core of our framework~\our, and largely determine the performance of LLMs as recommender systems~\cite{liu2023llmrec,liu2023chatgpt,kang2023llms-ctr,hou2023large}. Therefore, it is worth exploring how to choose a suitable LLM as the foundation for recommendation. In this section, we compare the differences between LLMs and traditional models on the recommendation performance, discuss how the different properties and tuning strategies of LLMs affect the recommendation results, present the limitations of LLMs as recommenders, and draw empirical conclusions through systematic experiments.

\subsection{Classification of LLMs}

In this paper, we consider language models that have a size larger that one billion as LLMs. In line with existing researches~\cite{zhao2023survey,wu2023survey,lin2023survey,li2023survey,fan2023survey}, LLMs can be categorized into different classes from several perspectives. 
Most typically, LLMs can be divided into open-source and closed-source models in terms of the public availability. When it comes to leveraging LLMs as the foundation model in recommender systems, tuning strategies can adjust LLMs towards specific recommendation tasks. From the perspective of the model architecture, various LLMs can also be categorized into types of encoder-decoder, causal decoder, and prefix decoder~\cite{zhao2023survey}. For the same framework of a LLM, it is widely acknowledged that the parameter scale and context length are two key factors that jointly determine capabilities of LLMs~\cite{lin2023rella,hou2023large}. 
\ignore{On the one hand, the scale of parameters determines how complex calculations the model can support. Research has shown that only when the parameters exceed a certain scale will LLMs emerge with abilities that small models do not possess~\cite{zhao2023survey}. On the other hand, LLMs also need a sufficiently large context length to support practical applications in addition to the parameter scale~\cite{lin2023rella,hou2023large}. The length of text input determines how much ``memory'' the language model has, and has become the core standard of the new generation of LLMs. }
To explore the recommendation performance \wrt different variants of LLMs, we focus on five aspects, \ie \emph{public availability}, \emph{tuning strategies}, \emph{model architecture}, \emph{parameter scale} and \emph{context length} as follows.

\subsubsection{Public Availability} According to whether the model checkpoints can be publicly obtained, existing LLMs can be divided into open-source models and closed-source models. In what follows, we will explain that both the two categories of LLMs can be leveraged as recommender systems. 

\paratitle{Open-source models} refer to LLMs whose model checkpoints can be publicly accessible. As shown in Table~\ref{tab:overall-instant}, researchers often use recommendation data to fine-tune open-source models for performance improvement. As the representative of open-source models, LLaMA~\cite{touvron2023llama} and its variants like Vicuna~\cite{vicuna2023} are widely used when leveraging LLMs for recommender systems~\cite{zheng2023LC-Rec,li2023e4srec}. Parameter-Efficient Fine-Tuning~(PEFT) strategies such as Low-Rank Adaptation~(LoRA)~\cite{hu2021lora} are frequently adopted for recommendation data considering the trade-off between effect and efficiency~\cite{liao2023llara,bao2023tallrec}. Other models such as the Flan-T5~\cite{chung2022flan-t5} series from Google Inc., \revision{ChatGLM~\cite{zhipu2024chatglm} from Zhipu AI and Qwen~\cite{yang2024qwen2} from Alibaba} are also popular in the field of recommender systems. The publicly available checkpoints of open-source models provide flexibility for LLMs to modify parameters \revision{tailored for recommendation tasks}.

\paratitle{Closed-source models} refer to LLMs whose model checkpoints can not be publicly accessible. For the closed-source LLMs leveraged as recommender systems, researchers generally study the zero-shot recommendation ability in cold-start scenarios. The most typical closed-source model is the ChatGPT series from OpenAI. The subsequent \revision{GPT-4~\cite{openai2023GPT4} and GPT-4o~\cite{openai2024gpt4o} have} stronger capabilities compared to ChatGPT, but they are still not open-source. In this paper, ChatGPT refers to the API function of ``gpt3.5-turbo-4k-0613'' unless specified. Without checkpoints, OpenAI provides several ways for researchers and users to improve the model performance on specific tasks, such as plugins for website browsing~\cite{lewis2020rag} and interfaces for fine-tuning ChatGPT~\cite{li2023ChatGPT-news-finetune}. However, the flexibility of closed-source models as recommender systems is still limited due to the expensive price and black-box parameters. Faced with this challenge, existing literature has explored to inject knowledge of recommender systems into closed-source models by means of prompt design~\cite{yao2023DOKE,wang2023zero}, retrieval enhancement~\cite{huang2023recommender,wang2023recmind,shu2023rah} and combination of traditional recommendation models~\cite{wang2023rethinkingCRS,gao2023chat}.

\subsubsection{Tuning Strategies}
During the deployment of LLMs in recommender systems, we can also classify existing work depending on whether LLMs are fine-tuned, as well as the various fine-tuning strategies employed, \ie \emph{prompting without tuning}, \emph{fine-tuning} and \emph{instruction tuning}.

\paratitle{Prompting without tuning} means evaluating the zero-shot recommendation ability of LLMs, which is generally used for closed-source models such as ChatGPT. By designing prompt templates, LLMs without fine-tuning can be used for recommendation tasks such as CTR predictions~\cite{di2023evaluating,kang2023llms-ctr}, sequential recommendation~\cite{hou2023large,wang2023recmind}, and Conversational Recommender Systems~(CRS)~\cite{he2023zero-shot-CRS,wang2023rethinkingCRS,jin2023core}. In this case, the user interest is expressed explicitly~(\eg ratings and reviews) or implicitly~(interacted items of users), and the limited candidate items can be recalled by traditional models, while prompting strategies such as In-Context Learning~(ICL) and Chain-of-Thoughts~(CoT) are used. Inspired by Artificial Intelligence Generated Content~(AIGC), it is worth noting that the excellent generation ability of LLMs provides opportunities for generative recommendation~\cite{liu2023genre-news,wang2023GeneRec}. Without providing candidate items, generative language models can generate the desired items that users need based on recommendation requirements, which can also be generalized into our framework. 

\paratitle{Fine-tuning} LLMs means that using recommendation data to fine-tune LLMs as recommender systems. Considering cost and efficiency, researchers often use parameter-efficient fine-tuning~(\eg LoRA~\cite{hu2021lora}) to adapt to recommendation scenarios~\cite{bao2023tallrec,liao2023llara,zheng2023LC-Rec}. As for LLMs, open-source models based on LLaMA~\cite{touvron2023llama} are widely used, including but not limited to LLaMA, LLaMA2~\cite{touvron2023llama2}, LLaMA3~\cite{dubey2024llama3} and Vicuna~\cite{vicuna2023} with different parameter sizes. Based on whether candidate items are provided, existing work on fine-tuning LLMs can be further divided into two kinds of recommendation tasks. On the one hand, researchers have explored fine-tuning models for recommendation that provide candidate items such as rating, re-ranking and CTR predictions. Specifically, the fine-tuning interface of the closed-source model ChatGPT has brought new breakthroughs to the research of LLMs, and there have been attempts to fine-tune ChatGPT for recommendation tasks~\cite{li2023ChatGPT-news-finetune}. On the other hand, LLMs can also be fine-tuned for recall tasks of recommender systems by retrieving candidates from the whole item pool~\cite{bao2023bi,lin2023transrec,petrov2023gptrec,li2023e4srec}. Through well-designed indexing, alignment, and retrieval strategies, generating recommendation items without providing lengthy candidate sequences is suitable for practical application scenarios~\cite{hua2023index,chen2024SIIT}.

\paratitle{Instruction tuning} means providing template instructions of recommenders as prompts to tuning targeted LLMs, generally involving multiple recommendation tasks~\cite{zhang2023recommendation,geng2022p5,qiu2023controlrec,chu2023RecSysLLM,tan2024IDGenRec}. However, some fine-tuning methods~(\eg TALLRec~\cite{bao2023tallrec}) also involve the form of instructions. In order to avoid ambiguity, we consider the instructions of a single task template as fine-tuning settings, and the instructions of multiple tasks as instruction-tuning settings. As the backbone of recommender systems, researchers generally use T5~\cite{raffel2020T5} or Flan-T5~\cite{chung2022flan-t5} for various recommendation scenarios such as rating, ranking, retrieving, explanation generation and news recommendation. Besides Flan-T5, RecRanker~\cite{luo2023recranker} is also proposed to instruction tuning LLaMA2~\cite{touvron2023llama2} as the ranker for top-$k$ recommendation. By instantiating appropriate instructions for each recommendation task, this setting can also be summarized into our framework.

\subsubsection{Model Architecture} For the architecture design of LLMs leveraged as recommender systems, we consider three mainstream architectures as summarized in~\cite{zhao2023survey}, \ie \emph{encoder-decoder}, \emph{prefix decoder}, and \emph{causal decoder}. The recommendation scenarios for each framework are introduced. 

\paratitle{The encoder-decoder architecture} adopts two Transformer blocks as the encoder and decoder, which is the basis of BERT~\cite{devlin2019bert}. In line with existing literature on LLM-based recommender systems, the bidirectional property of the encoder-decoder architecture allows LLMs to easily customize encoders and decoders towards recommendation~(\eg dual-encoder considering both ids and text~\cite{qiu2023controlrec}), and conveniently adapt to multiple recommendation tasks~\cite{chu2023RecSysLLM,li2023pbnr,geng2022p5,zhang2023recommendation,qiu2023controlrec}. 

\paratitle{The prefix decoder architecture} is also the decoder-only architectures, which is known as non-causal decoder. It can bidirectionally encode the prefix tokens like the encoder-decoder architecture, and perform unidirectional attention on the generated tokens like the causal decoder architecture. One of the representative LLMs based on the prefix decoder architecture is ChatGLM~\cite{zhipu2024chatglm}, and researchers have attempted to explore the recommendation performance of ChatGLM as one of the benchmarks in related work~\cite{liu2023llmrec}. \revision{Note that the variants of ChatGLM, \ie ChatGLM2 and ChatGLM3 employ the causal decoder architecture.}

\paratitle{The causal decoder architecture} has been widely adopted in various LLMs, and the series of GPT~\cite{brown2020language} and LLaMA~\cite{touvron2023llama} are the most representative models. It uses the unidirectional attention mask, and only decoders are deployed to process both the input and output tokens. Due to the popularity of the causal decoder architecture, most LLM-based recommender systems employ this framework to adapt to different recommendation tasks such as CTR predictions and sequential recommendation.



\subsubsection{Parameter Scale} To meet the diverse needs of different users, the most typical variant of LLMs is the parameter scale. In general, open-source models have multiple parameter sizes to choose from, and larger parameter sizes generally mean better capabilities~\cite{touvron2023llama,zhao2023survey}. But meanwhile, the corresponding computational and spatial complexity will also increase. Considering the memory and efficiency issues for experiments, researchers in the field of LLM-based recommender systems generally use LLMs with parameters no more than 10B~(\revision{B is short for billion and the same below}), while the performance of LLMs with larger parameters remains to be further explored in the field of recommender systems~\cite{kang2023llms-ctr}.

\subsubsection{Context Length} Another property closely related to user needs is the length of the input context. The inability to handle inputs with longer contexts means that decision-making cannot be accurately developed, thereby limiting the model capabilities~\cite{zhao2023survey,touvron2023llama2}. When the input exceeds the limited length, the input will be truncated, so sufficient context length is crucial for the user experience~\cite{hou2023large,dai2023uncovering,ma2023STELLA}. However, different lengths of context inputs imply different model architectures and parameters. When expanding the context length of a model, it often leads to higher time and memory complexity. To address the length limitation of LLMs, existing methods either selectively discard previous contexts using sliding windows~\cite{xiao2023StreamingLLM}, or only sample a portion of the context for retrieval augmentation~\cite{lewis2020rag,lin2023rella}, or employ small models without emergence ability. Despite recent strategies, the limitation of context length has not yet been truly resolved. Considering economic and efficiency issues, existing LLMs including open-source and closed-source models only provide a limited number of context length options. In this paper, we mainly focus on several classic lengths, \ie 2K, 4K, 8K, 16K, 32K \revision{and 128K}~(K is the abbreviation for one thousand, the same below).

\subsection{Research Questions and Experimental Setup} 

\subsubsection{Research Questions} In this section, we conduct experiments to verify the impact of different factors of LLMs on recommendation results focusing on the following four Research Questions~\revision{(RQ)}.

\begin{itemize}
    \item RQ1: What are the performance differences between LLMs and traditional recommenders?
    \item RQ2: How do different attributes of LLMs, including the public availability, model architectures, parameter scales, and context lengths affect the performance and inference time?
    \item RQ3: What are the similarities and differences in recommendation results with different tuning strategies? Is the LLM after instruction tuning more suitable for recommendation?
    \item RQ4: What are the limitations of leveraging LLMs as recommender systems?
\end{itemize}

\revision{
\subsubsection{Experimental Setup} This section considers two prevalent recommendation scenarios based on the tuning status of LLMs. To address RQ1, RQ2, and RQ4, we evaluate the zero-shot ranking efficacy of LLMs by using prompts as illustrated in Fig.~\ref{fig:basic-prompt-ranking}. In the context of RQ3, we fine-tune LLaMA-based LLMs to act as CTR predictors, with the corresponding data sample presented in Fig.~\ref{fig:basic-prompt-ctr}. Details regarding the experimental configurations are available in Section~\ref{sec:overall-exp-settings}.}

\subsubsection{Evaluated Models}

For empirical analysis, we assess the following baselines and LLMs.

\begin{itemize}
    \item \textbf{Random}: Random baseline recommends the $k$~(k=20 in this section) candidate items in a random order, which is the basic situation to evaluate the metric values of each dataset. 
    \item \textbf{Pop}: Pop ranks the candidate items based on their interaction times in the training set. We consider it as the fully-trained method since it uses the statistical information of datasets.
    \item \textbf{BPR}~\cite{rendle2009bpr}: BPR is one of the typical traditional models that utilize matrix factorization for recommendation. It is trained in the pair-wise paradigm \aka BPR training loss without considering temporal information.
    \item \textbf{SASRec}~\cite{kang2018sasrec}: SASRec is a sequential recommendation model based on the backbone of the classic self-attention network \aka Transformer~\cite{vaswani2017bert}, and achieves comparable performance among sequential models. It utilizes the unidirectional attention mask for model training and achieves comparable performance among sequential models. 
    \item \textbf{ChatGPT}: ChatGPT is a closed-source large-scale pre-trained language model developed by OpenAI, and we adopt the version on June 13, 2023, the same as GPT-4. We explore two context lengths of ChatGPT, \ie 4K and 16K.
    \ignore{It is part of the GPT~(Generative Pre-Trained Transformer) series of models aimed at understanding and generating human-like language by learning a large amount of textual data. Note that OpenAI has released interfaces of ChatGPT on March 1 and June 13, 2023, respectively. Considering the up-to-date requirements, we adopt the version on June 13, 2023 for recency, the same as GPT-4.}
    \item \textbf{GPT-4}~\cite{openai2023GPT4}: \revision{GPT-4 is a closed-source LLM with the context length of 8K. Building upon the success of ChatGPT, represents the next evolutionary leap in LLMs by OpenAI, delivering unprecedented capabilities in natural language processing.} 
    \item \revision{\textbf{GPT-4o}~\cite{openai2024gpt4o}: GPT-4o is a state-of-the-art closed-source LLM developed by OpenAI, released in 2024. It features a 128K token context window, significantly enhancing its ability to handle extended conversations and complex tasks requiring long-term memory. }
    \item \revision{\textbf{GPT-4o mini}: GPT-4o mini is a cost-efficient version of the GPT-4o model~\cite{openai2024gpt4o}, designed to deliver high-performance natural language processing capabilities with reduced computational requirements.}
    \item \revision{\textbf{OpenAI o1 mini}: OpenAI o1 mini is a closed-source lightweight reasoning model designed for applications that prioritize reasoning over broad factual knowledge, offering faster response time than GPT-4o. Both GPT-4o mini and OpenAI o1 mini support a 128K token context window.}
    \item \textbf{Flan-T5}~\cite{chung2022flan-t5}: Flan-T5 is an open-source language model based on the encoder-decoder architecture T5 released by Google~\cite{raffel2020T5}. Flan-T5 is extended from T5 by a multi-task fine-tuning paradigm \ie instruction tuning to enhance the generalization of different tasks. There are multiple variants of Flan-T5 in terms of parameters, including Flan-T5-Small~(80M), Flan-T5-Base~(250M), Flan-T5-Large~(780M), Flan-T5-XL~(3B) and Flan-T5-XXL~(11B). Since the first three models are too small to meet the requirements of LLMs discussed in this paper~(1B, B is short for billion and the same below), we consider the Flan-T5-XL and Flan-T5-XXL for comparison. All variants of Flan-T5 have a 512-token context length limit.
    \item \textbf{ChatGLM}~\cite{zeng2022glm}: ChatGLM is an open-source bilingual dialogue LLM that supports both Chinese and English, based on the General Language Model~(GLM)~\cite{zeng2022glm} with the prefix decoder architecture. The team released the second version ChatGLM2 and the third version ChatGLM3 in June 2023 and October 2023, respectively. \revision{Note that ChatGLM2 and ChatGLM3 have replaced the model architecture with a casual encoder. All three versions of open-source ChatGLM have a parameter scale of 6B. As for the context length, ChatGLM-6B supports 2K tokens while ChatGLM2-6B can cover 32K tokens at one time. Moreover, ChatGLM3-6B has both 2K and 32K versions.} 
    \item \textbf{LLaMA}~\cite{touvron2023llama}: LLaMA is an open-source language model introduced by MetaAI from the causal decoder architecture with four sizes~(7B, 13B, 33B and 65B). Due to its outstanding performance and low computational cost, LLaMA has received much attention from researchers so far.
    \item \textbf{Vicuna}~\cite{vicuna2023}: Vicuna~\cite{vicuna2023} is one of the most popular variants by extending LLaMA \revision{, and it has a context length of 4K tokens with two parameter scales~(7B and 13B).}
    \item \textbf{LLaMA2}~\cite{touvron2023llama2}: LLaMA2 is the next generation of LLaMA released in July, 2023. We also conduct experiments of the chatting version of LLaMA2~(\ie LLaMA2-chat)~\cite{touvron2023llama2}, and it is specifically tuned for the dialogue scenario by Reinforcement Learning with Human Feedback~(RLHF). \revision{Both LLaMA2 and LLaMA2-chat have three parameter scales~(7B, 13B and 70B), and the context length is upgraded from 2K in LLaMA to 4K in LLaMA2. }
    \item \revision{\textbf{LLaMA3}~\cite{dubey2024llama3}: LLaMA3 is an evolution in the LLaMA series developed by Meta, released in 2024. It features a dense Transformer~\cite{vaswani2017bert} architecture with up to 405 billion parameters and supports a context window of 128K tokens, enabling it to handle long and complex text sequences effectively. We mainly use two parameter scales of LLaMA3~(8B and 70B).}
    \item \revision{\textbf{Qwen2.5}~\cite{yang2024qwen2}: Qwen2.5 is a highly advanced series of LLMs developed by Alibaba, featuring a context length of up to 128K tokens and supporting outputs of up to 8K tokens, with model sizes ranging from 0.5B to 72B parameters, and we employ three parameter scales~(7B, 32B and 72B).}
    \item \revision{\textbf{DeepSeek-V3}~\cite{liu2024deepseekv3}: DeepSeek-V3 is an open-source LLM developed by DeepSeek, featuring a Mixture-of-Experts (MoE) architecture with 671 billion total parameters, of which 37 billion are activated per token. It also supports a 128K token context window.}
    \item \revision{\textbf{DeepSeek-R1}~\cite{liu2024deepseekv3}: DeepSeek-R1 is a reasoning model developed by DeepSeek with long-chain reasoning capabilities and transparent slow-thinking process. DeepSeek-R1 is expected to be open-sourced in subsequent iterations, and we treat it as an open-source model in Table~\ref{tab:overall-llms}. The results are derived from calls made through the official platform~\footnote{The URL of DeepSeek API: \url{https://chat.deepseek.com/}}.}
\end{itemize}


\begin{table}[t]
\caption{Overall performance of different models on recommendation. We consider fully-trained settings for traditional models and zero-shot settings for LLMs. ``IT'' denotes the average inference time for each user measured in seconds~(s). The best results are highlighted in \textbf{bold}, while the runner-up results are {\ul underlined}.} 
\centering
\resizebox{0.99\textwidth}{!}{
\begin{tabular}{@{}ccccccccccc@{}}
\toprule
 &
   &
   &
  \multicolumn{4}{c}{\textbf{MovieLens-1M}} &
  \multicolumn{4}{c}{\textbf{Amazon Books}} \\ \cmidrule(l){4-11} 
\multirow{-2}{*}{\textbf{Model}} &
  \multirow{-2}{*}{\textbf{\begin{tabular}[c]{@{}c@{}}Context\\ Length\end{tabular}}} &
  \multirow{-2}{*}{\textbf{\begin{tabular}[c]{@{}c@{}}Param.\\ Size\end{tabular}}} &
  \textbf{Coverage} &
  \textbf{NDCG@1} &
  \textbf{NDCG@10} &
  \textbf{IT (s)} &
  \textbf{Coverage} &
  \textbf{NDCG@1} &
  \textbf{NDCG@10} &
  \textbf{IT (s)} \\ \hline
\multicolumn{11}{c}{\cellcolor[HTML]{ECF4FF}\textit{Fully-trained Settings for Traditional Models}} \\ \hline
\textbf{Random} &
  - &
  0 &
  \textbf{1.0000} &
  0.0300 &
  0.2081 &
  0.01 &
  \textbf{1.0000} &
  0.0350 &
  0.2628 &
  0.01 \\ \midrule
\textbf{Pop} &
  - &
  1 &
  \textbf{1.0000} &
  0.1800 &
  0.4841 &
  0.03 &
  \textbf{1.0000} &
  0.1000 &
  0.2672 &
  0.03 \\ \midrule
\textbf{BPR} &
  - &
  \textless 1M &
  \textbf{1.0000} &
  0.2550 &
  0.5743 &
  0.04 &
  \textbf{1.0000} &
  0.2950 &
  0.6236 &
  0.04 \\ \midrule
\textbf{SASRec} &
  - &
  \textless 1M &
  \textbf{1.0000} &
  \textbf{0.6400} &
  \textbf{0.7916} &
  1.07 &
  \textbf{1.0000} &
  \textbf{0.6800} &
  \textbf{0.8305} &
  1.49 \\ \hline
\multicolumn{11}{c}{\cellcolor[HTML]{ECF4FF}\textit{Zero-shot Settings for LLMs}} \\ \hline
\multicolumn{11}{c}{Closed-source LLMs} \\ \midrule
 &
  4K &
  - &
  0.9583 &
  0.1817 &
  0.3985 &
  - &
  0.9850 &
  0.2467 &
  0.4276 &
  - \\ \cmidrule(l){2-11} 
\multirow{-2}{*}{\textbf{ChatGPT}} &
  16K &
  - &
  0.9600 &
  0.1500 &
  0.3735 &
  - &
  0.9800 &
  0.2400 &
  0.4032 &
  - \\ \midrule
\textbf{GPT-4} &
  8K &
  - &
  0.9900 &
  0.3100 &
  0.5828 &
  - &
  \textbf{1.0000} &
  0.3300 &
  0.5631 &
  - \\ \midrule
\textbf{\revision{OpenAI o1 mini}} &
  \revision{128K} &
  \revision{-} &
  {\ul \revision{0.9950}} &
  \revision{0.3200} &
  \revision{0.5729} &
  \revision{-} &
  \revision{0.9900} &
  \revision{0.2300} &
  \revision{0.4366} &
  \revision{-} \\ \midrule
\textbf{\revision{GPT-4o mini}} &
  \revision{128K} &
  \revision{-} &
  \revision{0.9850} &
  \revision{0.3200} &
  \revision{0.5453} &
  \revision{-} &
  \revision{0.9700} &
  \revision{0.2950} &
  \revision{0.5108} &
  \revision{-} \\ \midrule
\textbf{\revision{GPT-4o}} &
  \revision{128K} &
  \revision{-} &
  {\ul \revision{0.9950}} &
  {\ul \revision{0.3850}} &
  {\ul \revision{0.6222}} &
  \revision{-} &
  \textbf{\revision{1.0000}} &
  \revision{0.3450} &
  \revision{0.5938} &
  \revision{-} \\ \midrule
\multicolumn{11}{c}{Open-source LLMs with the Encoder-decoder Architecture} \\ \midrule
 &
   &
  3B &
  0.0050 &
  0.0000 &
  0.0016 &
  3.51 &
  0.0000 &
  0.0000 &
  0.0000 &
  4.33 \\ \cmidrule(l){3-11} 
\multirow{-2}{*}{\textbf{Flan-T5}} &
  \multirow{-2}{*}{0.5K} &
  11B &
  0.0050 &
  0.0000 &
  0.0016 &
  5.21 &
  0.0000 &
  0.0000 &
  0.0000 &
  8.28 \\ \midrule
\multicolumn{11}{c}{Open-source LLMs with the Prefix Decoder Architecture} \\ \midrule
\textbf{ChatGLM} &
  2k &
  6B &
  0.7750 &
  0.0300 &
  0.1945 &
  19.12 &
  0.7000 &
  0.0350 &
  0.2026 &
  19.52 \\ \midrule
\multicolumn{11}{c}{Open-source LLMs with the Causal Decoder Architecture} \\ \midrule
\textbf{ChatGLM2} &
  32K &
  6B &
  0.1900 &
  0.0450 &
  0.0885 &
  11.06 &
  0.1600 &
  0.0250 &
  0.0680 &
  13.62 \\ \midrule
 &
  2k &
  6B &
  0.6900 &
  0.0950 &
  0.2762 &
  7.95 &
  0.6550 &
  0.0450 &
  0.2273 &
  10.79 \\ \cmidrule(l){2-11} 
\multirow{-2}{*}{\textbf{ChatGLM3}} &
  32K &
  6B &
  0.7750 &
  0.0700 &
  0.2579 &
  14.06 &
  0.7050 &
  0.0550 &
  0.2068 &
  16.89 \\ \midrule
 &
   &
  7B &
  0.2700 &
  0.0350 &
  0.1068 &
  9.11 &
  0.2650 &
  0.0200 &
  0.0992 &
  9.35 \\ \cmidrule(l){3-11} 
 &
   &
  13B &
  0.2500 &
  0.0300 &
  0.1028 &
  9.51 &
  0.2250 &
  0.0250 &
  0.0867 &
  9.94 \\ \cmidrule(l){3-11} 
 &
   &
  33B &
  0.3900 &
  0.0400 &
  0.1328 &
  92.88 &
  0.2950 &
  0.0350 &
  0.1015 &
  106.61 \\ \cmidrule(l){3-11} 
\multirow{-4}{*}{\textbf{LLaMA}} &
  \multirow{-4}{*}{2k} &
  65B &
  0.8300 &
  0.0450 &
  0.1913 &
  171.39 &
  0.3750 &
  0.0500 &
  0.1253 &
  182.51 \\ \midrule
 &
   &
  7B &
  0.2650 &
  0.0500 &
  0.1078 &
  8.17 &
  0.4400 &
  0.0550 &
  0.1559 &
  11.12 \\ \cmidrule(l){3-11} 
\multirow{-2}{*}{\textbf{Vicuna}} &
  \multirow{-2}{*}{2k} &
  13B &
  0.4100 &
  0.0550 &
  0.1507 &
  9.26 &
  0.4800 &
  0.0700 &
  0.1813 &
  12.72 \\ \midrule
 &
   &
  7B &
  0.2350 &
  0.0500 &
  0.0888 &
  9.97 &
  0.5600 &
  0.0650 &
  0.1648 &
  10.01 \\ \cmidrule(l){3-11} 
 &
   &
  13B &
  0.4500 &
  0.0700 &
  0.1215 &
  15.64 &
  0.5100 &
  0.0750 &
  0.2150 &
  14.44 \\ \cmidrule(l){3-11} 
\multirow{-3}{*}{\textbf{LLaMA2}} &
  \multirow{-3}{*}{4K} &
  70B &
  0.7600 &
  0.1250 &
  0.2918 &
  24.38 &
  0.6600 &
  0.1150 &
  0.2912 &
  24.63 \\ \midrule
 &
   &
  7B &
  0.7950 &
  0.0900 &
  0.2744 &
  9.80 &
  0.7050 &
  0.1500 &
  0.3217 &
  9.86 \\ \cmidrule(l){3-11} 
 &
   &
  13B &
  0.8050 &
  0.1650 &
  0.3866 &
  14.80 &
  0.7500 &
  0.1650 &
  0.3411 &
  12.25 \\ \cmidrule(l){3-11} 
\multirow{-3}{*}{\textbf{\begin{tabular}[c]{@{}c@{}}LLaMA2\\ (chat)\end{tabular}}} &
  \multirow{-3}{*}{4K} &
  70B &
  0.9550 &
  0.2430 &
  0.4344 &
  23.07 &
  0.7850 &
  0.2100 &
  0.3827 &
  25.01 \\ \midrule
&
   &
  \revision{8B} &
  \revision{0.9250} &
  \revision{0.1850} &
  \revision{0.4234} &
  \revision{13.98} &
  \revision{0.9600} &
  \revision{0.1750} &
  \revision{0.3331} &
  \revision{14.04} \\ \cmidrule(l){3-11} 
\multirow{-2}{*}{\textbf{\revision{LLaMA3}}} &
  \multirow{-2}{*}{\revision{128K}} &
  \revision{70B} &
  \textbf{\revision{1.0000}} &
  \revision{0.3100} &
  \revision{0.5385} &
  \revision{22.29} &
  \revision{0.9900} &
  \revision{0.2300} &
  \revision{0.4144} &
  \revision{24.57} \\ \midrule
 &
   &
  \revision{7B} &
  \revision{0.6300} &
  \revision{0.0450} &
  \revision{0.2793} &
  \revision{2.72} &
  \revision{0.6250} &
  \revision{0.0400} &
  \revision{0.2654} &
  \revision{4.65} \\ \cmidrule(l){3-11} 
 &
   &
  \revision{32B} &
  {\ul \revision{0.9950}} &
  \revision{0.0450} &
  \revision{0.4234} &
  \revision{13.15} &
  \revision{0.9850} &
  \revision{0.0400} &
  \revision{0.3749} &
  \revision{20.50} \\ \cmidrule(l){3-11} 
\multirow{-3}{*}{\textbf{\revision{Qwen2.5}}} &
  \multirow{-3}{*}{\revision{128K}} &
  \revision{72B} &
  {\ul \revision{0.9950}} &
  \revision{0.0450} &
  \revision{0.4301} &
  \revision{24.89} &
  \revision{0.9850} &
  \revision{0.0400} &
  \revision{0.3756} &
  \revision{37.85} \\ \midrule
\textbf{\revision{DeepSeek-V3}} &
  \revision{128K} &
  \revision{671B} &
  \textbf{\revision{1.0000}} &
  \revision{0.3600} &
  \revision{0.5833} &
  \revision{4.48} &
  {\ul \revision{0.9950}} &
  {\ul \revision{0.4200}} &
  {\ul \revision{0.6296}} &
  \revision{5.25} \\ \midrule
\textbf{\revision{DeepSeek-R1}} &
  \revision{128K} &
  \revision{671B} &
  \revision{0.9850} &
  \revision{0.3000} &
  \revision{0.5270} &
  \revision{17.08} &
  \revision{0.9800} &
  \revision{0.2350} &
  \revision{0.4321} &
  \revision{24.33} \\ \bottomrule
\end{tabular}}
\label{tab:overall-llms}
\end{table}

\subsection{Observations and Discussion}

\subsubsection{LLMs Compared to Traditional Recommenders~(RQ1)} Compared to traditional models based on collaborative filtering of interacted data in fully-trained settings, we provide the cold-start recommendation performance of LLMs in zero-shot settings. In what follows, we introduce the empirical findings on the recommendation effect of different models from three aspects, \ie \emph{recommendation performance}, \emph{the impact of historical item sequences} and \emph{inference time}.

\paratitle{Recommendation performance of LLMs}. As shown in Table~\ref{tab:overall-llms}, we provide the fully-trained results of four traditional methods, as well as the zero-shot recommendation performance of various LLMs. For traditional recommenders, BPR~\cite{rendle2009bpr} based on collaborative filtering is significantly better than Pop based on popularity, and the sequential recommendation model SASRec~\cite{kang2018sasrec} combined with attention mechanism and temporal information is significantly better than BPR, which is consistent with the results in existing literature~\cite{rendle2009bpr,kang2018sasrec,zhou2020s3}. 
It is worth noting that for the 20 candidate items, LLMs that rely on natural languages cannot completely recall all items in most cases, and several models with poor abilities can only output a dozen items, greatly limiting the accuracy of recommendation results. Therefore, \revision{``coverage (recall@20)''} indicates the ability of LLMs to memorize, re-rank and output candidate items, and several approaches such as the re-generation method~\cite{li2023ChatGPT-news-finetune} and probability distribution outputs~\cite{yue2023llamarec} have been proposed to improve the recall performance. While for the traditional models, all candidate items can be recalled~(\ie coverage equals to 1).
Because the candidate items are not recalled completely, the recommendation effect of a few LLMs~(\eg Flan-T5 and ChatGLM) is not even as good as the random baseline. 
\revision{DeepSeek-R1 performs worse than DeepSeek-V3, indicating that the long-chain reasoning capabilities and slow-thinking process of LLMs are optimized for Mathematics and coding tasks, which is not necessarily suitable for the zero-shot recommendation scenario.}
\revision{It might be perplexing to observe that the Qwen2.5 model~\cite{yang2024qwen2} performs well on the NDCG@10 metric, yet all three parameter scales yield equally poor results on NDCG@1. This discrepancy arises because Qwen2.5 tends to provide lengthy explanations in its output, which significantly reduces the top-1 accuracy during output parsing.}
\revision{When evaluating different versions of the same model, a general trend is observed where newer versions tend to outperform their predecessors. For instance, LLaMA3~\cite{dubey2024llama3} typically surpasses LLaMA2~\cite{touvron2023llama2}, and ChatGLM3~\cite{zhipu2024chatglm} is generally superior to ChatGLM2. However, anomalies have been noted that ChatGLM outperforms ChatGLM2 on the Amazon Books dataset, and LLaMA-7B achieves a higher NDCG@10 score compared to LLaMA2-7B on the MovieLens-1M dataset. The underlying reason is that LLMs with parameters less than 10 billion often struggle to comprehend the prompts in recommendation tasks adequately, frequently resulting in outputs that do not meet the required format. We will explore this issue in detail through case studies in Section~\ref{sec:llm-limitation-case-study}.}
For most LLMs, the zero-shot recommendation performance is not as good as the baseline method Pop based on popularity of interactions in the dataset~\cite{liu2023chatgpt,hou2023large,dai2023uncovering}. However, the powerful LLMs like ChatGPT and LLaMA-70B~(chat)~\cite{touvron2023llama2} can achieve better results than Pop in zero-shot settings. 
\revision{Furthermore, GPT-4 and GPT-4o can perform better than the fully-trained matrix factorization model BPR on the MovieLens-1M dataset, and DeepSeek-V3 outperforms BPR on two datasets}, indicating the potential of LLMs to serve as the backbone of recommender systems.
In addition, the significant differences between the results of LLMs demonstrate the importance of selecting appropriate LLMs for downstream recommendation tasks~\cite{kang2023llms-ctr,liu2023llmrec,liu2023chatgpt}.

\begin{figure}[t]
\centering  
\subfigure[MovieLens-1M]{   
\begin{minipage}{0.45\textwidth}
\centering    
\includegraphics[scale=0.4]{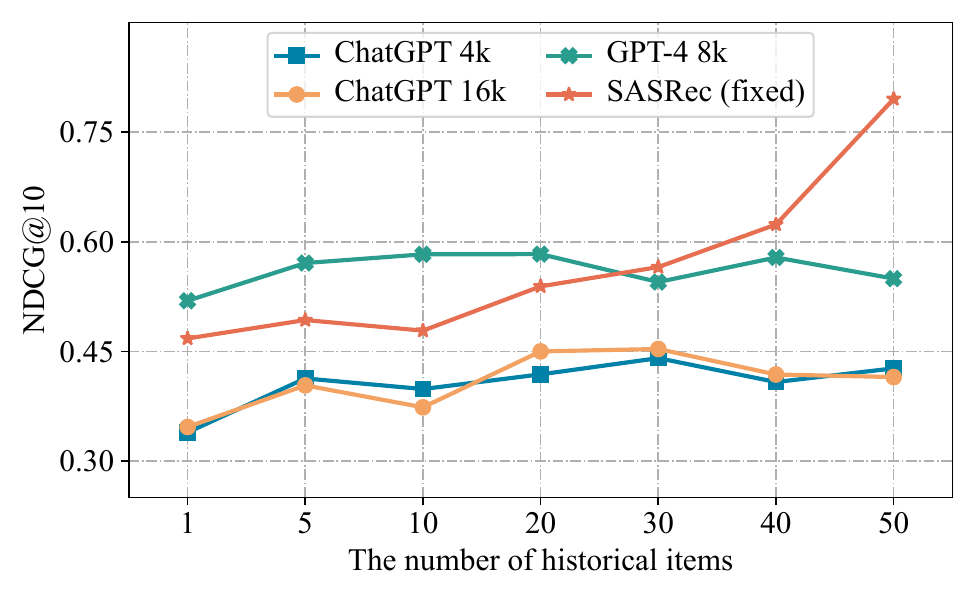}  
\end{minipage}
}
\subfigure[Amazon Books]{ 
\begin{minipage}{0.45\textwidth}
\centering    
\includegraphics[scale=0.4]{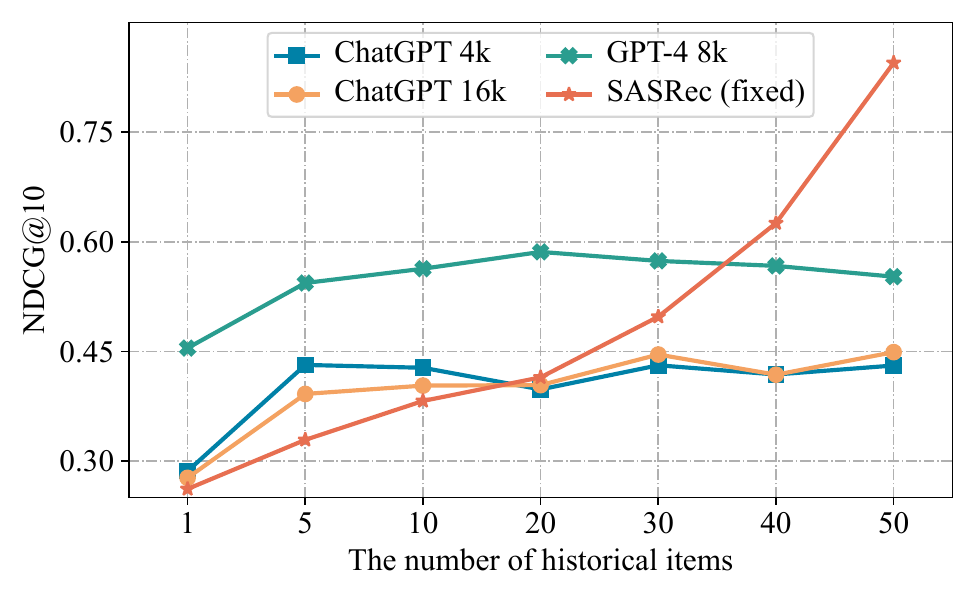}
\end{minipage}
}
\caption{The recommendation performance of LLMs \wrt the number of historical item sequences. \revision{As in~\ref{fig:basic-prompt-ranking}, we use recently interacted items to model user interest in the zero-shot ranking task. To compare the impact of historical item sequences on LLMs, we illustrate the ranking performance of the traditional sequential recommender SASRec~\cite{kang2018sasrec} and three closed-source LLMs with different numbers of historical items.}}    
\label{fig:fig-llms-context-length}    
\end{figure}

\paratitle{The impact of historical item sequences}. As for ranking tasks in Fig.~\ref{fig:basic-prompt-ranking}, the recently interacted historical items are used as the user representations. However, there is no standard value for the number of items that represent users. To analyze the impact of historical item sequences on the recommendation performance, we conduct experiments to explore the recommendation effect of the sequential recommender SASRec~\cite{kang2018sasrec} and the closed-source LLM ChatGPT with different numbers of historical interactions. For the fully-trained SASRec, the maximum length of the historical item sequence will affect the model framework and prediction results~\cite{kang2018sasrec,hou2022towards}. To ensure the model uniformity, we fix the model checkpoint with the historical item sequence of 50 as the ``SASRec (fixed)'' for comparison, and evaluate the recommendation performance~(NDCG@10~\cite{NDCG}) with different numbers of historical items. For ChatGPT and GPT-4, we get zero-shot results with different items to verify whether the powerful LLMs can deal with the long context for recommendation. In Fig.~\ref{fig:fig-llms-context-length}, with the increasing number of historical items, the results of SASRec improve steadily, while the performance of LLMs changes little. Consistent conclusions on two datasets can be drawn that even if LLMs can accept more historical items for user representations, increasing the number of historical items does not bring significant gains in the recommendation performance. The performance trends of LLMs show that the increased historical item sequence is not fully utilized by the language model. \revision{This phenomenon may be attributed to the fact that the long-sequence capabilities of LLMs have not been specifically optimized for zero-shot recommendation tasks. If targeted optimizations were applied to enhance the ability of LLMs to handle task-specific prompts, it is reasonable to expect that improvements in performance could be achieved.}
To further improve the mining of user interest for LLMs, approaches such as retrieval augmentation~\cite{lin2023rella} and prompting strategies~\cite{wang2023recmind,yao2023DOKE} can be used, which will be analyzed in Section~\ref{sec:prompt}.

\paratitle{Inference time of LLMs}. In the actual deployment of recommender algorithms, the inference efficiency is the decisive factor for industrial applications~\cite{sun2024PRM-KD,wang2023key-value}. In general, the inference time of models is closely related to the size of parameters. As shown in the last column of Table~\ref{tab:overall-llms}, for the lightweight traditional recommenders, the inference time of SASRec is about 1 second for each user. However for LLMs, except that closed-source models cannot accurately obtain inference time due to limitations of the API, the inference time of open-source models takes nearly 10 or more seconds for one prediction, leading to an unacceptable time delay in practical applications.

\begin{tcolorbox} [title = {Observations on the Overall Recommendation Performance of LLMs}, left=0mm, right=8mm]
    \begin{itemize}
        \item In zero-shot scenarios, LLMs have cold-start capabilities, and GPT-4o even surpasses collaborative filtering models. However, all LLMs are inferior to fully-trained sequential recommendation models.
        \item \revision{The long-chain reasoning capabilities and slow-thinking process of LLMs are optimized for Mathematics and coding tasks, which is not necessarily suitable for the zero-shot recommendation scenario.}
        \item Even if LLMs can accept more historical items for user representations, increasing the number of historical items does not bring significant gains in the recommendation performance, while traditional recommenders can better utilize longer sequences. \revision{It also shows that the user interest should be effectively modeled rather than blindly stacking historical items.}
        \item The inference time of LLMs for recommendation is unacceptable for applications.
    \end{itemize}
\end{tcolorbox}

\subsubsection{LLMs on Recommendations \wrt Four Aspects~(RQ2)} For different LLMs, differences in public availability and model architecture will lead to different recommendation scenarios, results and inference time~\cite{kang2023llms-ctr,liu2023llmrec}. For the same LLM, the parameter scale and context length also affect the efficiency and effectiveness of language models~\cite{zhao2023survey}. Therefore, we explore the impact of different LLMs on recommendations from four aspects, namely \emph{public availability}, \emph{model architecture}, \emph{parameter scale} and \emph{context length} as follows.

\paratitle{Public availability}. As shown in Table~\ref{tab:overall-llms}, closed-source models achieve significantly better results than the open-source models in the cold-start scenario, but they cannot outperform fully-trained sequential models. 
The results of the closed-source LLMs can be further improved through approaches like retrieval augmentation, which we will discuss in Section~\ref{sec:interest}. 
In terms of LLMs, ChatGPT with zero-shot settings has comparable recommendation performance with the fully-trained Pop especially on the sparse Amazon Books dataset, indicating the fundamental ability of LLMs on recommendation tasks. Furthermore, the upgraded GPT-4 \revision{and GPT-4o exceed} ChatGPT by a large margin due to its strong zero-shot generalization ability. The superior zero-shot performance of GPT sheds lights on leveraging LLMs for recommendation. Although \revision{DeepSeek-V3} has the comparable recommendation performance with \revision{GPT-4o}, the open-source models usually get poor results compared to GPT-4o in the zero-shot settings. The reason is that open-source models lack comprehensive cold-start capabilities, and their strength lies in the ability to integrate domain knowledge through strategies such as prompt tuning. In line with previous studies on LLMs~\cite{hou2023large,kang2023llms-ctr,ma2023STELLA,liu2023chatgpt}, employing a closed-source model in cold-start scenarios yields better results, while an open-source model is more flexible and easy to use when tuning is needed~\cite{bao2023tallrec,zheng2023LC-Rec,li2023e4srec,luo2023recranker,liu2023once}.

\paratitle{Model architecture}. For the \emph{model architecture}, Flan-T5~\cite{chung2022flan-t5} based on the encoder-decoder architecture has almost no ability to recommend items in the cold-start setting, as its training corpus does not involve specialized instructions of our task. Trained with prompts of recommendations, the encoder-decoder architecture is suitable for prompt tuning and instruction tuning~\cite{zhang2023recommendation,geng2022p5,chu2023RecSysLLM}. Similarly, the first and first version of ChatGLM~\cite{zeng2022glm} performs poorly on the zero-shot ranking task, and is not as good as Vicuna and LLaMA2 based on the causal decoder. However, the third version of ChatGLM, \ie ChatGLM3 has comparable recommendation performance with Vicuna~\cite{vicuna2023} and LLaMA2~\cite{touvron2023llama2}, which further indicates the importance of selecting an advanced foundation model. In terms of the series of LLaMA models~\cite{touvron2023llama}, \revision{LLaMA3 achieves the best performance}, and Vicuna is better than LLaMA, which is related to their training data and release time. Furthermore, the chat version of LLaMA2, \ie LLaMA2-chat is a series that uses conversational dialog instructions to fine-tune LLaMA2~\cite{touvron2023llama2}, which is more suitable for our tasks in ranking settings. Therefore, the results of LLaMA2-chat are significantly improved compared with LLaMA2, and LLaMA2-70B-chat \revision{and LLaMA3-70B} achieve better performance than the closed-source model ChatGPT. Generally speaking, researchers prefer to study recommendation tasks based on the causal decoder framework such as LLaMA, and the {latest} version of LLaMA has a better generalization ability. 

\begin{figure}[t]
\centering  
\subfigure[MovieLens-1M]{   
\begin{minipage}{0.48\textwidth}
\centering    
\includegraphics[scale=0.5]{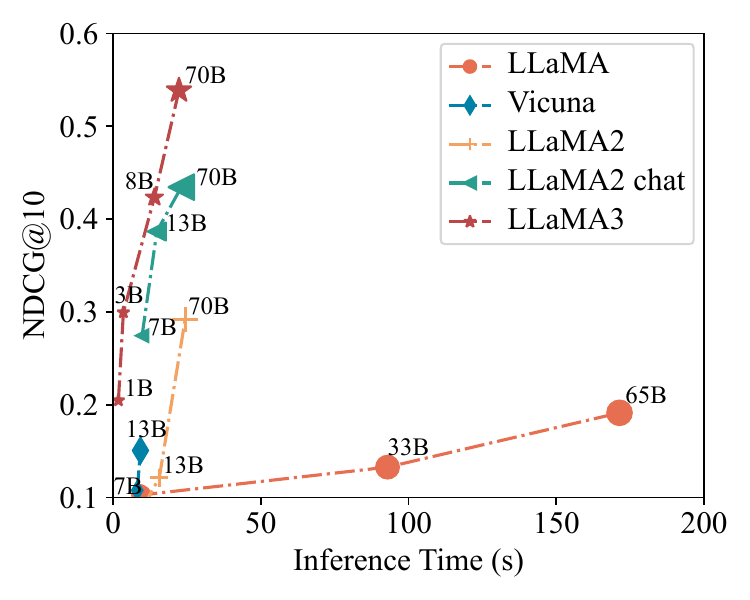}  
\end{minipage}\label{fig:fig-llms-scatter-params-ml-1m} 
}
\subfigure[Amazon Books]{ 
\begin{minipage}{0.48\textwidth}
\centering    
\includegraphics[scale=0.5]{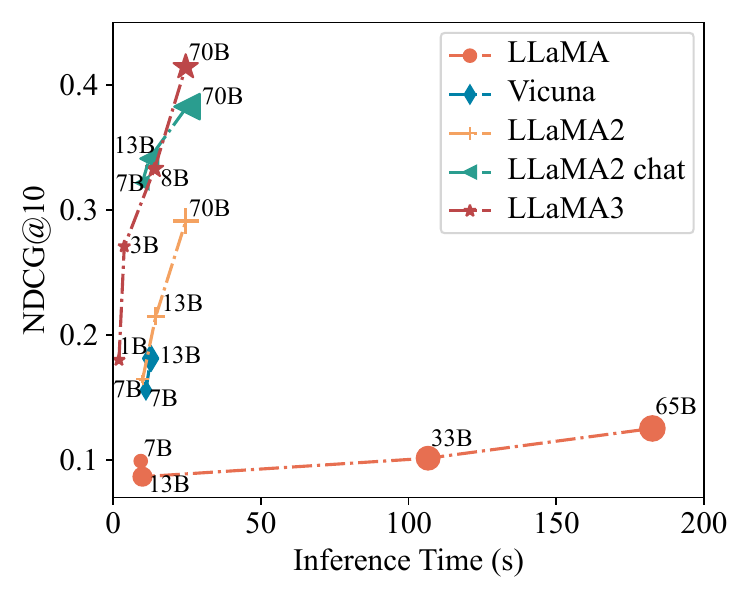}
\end{minipage}\label{fig:fig-llms-scatter-params-amazon} 
}
\caption{\revision{The recommendation performance and inference time of LLMs \wrt the parameter scale. We compare the different parameter scales of LLMs, and the larger the symbol in the figure, the larger the parameter scale.}}    
\Description{The recommendation performance and inference time of LLMs \wrt the parameter scale. We compare the different parameter scales of LLMs, and the larger the symbol in the figure, the larger the parameter scale.}
\label{fig:fig-llms-scatter-params}    
\end{figure}

\begin{figure}[t]
\centering  
\subfigure[MovieLens-1M]{   
\begin{minipage}{0.48\textwidth}
\centering    
\includegraphics[scale=0.5]{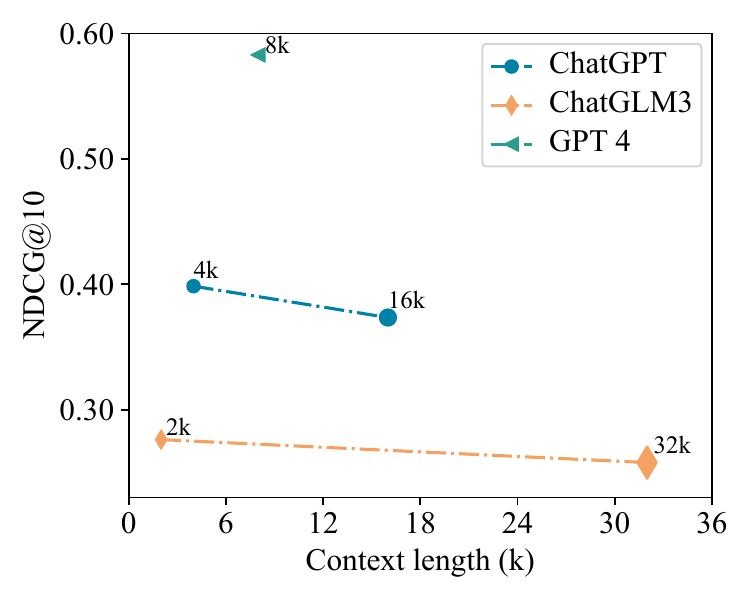}  
\end{minipage}\label{fig:fig-llms-scatter-context-ml-1m} 
}
\subfigure[Amazon Books]{ 
\begin{minipage}{0.48\textwidth}
\centering    
\includegraphics[scale=0.5]{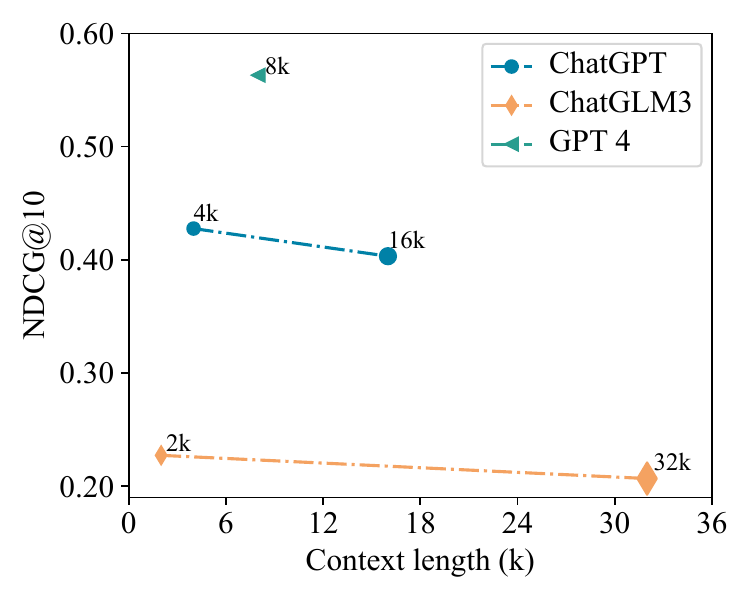}
\end{minipage}\label{fig:fig-llms-scatter-context-amazon} 
}
\caption{The recommendation performance of LLMs \wrt the context length. We compare the different context length of LLMs, and the larger the symbol in the figure, the longer the context length.}    
\label{fig:fig-llms-scatter-context}    
\end{figure}

\paratitle{Parameter scale}. It is widely recognized that the larger the parameter size, the more powerful the LLMs~\cite{zhao2023survey,kang2023llms-ctr,hou2023large}, the same applies in the field of recommender systems. To compare the effect and efficiency of LLMs \wrt the parameter scale, we compare the recommendation performance and inference time of LLaMA~\cite{touvron2023llama}, Vicuna~\cite{vicuna2023}, LLaMA2~\cite{touvron2023llama2}, LLaMA2-chat \revision{}{and LLaMA3~\cite{dubey2024llama3}} at different parameter scales in Fig.~\ref{fig:fig-llms-scatter-params}. As the scale of parameters enlarges, the recommendation performance and inference time of LLMs steadily increases, and both datasets~(Fig.~\ref{fig:fig-llms-scatter-params-ml-1m} and Fig.~\ref{fig:fig-llms-scatter-params-amazon}) have consistent conclusions. Therefore, it is necessary to consider the trade-off between performance and efficiency when choosing the parameter scale. Moreover, the performance improvement on the increasing scale of \revision{}{LLaMA2 and LLaMA3} is more significant than that of LLaMA, \revision{}{indicating that the scaling law of LLMs has to do with the capabilities of the base model}. 

\paratitle{Context length}. Different LLMs have different maximum input limitations~\cite{zhao2023survey,touvron2023llama,touvron2023llama2}, and a longer context input means LLMs can accommodate more historical items for recommendation. However, it remains to be explored whether the maximum input length of LLMs will affect the recommendation results when the length limitation is not exceeded. Therefore, we conduct experiments to investigate the differences in the recommendation performance between the two length versions of ChatGPT~(4K and 16K) and ChatGLM3~(2K and 32K). As shown in Fig.~\ref{fig:fig-llms-scatter-context}, expanding the length limitation of LLMs does not necessarily mean the better recommendation performance, while there is a slight decrease in NDCG@10. Furthermore, when the maximum input of LLMs remains unchanged, increasing the historical input of users results in insignificant gains as shown in Fig.~\ref{fig:fig-llms-context-length}. Therefore, the key to the recommendation problem is to enable LLMs to effectively utilize the information within the limited context~\cite{lin2023rella,yao2023DOKE}, and a suitable context length selection for LLMs as recommender systems is worthy of deep consideration.

\begin{tcolorbox} [title = {Observations on Recommendations \wrt Four Aspects of LLMs}, left=0mm, right=8mm]
    \begin{itemize}
        \item As for the public availability, closed-source models generally outperform the open-source models in the recommendation performance, but have poorer flexibility.
        \item As for the model architecture, different frameworks are adapted to different recommendation tasks and fine-tuning strategies, while LLMs with the casual decoder architecture are still mainstream.
        \item As for the parameter scale, the larger the parameter scale, the better the recommendation performance.
        \item As for the context length, a longer maximum context length leads to worse recommendation results. 
    \end{itemize}
\end{tcolorbox}

\begin{table}[t]
\caption{Overall performance of LLMs on CTR predictions. There are three settings, \ie zero-shot setting without fine-tuning, parameter-efficient fine-tuning~(PEFT) setting with a few parameters tuned, and fine-tuning~(FT) setting with all parameters tuned.}
\small
\begin{tabular}{@{}ccccccc@{}}
\toprule
\textbf{Dataset} & \multicolumn{3}{c}{\textbf{MovieLens-1M}}              & \multicolumn{3}{c}{\textbf{Amazon Books}}              \\ \midrule
\textbf{Model}   & \textbf{Zero-shot} & \textbf{PEFT}   & \textbf{FT}     & \textbf{Zero-shot} & \textbf{PEFT}   & \textbf{FT}     \\ \midrule
LLaMA-7B       & 0.4683       & 0.5479 & 0.6658 & 0.6488 & 0.8262 & 0.8469 \\
Alpaca-LoRA-7B & 0.5264       & 0.5767 & 0.6702 & 0.6558 & 0.8440 & 0.8533 \\
LLaMA2-7B      & {\ul 0.5284} & 0.6133 & 0.6457 & 0.6754 & 0.8550 & 0.8542 \\
LLaMA2-chat-7B   & 0.5255             & {\ul 0.6275}    & {\ul 0.6731}    & {\ul 0.7174}       & {\ul 0.8561}    & {\ul 0.8660}    \\
LLaMA3-8B        & \textbf{0.5724}    & \textbf{0.6577} & \textbf{0.6927} & \textbf{0.7216}    & \textbf{0.8610} & \textbf{0.8746} \\ \bottomrule
\end{tabular}
\label{tab:llm-finetune-ctr}
\end{table}


\subsubsection{Comparisons of Tuning Strategies for LLMs~(RQ3)} Due to the fact that LLMs are not customized to recommender systems, it is insufficient to only consider the zero-shot recommendation performance in cold-start scenarios~\cite{haleem2022ChatGPTera,kang2023llms-ctr,liu2023llmrec,luo2023recranker}. In order to explore the impact of different training strategies of LLMs on recommendations, we compare the CTR prediction performance of \revision{five} LLaMA-based LLMs on two datasets. Specifically, we consider three training settings of LLMs, \ie the zero-shot setting without fine-tuning, Parameter-Efficient Fine-Tuning~(PEFT) setting~(we use the LoRA~\cite{hu2021lora} here) with a few parameters tuned, and Fine-Tuning (FT)
setting with all parameters tuned, and summarize empirical conclusions from three aspects: \emph{overall performance of different settings}, \emph{the impact of instruction tuning}, and \emph{the impact of training data}.

\begin{figure}[t]
\centering  
\subfigure[MovieLens-1M]{   
\begin{minipage}{0.48\textwidth}
\centering    
\includegraphics[scale=0.37]{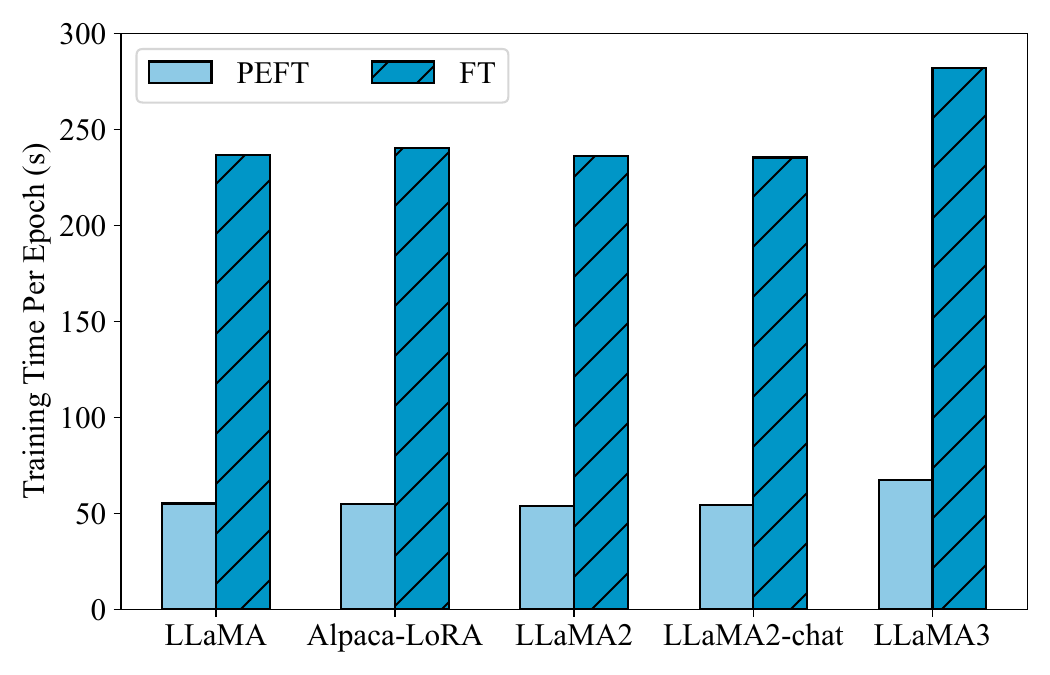}  
\end{minipage}\label{fig:fig-llms-training-time-ml}
}
\subfigure[Amazon Books]{ 
\begin{minipage}{0.48\textwidth}
\centering    
\includegraphics[scale=0.37]{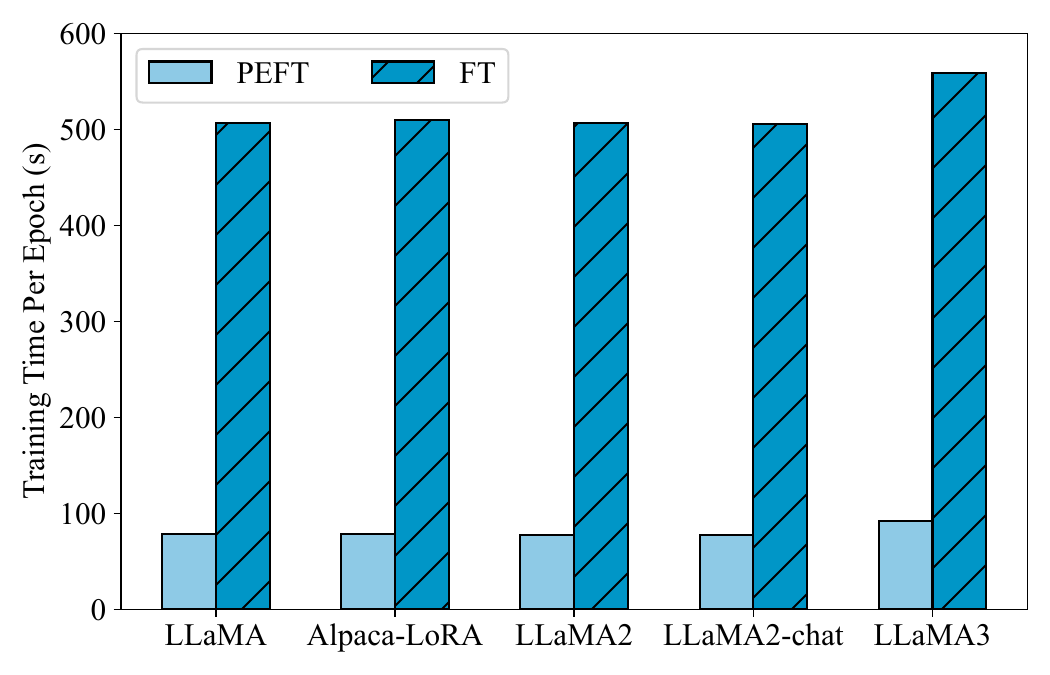}
\end{minipage}\label{fig:fig-llms-training-time-amazon}
}
\caption{Comparison of training time \wrt two fine-tuning strategies, \ie PEFT and FT. \revision{The training time of LLMs is averaged over all training epochs, measured in seconds~(s). Five LLaMA-based models are compared.}}    
\label{fig:fig-llms-training-time}    
\end{figure}

\paratitle{Overall performance of different settings}. As shown in Table~\ref{tab:llm-finetune-ctr}, we can see that the results of fine-tuning LLMs~(PEFT and FT) on only 256 samples are significantly better than the zero-shot performance in cold-start scenarios, and empirical findings are consistent across four LLMs on both datasets. Furthermore, considering the two kinds of fine-tuning strategies, the performance of the fine-tuning setting is even better than that of the PEFT setting since more parameters are tuned~\cite{hu2021lora}. In addition to recommendation effects, training efficiency is also a performance that deserves attention. Therefore, we compare the training time of the two fine-tuning strategies on two datasets. As shown in Fig.~\ref{fig:fig-llms-training-time}, the time for PEFT is significantly less than that for fine-tuning all parameters, which is in agreement with the existing literature~\cite{hu2021lora,zhao2023survey}. Therefore, the balance between efficiency and effectiveness needs to be further considered.

\begin{table}[t]
\centering
\begin{minipage}[t]{0.48\textwidth}
\centering
\small
\caption{\revision{Impact of LoRA hyper-parameters ($r$ and $\alpha$ represents the rank and scaling factor of LoRA, respectively) on the recommendation performance of LLaMA3-8B. Here, $r$ typically ranges from 4 to 64, and $\alpha$ is usually set to twice the value of $r$.}}
\label{tab:hyper-param-r-alpha}
\begin{tabular}{@{}cccc@{}}
\toprule
\textbf{\( r \)} & \textbf{\( \alpha \)} & {\textbf{MovieLens-1M}} & {\textbf{Amazon Books}} \\
\midrule
4  & 8   & 0.6378 & 0.8612 \\
8  & 16  & 0.6380 & 0.8610 \\
16 & 32  & 0.6592 & 0.8672 \\
32 & 64  & \textbf{0.6746} & {\ul 0.8779} \\
64 & 128 & {\ul 0.6745} & \textbf{0.8890} \\
\bottomrule
\end{tabular}
\end{minipage}
\hfill 
\begin{minipage}[t]{0.48\textwidth}
\centering
\caption{\revision{Impact of the dropout rate, a hyper-parameter in LoRA, on the performance of LLaMA3-8B fine-tuned as a CTR recommendation model on the MovieLens-1M and Amazon Books datasets.}}
\label{tab:hyper-param-dropout}
\small
\begin{tabular}{@{}ccc@{}}
\toprule
\textbf{Dropout} & \textbf{MovieLens-1M} & \textbf{Amazon Books} \\ \midrule
0                & 0.6375                & 0.8593                \\
0.1              & {\ul 0.6528}          & {\ul 0.8598}          \\
0.2              & \textbf{0.6577}       & 0.8548                \\
0.3              & 0.6388                & 0.8516                \\
0.4              & 0.6429                & \textbf{0.8610}       \\
0.5              & 0.6473                & 0.8420                \\ \bottomrule
\end{tabular}
\end{minipage}
\label{tab:hyper-params}
\end{table}



\begin{figure}[t]
\centering  
\subfigure[MovieLens-1M]{   
\begin{minipage}{0.48\textwidth}
\centering    
\includegraphics[scale=0.4]{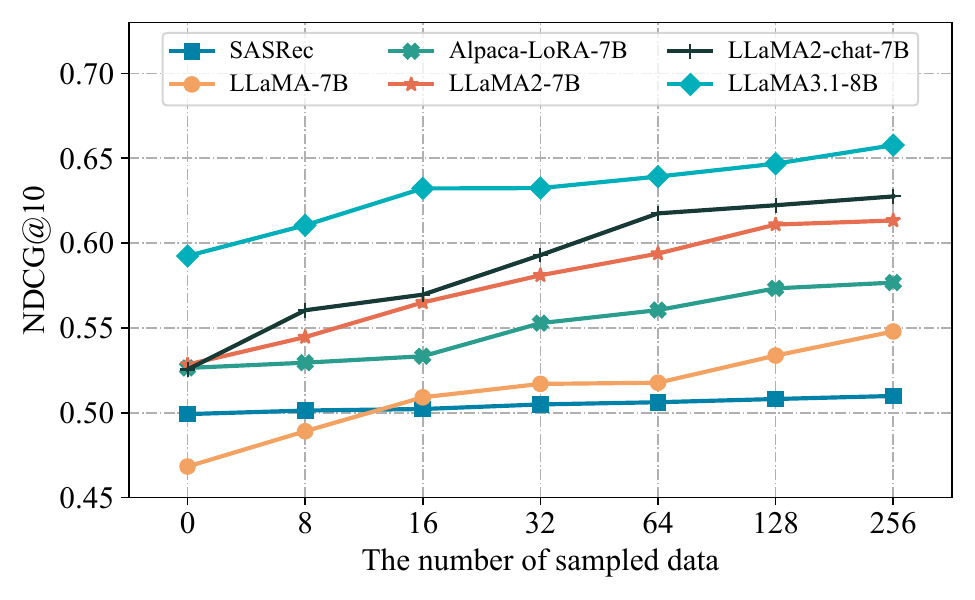}  
\end{minipage}\label{fig:fig-llms-training-number-ml}
}
\subfigure[Amazon Books]{ 
\begin{minipage}{0.48\textwidth}
\centering    
\includegraphics[scale=0.4]{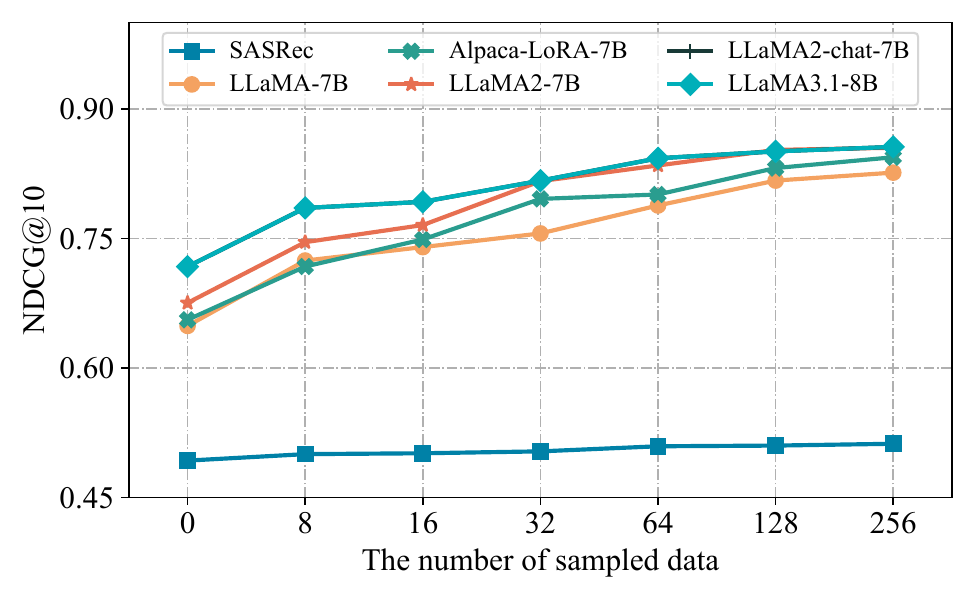}
\end{minipage}\label{fig:fig-llms-training-number-amazon}
}
\caption{The fine-tuned performance of LLMs \wrt the number of sampled data, and LoRA is utilized. \revision{For LLaMA-based LLMs, we analyze the effect of sampled data on the recommendation performance (measured by NDCG@10) of few-shot learning, and compare with the traditional recommender SASRec~\cite{kang2018sasrec}.}}    
\Description{The fine-tuned performance of LLMs \wrt the number of sampled data, and LoRA is utilized. For LLMs, we analyze the effect of sampled data on the recommendation performance (measured by NDCG@10) of few-shot learning, and compare with SASRec~\cite{kang2018sasrec}.}
\label{fig:fig-llms-training-number}    
\end{figure}

\paratitle{The impact of instruction tuning}. Before fine-tuning LLMs using the recommendation data, we are also concerned about whether the instruction tuning of LLMs using general data can further enhance the subsequent fine-tuning on recommendations. Therefore, we select two pairs of LLMs as controls, namely LLaMA and Alpaca-LoRA, as well as LLaMA2 and LLaMA2-chat. As for the Alpaca-LoRA model, it is the tuned version of LLaMA~\cite{touvron2023llama} using the the LoRA strategy~\cite{hu2021lora} on the Alpaca dataset~\cite{alpaca}. As for the LLaMA2-chat model, it is also the updated version of LLaMA2 using the RLHF strategy~\cite{touvron2023llama2} on the conversational dataset. As illustrated in Table~\ref{tab:llm-finetune-ctr} and Fig.~\ref{fig:fig-llms-training-number}, the recommendation results of Alpaca-LoRA are better than that of LLaMA, and the performance of LLaMA2-chat outforms that of LLaMA2, indicating the effectiveness of performing fine-tuning on general instructions. That is to say, enhancing the general knowledge of language models can also improve the domain capabilities of LLMs on specified recommendation tasks~\cite{bao2023tallrec}.

\revision{\paratitle{The impact of hyper-parameters of LoRA}. As depicted in Table~\ref{tab:hyper-param-r-alpha} and Table~\ref{tab:hyper-param-dropout}, we analyze the CTR prediction performance of LLaMA3-8B with different hyper-parameters of LoRA. We can find that increasing the rank (\( r \)) and scaling factor (\( \alpha \)) generally enhances the recommendation performance, likely due to the increased model capacity with more parameters tuned. Additionally, the impact of the dropout rate varies between datasets. For MovieLens-1M, a dropout rate of 0.2 yields the highest score of 0.6577, indicating that moderate dropout enhances generalization by preventing overfitting. While for Amazon Books, a dropout rate of 0.4 results in the best performance (0.8610), suggesting that the optimal dropout rate is context-dependent. These findings underscore the importance of meticulously tuning LoRA hyper-parameters to optimize performance of LLMs in CTR recommendation tasks.}

\paratitle{The impact of training data}. For fine-tuning LMs with few-shot learning, we conduct experiments to explore the impact of training samples on the recommendation performance. As shown in Fig.~\ref{fig:fig-llms-training-number}, we visualize the fine-tuning results corresponding to different numbers of training samples with the LoRA strategy, and also consider the traditional recommender SASRec~\cite{kang2018sasrec}. Despite a few sampled data, LLMs such as LLaMA-7B can quickly adapt to the task of CTR predictions, achieving noticeable improvements as the number of training samples increases. With only a small number of training samples, we can stimulate the appreciable recommendation ability of LLMs, reflecting their remarkable emergence ability. However, when the training samples increase from 0 to 256, the performance of the traditional recommendation model SASRec still remains 0.5. Experimental results show that LLMs have advantages over conventional recommenders on abilities of few-shot learning and adapting~\cite{bao2023tallrec}. In addition, the comparison without sampled data in Fig.~\ref{fig:fig-llms-training-number} also highlights the zero-shot cold-start ability of LLMs. 

\begin{tcolorbox} [title = {Observations on Tuning Strategies of LLMs}, left=0mm, right=8mm]
    \begin{itemize}
        \item For LLM-based recommendations, few-shot training results are better than the zero-shot performance, and fine-tuning all parameters is more effective but less efficient than parameter-efficient fine-tuning. 
        \item The instruction tuning using general data can further enhance the fine-tuning results of LLMs on specified recommendation tasks.
        \item \revision{As for parameter-efficient fine-tuning LLMs as recommenders with LoRA, increasing the rank (\( r \)) and scaling factor (\( \alpha \)) of LoRA generally enhances the recommendation performance due to the increased model capacity with more parameters tuned.}
        \item In few-shot training scenarios, LLMs are more capable of adapting to recommendation tasks compared to traditional recommendation models.
    \end{itemize}
\end{tcolorbox}

\begin{figure}[t]
\centering
\subfigure[\revision{Case study of the position bias when leveraging GPT-4~\cite{openai2023GPT4} as the zero-shot re-ranker on the MovieLens-1M dataset. LLMs tend to select items in front while the ground-truth movie appears at the end of the candidate list. GPT-4 is affected by the position bias of the candidate list.}]{
    \includegraphics[width=0.46\linewidth]{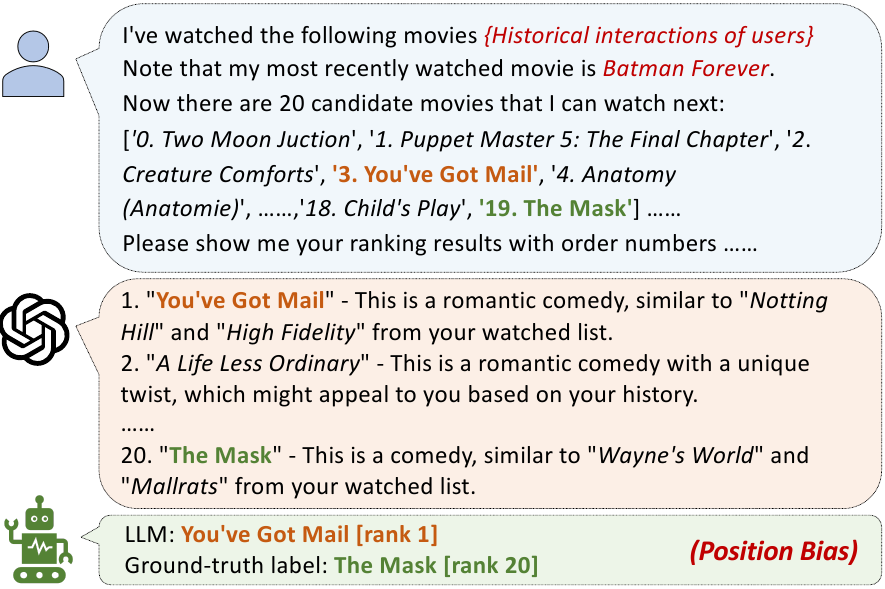}
    \label{fig:llm-limitation-position-bias}
}
\hspace{0.0015\linewidth}
\subfigure[\revision{Case study of the lack of domain knowledge when leveraging GPT-4~\cite{openai2023GPT4} as the zero-shot re-ranker on the MovieLens-1M dataset. ``Dick'' is a comedy film about two teenager girls, which is relevant to the movie ``Showgirls'', but GPT-4 ranks ``Dick'' at the end of the list.} ]{
    \includegraphics[width=0.46\linewidth]{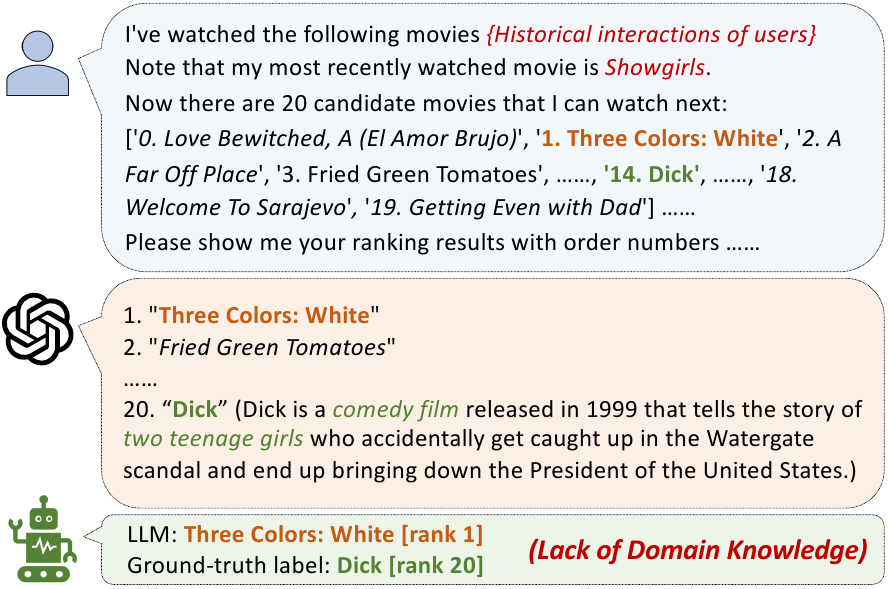}
    \label{fig:llm-limitation-lack-domain}
}
\\
\subfigure[\revision{Case study of misunderstanding output formats when leveraging DeepSeek-R1~\cite{liu2024deepseekv3} as the zero-shot re-ranker on the Amazon Books dataset. The candidate list provides 20 books while DeepSeek-R1 recommends 27 books, of which 10 are from historical interactions.}]{
    \includegraphics[width=0.46\linewidth]{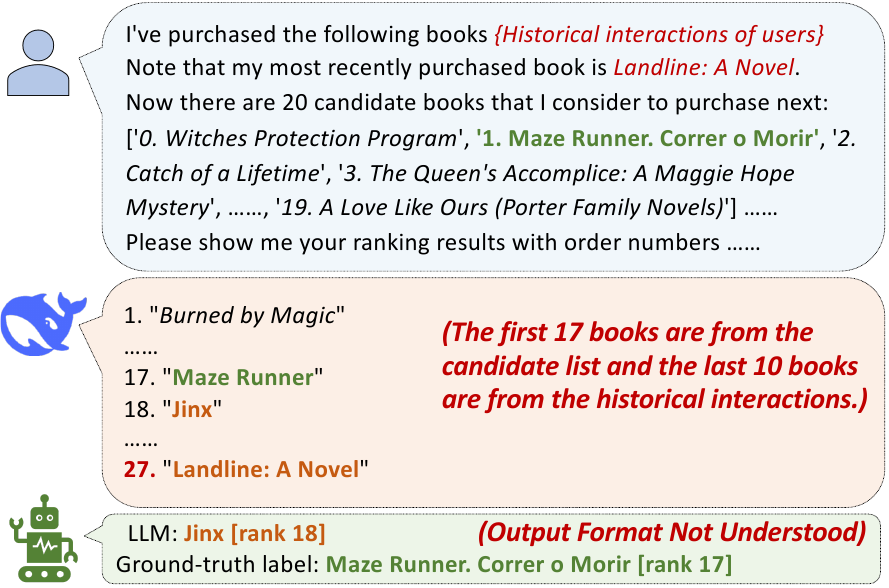}
    \label{fig:llm-limitation-output}
}
\hspace{0.0015\linewidth}
\subfigure[\revision{Case study of hallucination when leveraging ChatGLM3-6B~\cite{zhipu2024chatglm} as the zero-shot re-ranker on the Amazon Books dataset. ChatGLM3-6B ranks the purchased book first, leaves out the ground-truth book, and generates two Chinese books not within the candidate list.}]{
    \includegraphics[width=0.46\linewidth]{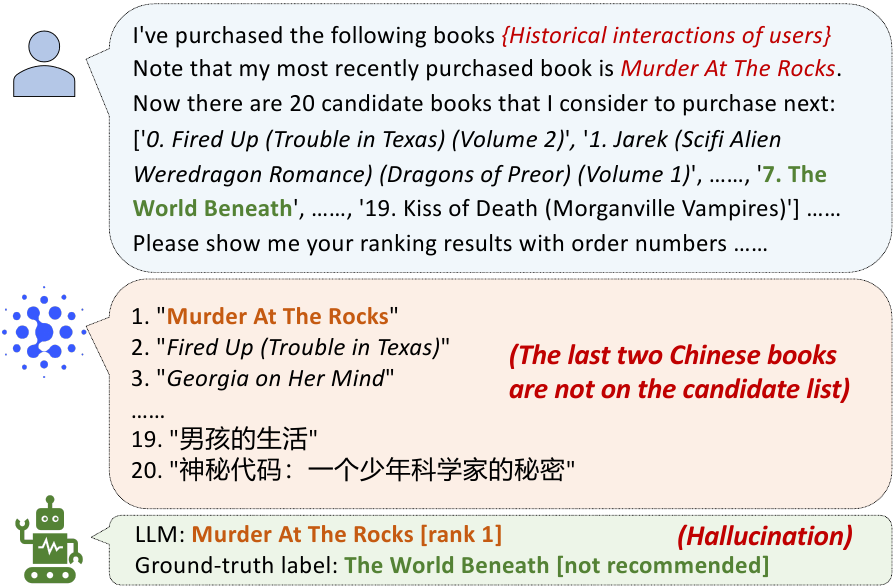}
    \label{fig:llm-limitation-hallu}
}
\caption{\revision{Case study of limitations when leveraging LLMs as zero-shot recommender systems.}
\Description{Case study of limitations when leveraging LLMs as zero-shot recommender systems.}}
\label{fig:fig-llm-limitation}
\end{figure}

\subsubsection{Case Study of Limitations~(RQ4)} \label{sec:llm-limitation-case-study}
Despite powerful capabilities of LLMs, there are also some limitations using LLMs as recommender systems. To illustrate the limitations of LLMs for recommendation, we choose four recommendation cases of LLMs as zero-shot re-ranker in Fig.~\ref{fig:fig-llm-limitation} to analyze the possible failure scenarios of LLMs. 

\paratitle{Position bias}. It is widely acknowledged that the generation results of LLMs have randomness, leading to instability in recommendation results, and the position bias is a typical manifestation. As shown in Fig.~\ref{fig:llm-limitation-position-bias}, the LLM places the ground-truth movie ``\emph{The Mask}'' at the bottom of ranking results, because ``\emph{The Mask}'' is at the end of the candidate item sequence. \ignore{That is to say, }When the order of candidate items is adjusted, the recommendation results will also change, indicating the position bias during recommendations. Existing literature has also found the position bias in LLM-based recommendations~\cite{hou2023large,ma2023STELLA,zhang2023recommendation}, and methods such as bootstrapping~\cite{hou2023large} and Bayesian probability framework~\cite{ma2023STELLA} have been proposed to calibrate unstable results. 

\paratitle{Lack of domain knowledge}. The lack of domain knowledge in recommendations may also lead to misunderstandings of LLMs. As shown in Fig.~\ref{fig:llm-limitation-lack-domain}, the LLM places the second movie in the candidate list, \ie ``\emph{Three Colors: White}'' at the first position in the re-ranking result, which still indicates the position bias. Furthermore, the LLM places the ground-truth item ``\emph{Dick}'' at the bottom of the ranking list. The possible reason is that relying solely on the movie title ``\emph{Dick}'', LLMs cannot infer that the movie is a comedy film about two teenage girls, which has a greater similarity to the recently watched movie ``\emph{Showgirls}'' by users. This case demonstrates that LLMs may make inappropriate decisions due to a lack of domain knowledge in the field of recommender systems. 

\revision{\paratitle{Misunderstanding output formats}. LLMs may exhibit limitations in retaining critical information regarding output formats from prompt instructions, potentially leading to responses that deviate from the intended requirements. As illustrated in Fig~\ref{fig:llm-limitation-output}, while the input prompt explicitly requires the model to re-rank only 20 candidate books, the model generates 27 outputs, including 10 books from the historical interaction sequence. Furthermore, three books from the original candidate list are omitted in the output. This phenomenon directly contributes to the observed suboptimal performance in the coverage metric, as demonstrated in Table~\ref{tab:overall-llms}, where the output of LLMs frequently fails to achieve 100\% coverage of the expected results. }

\revision{\paratitle{Hallucination}. The hallucination problem in LLM-based recommender systems refers to the phenomenon where LLMs generate items that is not grounded in their training data or the provided input, often producing factually incorrect outputs. As depicted in Fig.~\ref{fig:llm-limitation-hallu}, the LLM ranks the recently purchased book at the top of recommended list, and leaves out the ground-truth book. Moreover, the LLM generates two Chinese books not within the candidate list. In~\our, hallucination can manifest as fabricated details of recommendation reasons, incorrect factual statements of items, or irrelevant content that deviates from the intended recommendation task. }


\begin{tcolorbox} [title = {Observations on Limitations of LLMs}, left=0mm, right=8mm]
    \begin{itemize}
        \item Despite capabilities, there are limitations when leveraging LLMs as recommenders.
        \item Results of LLMs have randomness, leading to instability such as the position bias.
        \item The lack of domain knowledge in recommendations can lead to misunderstandings.
        \item \revision{LLM-based recommender systems may suffer from the hallucination problem.}
    \end{itemize}
\end{tcolorbox}

%% file: sec-prompt.tex
\section{PROMPT ENGINEERING SPECIALIZED FOR RECOMMENDER SYSTEMS}
\label{sec:prompt}

Prompt is an important medium for interactions between humans and language models, and a well-designed prompt can better stimulate the powerful capabilities of LLMs~\cite{liu2023prompting-survey, le2021many}. Improving the performance of artificial intelligence by designing and improving prompts is known as prompt engineering\ignore{, which is also applicable in the field of recommender systems~\cite{yao2023DOKE}. } 
When leveraging LLMs as recommender systems, although the specific prompts in different studies may not be the same, the format is largely identical with only minor differences~\cite{hou2023large,gao2023chat,dai2023uncovering,wang2023recmind}. In our work, we concentrate on the general framework of prompt engineering specialized for recommender systems, and summarize four key elements of prompt formats. 
Firstly, suitable task description is the primary condition for constructing prompts~\cite{geng2022p5,wang2023recmind}.
\revision{As illustrated in Table~\ref{tab:task-description-prompt}, we list task-specific prompts for LLMs as recommender systems in our framework.}
Secondly, the characteristic of recommendation tasks lies in the mining and utilization of the user personalized interest, and user interest modeling is the essential aspect to reflect the true intentions and preferences of users~\cite{xi2023towards-kar,yao2023DOKE}. 
Thirdly, the purpose of recommender systems is to provide users with appropriate items, and the selection, matching, and generation of candidate items is a worthwhile issue to consider, \ie candidate items construction~\cite{lin2023rella}. 
Fourthly, prompting strategies are important for eliciting the planning and reasoning abilities of LLMs, which are the key to distinguishing LLMs from other language models~\cite{zhao2023survey}. 
In this section, we will discuss the key components of prompts in~\our. 

\input{sec-task}

\input{sec-interest}
\input{sec-candidate}

\subsection{Prompting Strategies}
\label{sec:prompting-strategies}
With the construction of prompts for recommendation tasks, prompting strategies are leveraged to further elicit the general abilities of LLMs in text comprehending and problem solving~\cite{zhao2023survey,liu2023prompting-survey,le2021many}. In this section, we first summarize prompting strategies specialized for recommender systems, and provide empirical findings on two research questions.


\subsubsection{Prompting Strategies Specialized for Recommender Systems}

The planning ability emerging from LLMs is the key to distinguishing them from other language models~\cite{zhao2023survey}. 
In general tasks with LLMs, researchers use methods such as Chain-of-Thought~(CoT) prompting to stimulate the logical reasoning ability of LLMs for solving complex problems~\cite{diao2023active-prompt,liu2022generated-knowledge-prompt}. As for~\our, leveraging LLMs for recommender systems can also employ prompting strategies to further improve the performance of recommendations~\cite{yao2023DOKE,liu2023recprompt}. In addition, due to the user and item settings of the recommendation task, the prompting strategy needs to be customized based on the specific user needs in recommender systems. In this paper, we concentrate on typical prompting strategies for recommendation tasks, \ie \emph{zero-shot prompting}, \emph{few-shot prompting}, \emph{recency-focused prompting}, \emph{role prompting}, \emph{chain-of-thought prompting} and \emph{self-prompting strategy}, and more advanced planning strategies in~\our~are left for future exploration~\cite{liu2023prompting-survey}.

\paratitle{Zero-shot prompting}
is the basic scenario and common form among various prompting strategies. Zero-shot prompting directly provides the context information and task description for LLMs without reference examples. The task description and prompt formats vary in line with different recommendation tasks~\cite{wang2023recmind,liu2023chatgpt,dai2023uncovering}.

\paratitle{Few-shot prompting} 
is opposite to zero-shot prompting, and it provides a few demonstrations in prompts to help LLMs better understand the user intention~\cite{brown2020language}.
While for recommendation tasks, the matching of provided examples with the current interest of personalized users determines the effectiveness of few-shot prompting. Therefore, it is worth discussing how to provide reliable context for subsequent recommendations~\cite{hou2023large}.

\paratitle{Recency-focused prompting} 
is first proposed by~\cite{hou2023large} based on the fact that the next predicted item in recommendations has a greater correlation with the recently interacted item than other historical items. Therefore, explicitly emphasizing the recently interacted items in prompts is also a practical prompting strategy. In our initial prompt, recency-focused prompting is used by default.

\paratitle{Role prompting} 
refers to playing a ``role-playing'' game with the language model. In the recommendation task, we can specify the role of LLMs as the recommender system to serve users in a targeted manner, adding auxiliary information to describe the role of recommenders in detailed prompts such as conversational recommender systems~\cite{he2023zero-shot-CRS,wang2023rethinkingCRS,jin2023core}. 

\paratitle{Chain-of-thought prompting}
is also known as the CoT prompting, which is widely applied in reasoning tasks such as question answering and mathematical inference~\cite{wei2022few-shot-CoT}. CoT prompting can elicit the ability of LLMs to solve problems step by step~\cite{kojima2022zero-shot-CoT}, or further explicitly decompose the reasoning and analysis process of the task using least-to-most prompting strategy~\cite{zhou2022least-to-most}. In the field of recommender systems, explicit steps can also be provided manually by researchers to assist in solving recommendation tasks~\cite{wang2023zero}. For~\our, we not only focus on the basic CoT prompts, but also explore the role of custom-designed step-by-step prompts. 


\paratitle{Self-prompting strategy}
means that the knowledge for answering questions can be obtained by prompting LLMs multiple times. LLMs can be required to generate relevant knowledge, providing necessary information for concepts in the original problem~\cite{liu2022generated-knowledge-prompt}. Meanwhile, the randomness and self consistency~\cite{wang2022self-consistency} generated by LLMs allow them to generate multiple inference chains, and then use the majority voting method on the results obtained from all chains as final predictions. When using LLMs to the re-ranking stage, there are many possible combinations of candidate items, and the recommendation results from the same LLM with the same prompt can be totally different. Therefore, bootstrapping with multiple trials is a scientific guarantee to reduce bias~\cite{hou2023large,ma2023STELLA,li2023ChatGPT-news-finetune}. 


\subsubsection{Research Questions and Experimental Setup}

In this section, we explore the construction and design of prompting strategies for~\our on the following two research questions:

\begin{itemize}
    \item \revision{RQ1: When prompting LLMs a zero-shot recommender systems, what are the performance differences of different prompting strategies?}
    \item \revision{RQ2: When fine-tuning LLMs as recommender systems, what are the performance differences of different prompting strategies?}
\end{itemize}

\begin{figure}[t]
    \centering
    \includegraphics[width=0.99\linewidth]{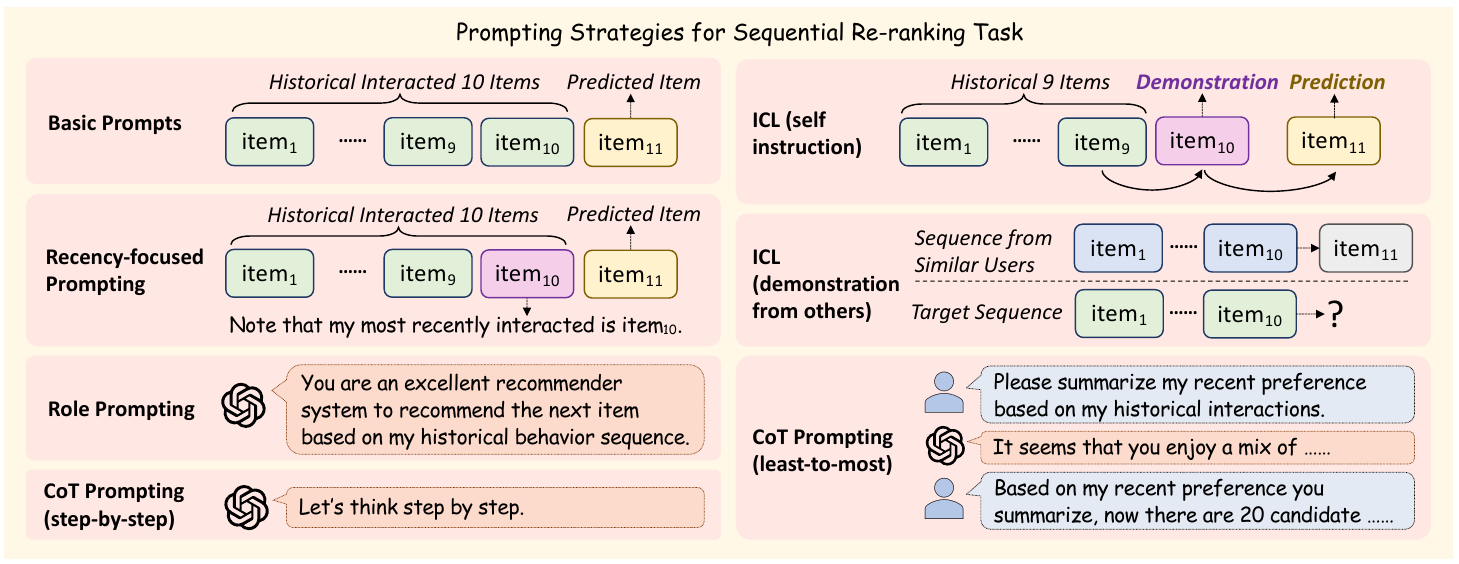}
    \caption{\revision{Examples of prompting strategies for LLMs in the scenario of the sequential re-ranking task based on Fig.~\ref{fig:basic-prompt-ranking}. ``ICL'' denotes In-Context Learning and ``CoT'' stands for Chain-of-Thoughts.}}
    \Description{
        Examples of prompting strategies for LLMs in the scenario of the sequential re-ranking task based on~\ref{fig:basic-prompt-ranking}. ``ICL'' denotes In-Context Learning and ``CoT'' stands for Chain-of-Thought.
    }
    \label{fig:fig-prompting-strategy}
\end{figure}

\revision{In the scenario of the sequential re-ranking task based on Fig.~\ref{fig:basic-prompt-ranking}, we conduct experiments with prompting strategies illustrated in Fig.~\ref{fig:fig-prompting-strategy}, and the LLM is ChatGPT: }

\begin{itemize}
    \item \textbf{Original}: it is the basic prompt used in our experiments as illustrated in Figure~\ref{fig:basic-prompt-ranking}.
    \item \textbf{\emph{w/o} Recency-focused prompting}: we remove prompts on the recently interacted item. 
    \item \textbf{\emph{w/} Role prompting}: we add the role-playing description to the original prompt. 
    \item \textbf{\emph{w/o} CoT~(step-by-step)}: we remove the magical spell ``let's think step by step'' in the original prompt to observe the performance differences.
    \item \textbf{\emph{w/} CoT~(least-to-most)}: we obtain recommendation results by prompting LLMs twice. First, LLMs are prompted to summarize the personal preference of users based on the recently interacted item sequence. Then in the second stage, we concatenate the summarized profile to the original prompt for recommendation. 
    \item \textbf{\emph{w/} ICL~(self instruction)}: we use the last interaction of the same user as the example for ICL~\cite{hou2023large}.
    \item \textbf{\emph{w/} ICL~(demonstration from others)}: we retrieve an example from historical sequences of other users for the current user, and add the demonstration to the original prompt. For the selection of examples, we encode recent items of users by textual representations, and search for similar user sequences based on the inner product between vectors.
\end{itemize}

\revision{In the scenario of fine-Tuning LLMs as the CTR predictor based on Fig.~\ref{fig:basic-prompt-ctr}, we employ LLaMA3-8B as the LLM and LoRA is utilized for PEFT. As for prompting strategies of fine-tuning LLaMA3-8B, we devise four prompts as follows:}

\begin{itemize}
    \item \revision{\textbf{Implicit prompting}: it is the basic prompt used in our experiments as illustrated in Fig.~\ref{fig:basic-prompt-ctr}. According to the rating threshold, we divide the items users interact with into likes and dislikes, and only provide the implicit feedback for LLMs as prompts.}
    \item \revision{\textbf{Explicit prompting}: compared to implicit prompting, explicit prompting does not divide items, and provides LLMs with explicit feedback of ratings scores. An example is provided in the third row of Table~\ref{tab:prompt-design-finetune}.}
    \item \revision{\textbf{Hybrid implicit and explicit prompting}: based on implicit prompting, hybrid prompting further adds ratings of the target user for each provided item. An example is provided in the fourth row of Table~\ref{tab:prompt-design-finetune}.}
    \item \revision{\textbf{CoT prompting}: we add the magical spell ``let's think step by step'' to the basic prompt of implicit prompting.}
\end{itemize}

\begin{table}[t]
\centering
\caption{Performance comparison on prompting strategies in Fig.~\ref{fig:fig-prompting-strategy} when leveraging ChatGPT as re-ranker. \revision{``\emph{w/o}'' denotes that we remove related descriptions compared to the original prompt, and ``\emph{w/}'' denotes that we add corresponding instructions to the prompt.}}
\small
\begin{tabular}{@{}ccccccc@{}}
\toprule
                                          & \multicolumn{3}{c}{\textbf{MovieLens-1M}}             & \multicolumn{3}{c}{\textbf{Amazon Books}}             \\ \cmidrule(l){2-7} 
\multirow{-2}{*}{\textbf{Method}}         & \textbf{N@1} & \textbf{N@10} & \textbf{N@20} & \textbf{N@1} & \textbf{N@10} & \textbf{N@20} \\ \midrule
\textbf{Original}                         & 0.1817          & 0.3985           & 0.4629           & 0.2467          & 0.4276           & 0.5054           \\ \midrule
\multicolumn{7}{c}{\cellcolor[HTML]{ECF4FF}\textit{Zero-shot Prompting}}                                                                                           \\ \midrule
\textbf{\emph{w/o} Recency-focused prompting}    & 0.1550          & 0.3781           & 0.4546           & 0.2300          & 0.4256           & 0.4978           \\ \midrule
\textbf{\emph{w/} Role prompting}                & {\ul 0.2750} & {\ul 0.4821}  & {\ul 0.5379}  & \textbf{0.2800} & \textbf{0.4814}  & \textbf{0.5489}  \\ \midrule
\textbf{\emph{w/o} CoT~(step-by-step prompting)} & 0.2400          & 0.4543           & 0.5193           & {\ul 0.2650} & 0.4439           & 0.5163           \\ \midrule
\textbf{\emph{w/} CoT~(least-to-most prompting)} & \textbf{0.2783} & \textbf{0.5099}  & \textbf{0.5603}  & 0.2617          & {\ul 0.4729}  & {\ul 0.5403}  \\ \midrule
\multicolumn{7}{c}{\cellcolor[HTML]{ECF4FF}\textit{Few-shot Prompting}}                                                                                            \\ \midrule
\textbf{\emph{w/} ICL~(self instruction)}         & 0.2000          & 0.4558           & 0.5171           & 0.2500          & 0.4265           & 0.5055           \\ \midrule
\textbf{\emph{w/} ICL~(demonstration from others)}         & 0.1900          & 0.4015           & 0.4684           & 0.2500          & 0.4039           & 0.4950           \\ \bottomrule
\end{tabular}
\label{tab:prompt-design-analysis}
\end{table}

\begin{table}[t]
\centering
\caption{\revision{Performance comparison of prompting strategies for fine-Tuning LLaMA3-8B on the CTR prediction task. The model is fine-tuned using LoRA, and the recommendation performance is evaluated using the accuracy metric on the MovieLens-1M (Movie) and Amazon Books (Book) datasets.}}
\small
\begin{tabular}{@{}p{0.68\textwidth}>{\centering\arraybackslash}p{0.12\textwidth}>{\centering\arraybackslash}p{0.12\textwidth}@{}}
\toprule
\textbf{Prompting   Strategy}                            & \textbf{Movie} & \textbf{Book} \\ \midrule
\textbf{Implicit Prompting} {\color[HTML]{656565}(Given the   user's preference and unpreference, identify whether the user will like the target item by answering "Yes." or "No.".)} &
  \multirow{2}{*}{\makecell{0.6577}} &
  \multirow{2}{*}{\makecell{0.8610}} \\ \midrule
\textbf{Explicit Prompting} {\color[HTML]{656565} (Given the   user's ratings of items they have interacted with (1-5, with 1 being   the lowest and 5 being the highest), identify $\ldots$)} &
  \multirow{2}{*}{\makecell{\textbf{0.6847}}} &
  \multirow{2}{*}{\makecell{\textbf{0.8990}}} \\ \midrule
\textbf{Hybrid Implicit and Explicit  Prompting} {\color[HTML]{656565} (Given the user's ratings of items they have interacted with, we consider ratings greater than or equal to 4 as preferences, and ratings less than 4 as   unpreferences, identify $\ldots$)} &
  \multirow{3}{*}{\makecell{0.6649}} &
  \multirow{3}{*}{\makecell{0.8978}} \\ \midrule
\textbf{CoT Prompting}   {\color[HTML]{656565} (Let’s think step by step.)} & {\ul 0.6704} & {\ul 0.8699} \\ \bottomrule
\end{tabular}
\label{tab:prompt-design-finetune}
\end{table}

\subsubsection{Observations and Discussion}

Findings on prompting strategies of LLMs are listed as follows:

(1) \emph{The impact of prompting strategies on prompting LLMs~(RQ1)}. As shown in Table~\ref{tab:prompt-design-analysis}, we compare different prompting strategies for ChatGPT as the zero-shot re-ranker, and report results based on the classification of zero-shot prompting and few-shot prompting. As for the zero-shot prompting strategies, we can see that removing the recency-focused prompting sentence~(\emph{w/o} recency-focused) largely decreases the recommendation performance, demonstrating the key role of recently interacted items in recommendation. Although we provide historical items in chronological order based on timestamps, LLMs still needs explicit guidance to understand the importance of recent items, indicating that heuristic knowledge in the recommendation field needs to be supplemented for LLMs. Meanwhile, adding the role prompting descriptions to the original prompt~(\emph{w/} role prompting) significantly improves the performance of zero-shot prompting, which shows that role-playing and expert-like prompts can better leverage the capabilities of LLMs in specific fields or tasks~\cite{zhao2023survey}. In addition, an important strategy for zero-shot prompting is the CoT, and we compare effects of two kinds of CoT prompting. When we remove the basic prompting sentence ``let's think step by step'' in the original prompt, the recommendation effect is improved, possibly due to the fact that following the step-by-step prompts is not conducive to the extraction of results from textual outputs. It also indicates that specific problem decomposition may be required for recommendation tasks rather than general prompts, and the superior results of summarizing recent interest~(least-to-most prompting strategy) confirm this finding. 
In contrast to zero-shot prompting, the typical representative of few-shot prompting is ICL based on contextual examples, and we study the ICL results considering demonstrations from the current user~(self) and other users~(others), respectively. It can be seen that using examples of the target user is better than demonstrations from other users, indicating the personalized needs of each user. However, the strategy of few-shot prompting has insignificant advantage compared to the zero-shot prompting, \revision{and more advanced planning strategies such as automatic prompting~\cite{li2024PAP-REC} and soft prompting~\cite{ramos2024PeaPOD} shed lights on LLM-based recommender systems.}

\revision{(2) \emph{The impact of prompting strategies on fine-tuning LLMs~(RQ1)}.
As shown in Table~\ref{tab:prompt-design-finetune}, we compare four prompting strategies for fine-Tuning LLaMA3-8B with LoRA on the CTR prediction task, highlighting the importance of prompt design in optimizing LLMs. Explicit prompting is the most effective strategy since it provides LLMs with quantifiable user preferences, enabling LLMs to capture nuanced user-item interactions. In contrast, Implicit prompting that relies on two-category data of user preferences gets poor results, suggesting that the lack of explicit guidance may hinder the capabilities of LLMs to infer user interest accurately.
The hybrid implicit and explicit prompting strategy, while combining elements of both approaches, does not surpass explicit prompting, indicating that it is not scientific to set the threshold of likes and dislikes artificially. CoT prompting with step-by-step reasoning demonstrates competitive performance, suggesting that incorporating reasoning pathways can enhance the decision-making process when fine-tuning LLMs. 
These results emphasize that the choice of prompting strategy is not merely a technical detail but a critical factor influencing the effectiveness of fine-tuning LLMs. Future work could explore more sophisticated hybrid approaches or adaptive prompting techniques that dynamically adjust based on the complexity of the input or the task at hand. Additionally, investigating the interplay between prompting strategies and model architectures could provide deeper insights into optimizing LLMs for recommendation tasks~\cite{tan2024IDGenRec,yin2024TTDS}.}

\begin{tcolorbox} [title = {Observations of Prompting Strategies}, left=0mm, right=8mm]
    \begin{itemize}
        \item Although provided with historical items in chronological order, LLMs still needs explicit guidance to understand the importance of recent items. 
        \item Role-playing and expert-like prompts can better leverage the capabilities of LLMs. 
        \item In chain-of-thought prompting, specific problem decomposition is required for recommendation tasks rather than general prompts.
        \item The few-shot prompting strategy has insignificant advantages in recommendations. 
        \item \revision{When fine-tuning LLMs as recommender systems, the explicit feedback~(\eg ratings) is better than the implicit feedback~(\eg interactions) to elicit capacities of LLMs.}
    \end{itemize}
\end{tcolorbox}

%% file: sec-task.tex
\subsection{Task Description} 
\label{sec:task-descriptions}

\begin{table}[t]
\caption{\revision{Task-specific prompts for LLMs as recommender systems, and we categorize them into two kinds of tasks, \ie accuracy-focused recommendation task and beyond-accuracy recommendation task.}} 
\small
\centering
\begin{tabular}{@{}>{\centering\arraybackslash}p{0.39\textwidth}p{0.58\textwidth}@{}}
\toprule
\textbf{Recommendation Task}  & \multicolumn{1}{c}{\textbf{Task-specific Prompts}}                                                                                                     \\ \midrule
\multicolumn{2}{c}{\cellcolor[HTML]{ECF4FF}\textit{Accuracy-focused Recommendation Task}}                                                                          \\ \midrule
\multirow{2}{*}{\makecell{Rating Prediction}}             & Based on the user's historical behavior and item features, predict the rating the user would give to the item~\cite{geng2022p5}.         \\ \midrule
\multirow{2}{*}{\makecell{Click-through Rate Prediction}} & Given the user's preference and unpreference, identify whether the user will like the target movie by answering "Yes." or "No."~\cite{bao2023tallrec}.   \\ \midrule
\multirow{2}{*}{\makecell{Sequential Recommendation}}     & A user has interacted with the following items in order, recommend the next item the user will interact with~\cite{hou2023large}.  \\ \midrule
\multirow{2}{*}{\makecell{Conversational Recommender System}} &
  Pretend you are a movie recommender system. I will give you a conversation between a user and you~\cite{he2023zero-shot-CRS}. \\ \midrule
\multicolumn{2}{c}{\cellcolor[HTML]{ECF4FF}\textit{Beyond-accuracy Recommendation Task}}                                                                           \\ \midrule
\multirow{2}{*}{\makecell{Explainable Recommendation}}    & As a movie recommender system, give reasons why the customer watches this movie Return of the Jedi~\cite{luo2023LLMXRec}. \\ \midrule
\multirow{2}{*}{\makecell{Fairness-aware Recommendation}} & Act as a fair recommender system balancing between popular and less-known movies to ensure fairness~\cite{deldjoo2024bias-llmrs}.          \\ \midrule
\multirow{3}{*}{\makecell{Diversified Recommendation}} &
  Your task is to re-rank this candidate list and provide a final top-$n$ recommendation list where the goal is to balance relevance and diversity~\cite{carraro2024DiverseReranking}. \\ \bottomrule
\end{tabular}\label{tab:task-description-prompt}
\end{table}

Task description of recommendation tasks for~\our~is a key part of prompts that makes LLM understand the goal of specific recommendation assignments. 
Typically, based on the different recommendation objectives, existing recommendation tasks can be divided into the following categories: point-wise optimization for a single item~(\eg binary cross entropy loss~\cite{rendle2010factorization}), pair-wise optimization for positive and negative items~(\eg Bayesian personalized ranking loss~\cite{rendle2009bpr}), and list-wise optimization for multiple items~(\eg variational auto-encoders~\cite{liang2018variational}). For~\our, we can adapt our work into multiple tasks by changing the description of prompts and the form of candidate items. That is to say, our framework can not only be used for re-ranking~\cite{hou2023large} and retrieving~\cite{li2023e4srec} tasks, but also support task forms such as point-wise prediction~\cite{kang2023llms-ctr}, pair-wise comparison~\cite{dai2023uncovering} and list-wise re-ranking formats. 
\revision{As illustrated in Table~\ref{tab:task-description-prompt}, we list task-specific prompts for LLMs as recommender systems.}


\paratitle{Point-wise recommendation}
~\cite{cheng2016wide} regards the recommendation task as a binary classification problem, such as the rating scoring task~\cite{geng2022p5,chu2023RecSysLLM} and click-through rate~(CTR) predictions~\cite{kang2023llms-ctr, bao2023tallrec}. It generates a score or likelihood of preference for a given user-item pair with feature engineering. The recommender system ranks items solely based on the individual characteristics or historical interactions of users without considering the comparison and mutual influence between items in the input ranking list. When utilizing LLMs for point-wise recommendation, the description mainly involves the user-item information with the range of score or answer list to limit output formats. 

\paratitle{Pair-wise recommendation}
~\cite{rendle2009bpr} involves comparing two items in pairs to determine the relative preference for a particular user. Instead of focusing on individual items, it evaluates pairs of items and calculates the semantic distance between them. This method often creates item pairs, calculates relative scores, and then ranks or recommends items based on pair-wise comparisons. The description of pair-wise recommendation in prompts consists of a positive item and a negative item, instructing LLMs to give the answer from a binary choice list~\cite{dai2023uncovering}.

\paratitle{List-wise recommendation}~\cite{liang2018variational}
involves optimizing an entire list of recommended items for a user. Instead of considering items individually or in pairs, this method treats the entire recommendation list as a single entity and captures the interior correlations within the list. It aims to create a ranked list of items that collectively maximizes user satisfaction. For LLM-based list-wise recommendations, the prompt contains a list of items and corresponding instructions, eliciting the ability of LLMs to explore the potential relationship within the item sequence~\cite{hou2023large,ma2023STELLA}. 

\paratitle{Matching}~\cite{bao2023bi,li2023e4srec}
in recommendation involves obtaining a small subset of candidates from the entire item pool for users, and candidate items are not provided in prompts at this stage. Due to the lack of domain knowledge in recommender systems, leveraging LLMs for item retrieval requires additional strategies such as model training, item indexing and grounding approaches~\cite{zheng2023LC-Rec}. The prompt for matching tasks includes the description of a user in different forms, \eg textual profiles and user-item interaction histories, in order to quickly match suitable items.



\paratitle{Ranking}
~\cite{luo2023recranker} in recommendation involves sorting or ordering items based on the predicted relevance or probability of interest to the target user. It aims to present the most relevant items at the top of the recommendation list. By analyzing historical user-item interactions, user preferences, item features, and the given cadidate item list, it requires LLMs to estimate the relevance of items and determine corresponding positions in the recommendation list~\cite{li2023ChatGPT-news,luo2023recranker,gao2023chat}. 

%% file: sec-interest.tex
\subsection{User Interest Modeling}
\label{sec:interest}

Compared to general tasks solved by LLMs, the characteristic of recommendation tasks lies in the mining and utilization of the user personalized interest~\cite{yao2023DOKE,shu2023rah}. Users are influenced by multiple factors when making recommendation decisions, including but not limited to their long-term preferences, short-term intentions, market popular tendencies and occasional environmental biases~\cite{cheng2016wide,he2017neural}. Among them, the most essential aspect is the interest of users, which is the purpose of recommender systems~\cite{he2020lightgcn,zhou2020s3,hou2022towards}. In this section, we first classify the user interest type~(Section~\ref{sec:prompt-user-interest-type}), outline the user representation forms~(Section~\ref{sec:prompt-user-representation-forms}), and summarize existing modeling methods~(Section~\ref{sec:prompt-modeling-methods}). Then, we discuss approaches to modeling the user interest~(Section~\ref{sec:prompt-rq-observe-short-term}), and conduct experiments to provide empirical analysis and key findings.

\subsubsection{User Interest Type}
\label{sec:prompt-user-interest-type}

In recommender systems, there are multiple ways to classify the user interest~\cite{rendle2009bpr,guo2017deepfm,zhou2020s3}. For example, based on the type of feedback between users and the platform, the user interest can be divided into the explicit interest~(\eg ratings) and implicit interest~(\eg clicks). Although the explicit feedback can reflect the true intentions of users, the sparse and expensive data limits its application scenarios. In general, we mainly explore user interest modeling in LLMs under implicit feedback scenarios. 
Another classification of the user interest is based on the time duration and stability, \ie \emph{short-term interest}, \emph{long-term interest}, and \emph{hybrid interest} as follows:

\paratitle{Short-term interest} refers to the sudden and accidental intentions or tendencies of personalized users in recent interactions, which is prone to change and can be influenced by environmental factors~\cite{lin2023rella,hou2023large}. That is to say, short-term interest is recent, temporary, and variable~\cite{zhang2023recommendation}.

\paratitle{Long-term interest} refers to the stable preferences of users towards certain content, themes, and elements, which is not easily changed in the short term and will continue to affect recommendation decisions of users~\cite{shu2023rah,wang2023recmind}. In contrast to short-term intentions, long-term interest is long-term, sustained and stable.

\paratitle{Hybrid interest} means the combination of long-term preferences and short-term intentions~\cite{friedman2023recllm,liu2023recprompt,zhiyuli2023bookgpt}. On the one hand, short-term intentions are dominated by long-term preferences~\cite{yue2023llamarec}. On the other hand, long-term preferences consist of short-term interest across periods~\cite{li2023pbnr,qiu2023controlrec}. 


\subsubsection{User Representation Forms}
\label{sec:prompt-user-representation-forms}

In~\our, user interest modeling should consider both the input form for the prompt template and the specific storage form for the interest memory~\cite{wang2023recmind}. Generally, there are three types of input contents for interest modeling in LLMs: \emph{historical item lists}, \emph{interest descriptions} and \emph{user embeddings}. Corresponding to different representation forms of the user interest, there are also various storage forms in the interest memory. 

\paratitle{Historical item lists} refer to representing personalized users based on the sequence of historical interacted items, which is widely used in session-based recommendation and sequential recommendation~\cite{hou2022towards,zhou2020s3,kang2018sasrec}. For token-based item indexing, the ID sequence of items is used for user modeling in LLMs, similar to traditional sequential recommendation models~\cite{geng2022p5}. However, without fine-tuning, LLMs cannot recognize the meaning of corresponding IDs. For description-based item indexing, existing researches generally concatenate the attributes~(\eg titles) of items in the temporal order, and then input the item sequence as text descriptions to LLMs as user representations~\cite{li2023recformer,zhiyuli2023bookgpt,hou2023large}. Only the item IDs that the user has interacted with need to be stored as the interest memory. The list of items can be stored in the form of \texttt{numpy} or \texttt{tensor} arrays. Despite the simplicity and effectiveness, text attributes such as titles cannot fully represent items with ambiguity~\cite{li2023recformer}. At the same time, item sequences inevitably contain noise information~\cite{wei2023llmrec}, and limited sequence lengths make it difficult to accurately express the user interest. 

\paratitle{Interest descriptions} refer to representing users based on textual descriptions, which is more applicable to the text input of LLMs~\cite{yao2023DOKE}. That is to say, we can use natural languages to model the user interest in the form of text and input them into LLMs~\cite{xi2023towards-kar}. To facilitate the storage and retrieval of textual descriptions, \emph{vector stores} are commonly used~\cite{touvron2023llama2}. In LLMs, vector stores in the form of key-value pairs can connect implicit vector embeddings with explicit textual descriptions, improving the retrieval efficiency and comprehension ability~\cite{li2023e4srec,petrov2023gptrec}. However, the key lies in how to mine and describe the user interest, and this is what we should make efforts to discuss.

\paratitle{User embeddings} refer to concatenating embeddings of users to the input of LLMs as user representations~\cite{zhang2023collm}, which is often used for efficient fine-tuning of language models. In this case, we assign each user a unique ID and corresponding embedding, and add the vector representations of users to the input of LLMs for fine-tuning. At the same time, user embeddings can also be trained from small-scale traditional recommendation models~\cite{zhang2023bridging}, thereby incorporating collaborative filtering features from other users. However, it is worth noting that vector embeddings of user representations are only suitable for scenarios where the domain knowledge can be injected into LLMs, and the black-box property of embeddings also poses challenges to the explainability. 


\subsubsection{Modeling Methods}
\label{sec:prompt-modeling-methods}

\begin{figure}[t]
    \centering
    \includegraphics[width=0.95\linewidth]{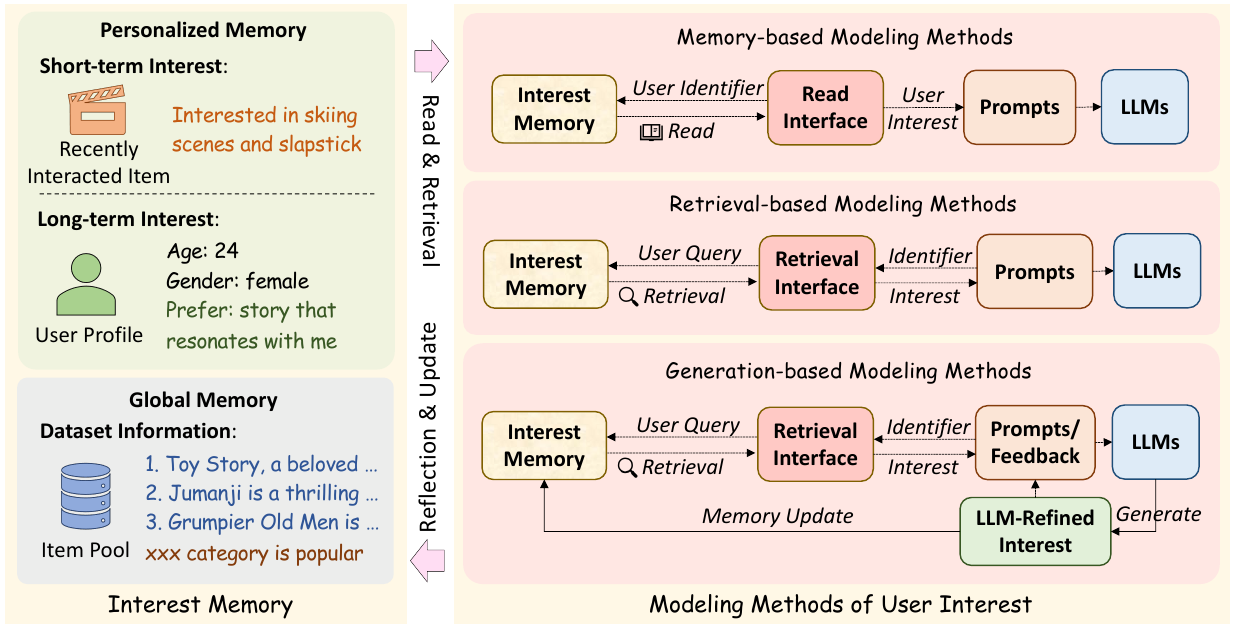}
    \caption{\revision{Comparison of three modeling methods of user interest, \ie memory-based methods, retrieval-based methods, and generation-based methods. Interest memory consists of global and personalized memory.}}
    \Description{
        Comparison of three modeling methods of user interest, \ie (1) memory-based methods, (2) retrieval-based methods, and (3) generation-based methods. Interest memory consists of global and personalized memory.
    }
    \label{fig:fig-user-interest-modeling}
\end{figure}

\revision{As illustrated in Fig.~\ref{fig:fig-user-interest-modeling}}, we classify methods for modeling interest as \ie \emph{memory-based methods}, \emph{retrieval-based methods} and \emph{generation-based methods}.

\paratitle{Memory-based methods} assign external memory to users for storing interest-related historical information~\cite{xi2023towards-kar,shu2023rah}. As for user representations, the encoded interest of users can be obtained from the memory~\cite{zhang2023bridging}, and LLMs are instructed to utilize the memorized interest with carefully designed prompts. The iterative updating of memories is the key to memory-based methods. As for personalized memory, there are three important operations~\cite{wang2023survey}: (1) \emph{memory reading} is to obtain contents of the interest memory based on the user identifier. (2) \emph{Memory writing} is to augment new interest of users based on the latest interactions between users and items. (3) \emph{Memory reflection} is the periodic examination and updating of existing issues in the memory. Strategies for memory reflection include but are not limited to self-summarization, self-correction, and reflection based on user feedback~\cite{wang2023enhancing}. It is memory reflection that makes memory not just a stack of historical records but a summary of user interest~\cite{yao2023DOKE}. Note that memory-based methods specifically refer to obtaining contents from the memory without further processing procedures.

\paratitle{Retrieval-based methods} add a module for personalized interest retrieval on top of the memory-based methods, utilizing retrieval strategies to obtain user interest from the personalized memory~\cite{wang2023recmind,lin2023rella,huang2023recommender}. As for the personalized query, there are generally three ways to form a query for retrieval: (1) candidate items as search queries~\cite{lin2023rella} for relevant items, (2) recently interacted items as search queries for similar items~\cite{zhang2023agentcf}, and (3) the user profile as search queries for personalized items~\cite{wang2023recmind}. For the criteria of retrieval, there are multiple trade-offs, including dimensions such as relevance, recency, and diversity with respect to the user interest~\cite{wang2023survey}.

\paratitle{Generation-based methods} utilize the generation capabilities of LLMs to summarize, infer, and derive comprehensive user interest~\cite{yao2023DOKE,wang2023zero,luo2023recranker}. Generally, a combination of memory-based and retrieval-based methods is required for generation-based methods. For generative language models, the user interest induced by generative retrieval can further stimulate the prompting ability of LLMs~\cite{liu2023genre-news,zhang2023generative}. The generative approach can also be used for data augmentation, leveraging the general knowledge of LLMs to augment textual descriptions of the user interest. In current recommendations, LLMs mainly utilize the retrieved relevant items to generate summarized interest descriptions, which can also serve as a strategy for the memory reflection~\cite{wang2023recmind,shu2023rah}.

\subsubsection{\revision{Research Questions and Observations for Modeling User Interest}}
\label{sec:prompt-rq-observe-short-term}

\revision{
In this section, we explore methods for user interest modeling in~\our~\wrt the following two research questions.}

\revision{
\begin{itemize}
    \item RQ1: For retrieval-based modeling methods, how to design query and memory to enrich the expression of user interest?
    \item RQ2: How do different methods of user interest modeling affect the recommendation results?
\end{itemize}}


\revision{In this section, we conduct experiments for the zero-shot re-ranking task as illustrated in Fig.~\ref{fig:basic-prompt-ranking}, and ChatGPT is employed as the zero-shot re-ranker.}

For RQ1, to analyze how to select recent and relevant items as the representation of user interest, we adopt the retrieval-based strategy to retrieve items from the long-term interest. 
Specifically, following the memory-based and retrieval-based strategy, the personalized query, \eg candidate items, recently interacted items or the user profile, is encoded into vectors in a unified form. 
The most relevant items are retrieved from memory based on semantic similarity. Meanwhile, we set weights based on the recency of interactions so that the recent items have a higher probability of being retrieved. 
Finally, the corresponding retrieved text is obtained based on key-value pairs in the memory. Here, we explore the design of the personalized query and memory of users for item retrieval, and study the influence of different variants as follows.

\begin{itemize}
    \item \textbf{Variants of the personalized query}: as for the personalized query, we utilize the interest of users for retrieving items, and compare two strategies for constructing the query.
    \begin{itemize}
        \item \textbf{Short-term interest}: we utilize the short-term interest for the memory retrieval of the long-term interest. The summarized text based on 10 recently interacted items is employed as the query for retrieving items.
        \item \textbf{Long- and short-term interest}: we first summarize the personalized profile of each user based on all items he or she has interacted, then concatenate the user profile and the summarized text based on 10 recently interacted items for the memory retrieval of the long-term interest. We utilize the long-term user profile and short-term interest for the memory retrieval of the long-term interest.
    \end{itemize}
    \item \textbf{Variants of the memory}: as for the personalized memory, we also consider two variants.
    \begin{itemize}
        \item \textbf{Global memory}: we utilize descriptions of items in the datasets as the memory. Since all users in a dataset share the same item space, the memory \wrt item descriptions is not the personalized memory but global memory. 
        \item \textbf{Personalized memory}: for each item that a user interacts with, we instruct LLMs to generate personalized descriptions based on the rating score or comment text as personalized memories. Therefore, the description of the same item varies among personalized users, which is different from the global memory.
    \end{itemize}
\end{itemize}

\revision{
For RQ2, to examine the impact of modeling forms on the user interest, we design 10 modeling forms according to short-term interest, long-term interest and hybrid interest as follows:}

\begin{itemize}
    \item \textbf{Variants of short-term interest modeling forms}: 
    \begin{itemize}
        \item \textbf{Recent items}: we use titles of the recent 10 items as the short-term interest. 
        \item \textbf{Personalized interest of recent items}: we construct the personalized memory for each user based on the user-item interactions, and use the personalized descriptions of recently interacted items as the short-term interest.
        \item \textbf{Recent items with personalized interest}: we employ both titles of the recent 10 items, and the personalized text of recent items to prompt LLMs as the short-term interest.
        \item \textbf{Recent items with LLM-generated short-term interest}: we utilize the least-to-most prompting strategy as mentioned in Section~\ref{sec:prompt}, which summarizes the interest text based on the recently interacted 10 items, then concatenate the summarized text and the 10 items. 
    \end{itemize}
    \item \textbf{Variants of long-term interest modeling forms}: 
    \begin{itemize}
        \item \textbf{Retrieved items from user history, recent items as query}: it retrieves 10 relevant items from the long-term interest. Since recent items have a higher probability to be retrieved. 
        \item \textbf{Retrieved personalized interest, recent items as query}: it also retrieves 10 relevant items from the long-term history like ``Retrieved items from user history, recent items as query'', but this variant uses the personalized descriptions of items from the personalized memory. 
        \item \textbf{LLM-generated summarization of retrieved personalized interest}: it employs the retrieved personalized text from long-term history in the last variant by summarizing the retrieved 10 texts into one description. Not providing 10 specific items for LLMs, this variant only summarizes the user interest through personalized text.
    \end{itemize}
    \item \textbf{Variants of hybrid interest modeling forms}: 
    \begin{itemize}
        \item \textbf{Recent and retrieved relevant items}: it uses both the retrieved 10 items from long-term history and the recent 10 items as the representation of user interest, considering both the long-term and short-term interest. It is worth noting that the retrieved items in this variant do not include the latest 10 items.
        \item \textbf{Retrieved personalized interest, profile and recent items as query}: it retrieves recent and relevant items from the long-term interest, and for the query, both the personalized user profile and short-term interest are utilized.
        \item \textbf{Recent items with LLM-generated long-term interest}: it concatenates the user profile summarized by LLMs and the recently interacted 10 items for interest modeling. 
    \end{itemize}
\end{itemize}

\begin{table}[t]
\centering
\caption{\revision{Performance comparison of user interest modeling forms across three modeling methods. The {\color[HTML]{9A0000} red} color indicates short-term interest modeling, {\color[HTML]{3166FF} blue} indicates long-term interest modeling, and {\color[HTML]{6200C9} purple} indicates hybrid interest modeling. Evaluation metrics N@1 and N@10 represent NDCG@1 and NDCG@10, respectively, and the same below. Following the experimental setup outlined in Table~\ref{tab:overall-llms}, we evaluate the zero-shot recommendation performance of ChatGPT on the MovieLens-1M and Amazon Books datasets.}}
\small
\begin{tabular}{@{}lcccc@{}}
\toprule
\multicolumn{1}{c}{}                                                            & \multicolumn{2}{c}{\textbf{MovieLens-1M}} & \multicolumn{2}{c}{\textbf{Amazon Books}} \\ \cmidrule(l){2-5} 
\multicolumn{1}{c}{\multirow{-2}{*}{\textbf{Modeling Form of User   Interest}}} & \textbf{N@1}             & \textbf{N@10}          & \textbf{N@1}             & \textbf{N@10}           \\ \midrule
\multicolumn{5}{c}{\cellcolor[HTML]{ECF4FF}\textit{Memory-based Modeling Methods}}                                                                    \\ \midrule
{\color[HTML]{9A0000} Recent items}                                             & 0.1817          & 0.3985         & 0.2467          & 0.4276         \\ \midrule
{\color[HTML]{9A0000} Personalized interest of recent items}                    & 0.2100          & 0.4380         & 0.2500          & 0.4733         \\ \midrule
{\color[HTML]{9A0000} Recent items with personalized interest}                  & 0.2350          & 0.4483         & 0.3000          & {\ul 0.5076}   \\ \midrule
\multicolumn{5}{c}{\cellcolor[HTML]{ECF4FF}\textit{Retrieval-based Modeling Methods}}                                                                 \\ \midrule
{\color[HTML]{3166FF} Retrieved relevant items from user history, recent items as query}       & 0.1950          & 0.4343          & 0.2850          & 0.4650          \\ \midrule
{\color[HTML]{6200C9} Recent and retrieved relevant items}                      & 0.2400          & 0.4441         & {\ul 0.3250}    & 0.4636         \\ \midrule
{\color[HTML]{3166FF} Retrieved personalized interest, recent items as query}                  & {\ul 0.2550}    & {\ul 0.4714}    & 0.3150          & \textbf{0.5241} \\ \midrule
{\color[HTML]{6200C9} Retrieved personalized interest, profile and recent items as query} & 0.2400          & 0.4563          & \textbf{0.3300} & 0.4712          \\ \midrule
\multicolumn{5}{c}{\cellcolor[HTML]{ECF4FF}\textit{Generation-based Modeling Methods}}                                                                \\ \midrule
{\color[HTML]{9A0000} Recent items with LLM-generated short-term interest}                    & \textbf{0.2783} & \textbf{0.5099} & 0.2617          & 0.4729          \\ \midrule
{\color[HTML]{6200C9} Recent items with LLM-generated long-term interest}                    & 0.2000          & 0.4466          & 0.2500          & 0.4925          \\ \midrule
{\color[HTML]{3166FF} LLM-generated summarization of retrieved personalized interest}          & 0.2000          & 0.4492          & 0.2500          & 0.4872          \\ \bottomrule
\end{tabular}
\label{tab:interest-model-form}
\end{table}

\begin{table}[t]
\centering
\caption{Performance comparison of query and memory designs for retrieval-based methods. We use retrieved relevant items from memory as user interest in prompts. The impact of query forms (short-term interest vs. long- and short-term interest) and memory forms (global memory vs. personalized memory) is analyzed.}
\small
\begin{tabular}{@{}cccccc@{}}
\toprule
\multirow{2}{*}{\textbf{Query}}                                                                         & \multirow{2}{*}{\textbf{Memory}} & \multicolumn{2}{c}{\textbf{MovieLens-1M}} & \multicolumn{2}{c}{\textbf{Amazon Books}} \\ \cmidrule(l){3-6} 
                                                                                                        &                                  & \textbf{N@1}     & \textbf{N@10}    & \textbf{N@1}     & \textbf{N@10}    \\ \midrule
Short-term interest                                                                      & Global memory                 & 0.2000              & 0.4295              & 0.2283              & 0.4144              \\
Long- and short-term interest                                                                    & Global memory                 & {\ul 0.2200}              & {\ul 0.4473}              & 0.2400              & 0.3980              \\
Short-term interest                                                                      & Personalized memory              & \textbf{0.2400}     & 0.4441              & {\ul 0.3250}              & {\ul 0.4636}              \\
Long- and short-term interest & Personalized memory              & \textbf{0.2400}     & \textbf{0.4563}     & \textbf{0.3300}     & \textbf{0.4712}     \\ \bottomrule
\end{tabular}
\label{tab:interest-query-memory}
\end{table}

\subsubsection{Observations and Discussion}
\revision{
As shown in Table~\ref{tab:interest-model-form} and Table~\ref{tab:interest-query-memory}, we summarize the following findings for modeling the user interest in~\our: 
}

(1) \emph{The design of query and memory for retrieval-based modeling methods~(RQ1)}.
We conduct an experiment to explore the retrieval strategies for recent and relevant items, and evaluate the selection of query and memory for item retrieval. As shown in Table~\ref{tab:interest-query-memory}, we compare four combinations with queries and memories. In terms of the query, combining the design of long-term and short-term interest yields better results than only the short-term interest. When selecting items that represent interest for recommendation, the query needs to consider both recent tendencies and long-term preferences of users. Our results have shown that the personalized long-term interest of users can have a positive effect on retrieval for interest modeling. In terms of the memory, the design of the personalized memory with user interest outperforms the global memory with general descriptions. This result is in line with features of personalized recommender systems~\cite{shu2023rah,wang2023recmind}, indicating the importance of personalization when designing the user interest for LLMs~\cite{zhang2023agentcf}.

\revision{(2) \emph{The impact of different modeling forms of the user interest~(RQ2)}.
In Table~\ref{tab:interest-model-form}, we compare the user interest modeling forms across three modeling methods (memory-based, retrieval-based and generation-based), and highlight three interest types (short-term, long-term and hybrid interest) with three colors. From the perspective of modeling methods, the recommendation performance of retrieval-based and generation-based methods is generally better than that of memory-based methods, indicating the importance of Retrieval-Augmented Generation~(RAG) from the interest memory. Meanwhile, on the MovieLens-1M dataset with dense interactions, generative summarization based on short-term interest can further improve the effect, while on the Amazon books dataset with sparse interactions, retrieval-based modeling methods achieve the best, indicating that the retrieval results cannot be completely replaced by summative sentences. On the other hand, long-term interest and short-term interest can complement each other, and combining the two may yield better results. Instead of concatenating the long-term and short-term sequences, one possible combination approach is to use the short-term interest as queries to retrieve the long-term interest, serving as the domain knowledge for LLMs to model user preferences.}

\revision{
\begin{tcolorbox} [title = {Observations of Modeling User Interest}, left=0mm, right=8mm]
    \begin{itemize}
        \item While selecting recent and relevant items, it is preferable to combine long-term and short-term user profiles with personalized descriptions for query and memory. 
        \item \revision{The recommendation performance of retrieval-based and generation-based methods is generally better than that of memory-based methods, indicating the importance of Retrieval-Augmented Generation~(RAG) from the interest memory.}
        \item \revision{Long-term interest and short-term interest can complement each other, and combining the two may yield better results.} Retrieving long-term preferences based on short-term intentions is a suitable way for forming hybrid interest descriptions, and more approaches can be further explored.
    \end{itemize}
\end{tcolorbox}}

%% file: sec-candidate.tex
\subsection{Candidate Items Construction}
\label{sec:candidate}

\subsubsection{Procedures of Candidate Items Construction}

\begin{figure}[htbp]
    \centering
    \includegraphics[width=0.95\linewidth]{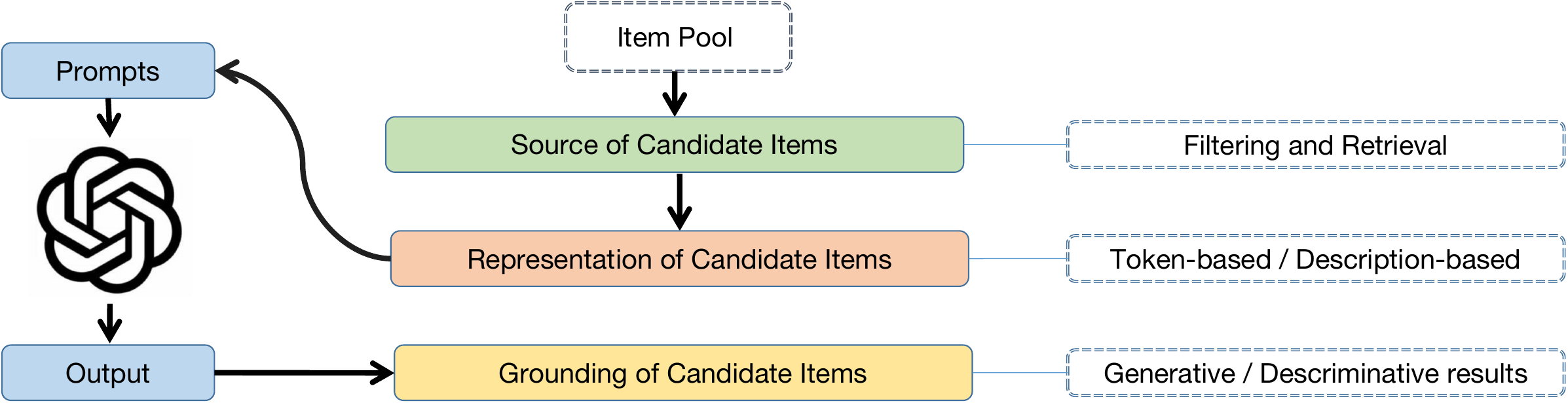}
    \caption{Main procedures of candidate items construction in ~\our, including the source, representation and grounding of candidate items.}
    \Description{Main procedures of candidate items construction in ~\our, including the source, representation and grounding of candidate items.}
    \label{fig:fig-candidate}
\end{figure}

When leveraging LLMs as recommender systems, a crucial problem is that LLMs tend to recommend items not in the dataset~\cite{geng2022p5}.
One of the approaches is to provide a limited number of candidate items for model selection. On the one hand, the source of candidate items determines the upper limit of personalized recommendation accuracy~\cite{hou2023large,zhang2023recommendation}. On the other hand, the form of candidate items reflects the specific recommendation task and exerts huge impacts on performance of LLMs for recommendation~\cite{dai2023uncovering}. 
As shown in Fig.~\ref{fig:fig-candidate}, we will discuss the main procedures of processing candidate items, including the source, representation and grounding of candidate items, and then conduct experiments on the effects of candidate items.

\paratitle{Source of candidate items}.
In practical industrial applications, recommendation is a multi-stage process, including two kinds of typical stages, \ie the recall stage and the re-ranking stage~\cite{bao2023bi,guo2017deepfm,rendle2009bpr}. The recall stage in recommender systems is used for preliminary filtering and selection from the entire set of item candidates, while the re-ranking stage is used for readjusting the selected candidates, which is more suitable for LLMs as the re-ranker~\cite{hou2023large,ma2023STELLA}. Therefore, the source of candidate items to be ranked is crucial for the results of LLMs for recommendation. Popular approaches for selection of candidate items include traditional recommendation models~\cite{yue2023llamarec} and retrieval algorithms~\cite{lin2023rella}, both of which provide a list of potentially recommended items. 

\paratitle{Representation of candidate items}.
In the construction of prompts, we need to use historical items to represent users and provide candidate items for LLMs to rank, both of which involve item indexing~\cite{geng2022p5,lin2023transrec}. However, there is a gap between the item representation in LLMs and recommendation, so the indexing of items between LLMs and recommender systems becomes the key and difficult point of~\our. There are three typical ways to index items: 1) token-based identifiers, 2) description-based identifiers and 3) hybrid identifiers~\cite{zheng2023LC-Rec,wang2024LETTER}. For token-based identifiers, researchers often use numerical IDs to identify items~\cite{geng2022p5}. Since token-based identifiers are widely utilized in traditional recommendation model, they are naturally treated as an item-indexing way for aligning LLMs with recommendation tasks. For description-based identifiers, researchers generally use the title or other textual attributes of the item as an identifier, and formalize it into natural language as input~\cite{dai2023uncovering,liu2023chatgpt,bao2023bi}. Due to the richness of natural language, description-based identifiers have high readability and flexibility. For hybrid identifiers, considering collaborative and textual information of users or items, LLMs can perceive both token-based and description-based identifiers~\cite{liao2023llara,li2023e4srec}. Current work using hybrid identifiers can be divided into different types. Some of them add item sequence with textual description and ID information respectively into different positions of prompts~\cite{zhang2023collm}, while others combine token-based and description-based embeddings together for each item in sequences through concatenation~\cite{li2023e4srec}. 

\paratitle{Grounding to candidate items}.
After constructing candidate items for recommendation, how to ground the output of LLMs into recommendation item lists is a crucial problem. Methods for grounding to candidate items are associated with different types of recommendation tasks. For discriminative recommendation, which provide candidate item lists in the prompts, it is easy for LLMs to follow instructions and generate the output right within the candidate item lists through elaborately designed prompts~\cite{zhiyuli2023bookgpt,liao2023llara}. For generative recommendation tasks, which do not explicitly provide candidate item list in prompts but require further processing for the output, recent work indicates effective approaches for mapping the output of LLMs into candidate item lists~\cite{zheng2023LC-Rec}. For example, GPT4Rec~\cite{li2023gpt4rec} first generates hypothetical ``search queries'' using titles of recently interacted items, and then retrieves similar items from the item pool by BM25 algorithm. E4SRec~\cite{li2023e4srec} directly utilizes the nearest neighbor search between the output of LLMs and the vectors in the item linear projection component for grounding to candidate items. In this case, the recommendation process will not suffer from length limitation problems since they choose not to include candidate item lists into the input of LLMs. 

As shown in Figure~\ref{fig:fig-framework}, our framework involves two mapping processes between the recommendation and language space: (1) when using LLMs to represent users, items in the recommendation space need to be used as historical sequences. (2) When re-ranking candidate items, it is required to provide candidate items in the recommendation space for LLMs. The key issue of semantic space alignment is the indexing and grounding of items. As for the item indexing, token-based sequences and description-based natural languages are two typical forms. As for the grounding methods, the generative method renders LLMs to output labels of all candidate items, and other mapping strategies such as logits distribution and similarity calculation can also be considered~\cite{yue2023llamarec,li2023e4srec}.

\subsubsection{Research Questions and Experimental Setup}
In this section, we explore the effect of candidate items on~\our, and provide empirical findings on the following research questions:
\begin{itemize}
    \item RQ1: Whether representations of items lead to similar recommendation performance?
    \item RQ2: Given a list of candidates recalled by traditional recommenders, can LLMs improve the recommendation results by further re-ranking the given candidates?
    \ignore{\item RQ3: How to select suitable grounding methods for improving recommendation performance?}
\end{itemize}

We explore the performance differences between the following two item representation methods. \revision{Basic prompts used in this section is illustrated in Fig.~\ref{fig:basic-prompt-ranking}, and ChatGPT is employed as the zero-shot re-ranker.}

\begin{itemize}
    \item \revision{\textbf{Token-based identifiers}}: we can assign serial numbers or indicators~(\eg ABCD) to the candidate items, and instruct LLMs to only output the re-ranked identifiers.
    \item \textbf{Description-based identifiers}: we instruct LLMs to directly output the complete text~{\eg titles} of candidate items, and the re-ranking results are parsed from the textual output.
\end{itemize}

Moreover, researchers tend to adopt a two-stage recommendation paradigm, in which conventional recommenders initially recall multiple candidates, and subsequently, LLMs evaluate and rank these candidates to provide enhanced recommendations. This approach effectively reduces the computational load of LLMs, enabling more efficient deployment. However, there is still no consensus regarding the effectiveness of LLMs in reranking the candidates retrieved by traditional recommendation models~\cite{hou2023large, ren2023RLMRec}. In order to address this, we conduct experiments to compare the impact of ChatGPT in re-ranking the original recommendation results generated by traditional recommenders. We evaluate three classical recommenders, including Pop, BPR~\cite{rendle2009bpr}, and SASRec~\cite{kang2018sasrec}. 

\begin{table}[t]
\centering
\caption{Performance comparison on the grounding forms of LLMs \wrt candidate items. \revision{Token-based identifiers represent items numerically, while the description-based identifiers denote textual titles.}}
\small
\begin{tabular}{@{}ccccccc@{}}
\toprule
\multirow{2}{*}{\textbf{Method}}       & \multicolumn{3}{c}{\textbf{MovieLens-1M}}             & \multicolumn{3}{c}{\textbf{Amazon Books}}             \\ \cmidrule(l){2-7} 
                                       & \textbf{N@1} & \textbf{N@10} & \textbf{N@20} & \textbf{N@1} & \textbf{N@10} & \textbf{N@20} \\ \midrule
\textbf{\revision{Token-based identifiers}}          & 0.2750          & 0.5423           & 0.5725           & 0.0600          & 0.1473           & 0.1685           \\
\textbf{Description-based identifiers} & \textbf{0.3100} & \textbf{0.5828}  & \textbf{0.6043}  & \textbf{0.1950} & \textbf{0.4546}  & \textbf{0.5079}  \\ \bottomrule
\end{tabular}
\label{tab:prompt-candidate-grounding}
\end{table}


\begin{table}[t]
\centering
\small
\caption{Performance comparison of traditional recommendation methods with and without ChatGPT as a re-ranker. \revision{``+ LLM'' denotes the recommendation results of LLMs by re-ranking the top-20 candidate items of traditional models. ``Impr.'' denotes the percentage improvement (or degradation) achieved by LLMs.}}
\begin{tabular}{@{}ccccccc@{}}
\toprule
\multirow{2}{*}{\textbf{Method}} & \multicolumn{3}{c}{\textbf{MovieLens-1M}}             & \multicolumn{3}{c}{\textbf{Amazon Books}}             \\ \cmidrule(l){2-7} 
                                 & \textbf{N@1} & \textbf{N@10} & \textbf{N@20} & \textbf{N@1} & \textbf{N@10} & \textbf{N@20} \\ \midrule
\textbf{Pop}                     & \textbf{0.0000} & \textbf{0.0071}  & \textbf{0.0110}  & \textbf{0.0000} & 0.0000           & 0.0012           \\
\textbf{Pop~(+ LLM)}             & \textbf{0.0000} & 0.0033           & 0.0086           & \textbf{0.0000} & \textbf{0.0014}  & \textbf{0.0014}  \\
\textbf{Impr.}                   & 0.00\%          & -53.52\%         & -21.82\%         & 0.00\%          & 1400.00\%        & 16.67\%          \\ \midrule
\textbf{BPR}                     & 0.0000          & 0.0084           & 0.0135           & 0.0000          & 0.0064           & 0.0076           \\
\textbf{BPR~(+ LLM)}             & \textbf{0.0050} & \textbf{0.0146}  & \textbf{0.0170}  & \textbf{0.0050} & \textbf{0.0082}  & \textbf{0.0107}  \\
\textbf{Impr.}                   & 5000.00\%       & 73.81\%          & 25.93\%          & 5000.00\%       & 28.13\%          & 40.79\%          \\ \midrule
\textbf{SASRec}                  & \textbf{0.0750} & \textbf{0.1600}  & \textbf{0.1836}  & 0.0450          & 0.1129           & 0.1318           \\
\textbf{SASRec~(+ LLM)}          & 0.0700          & 0.1475           & 0.1736           & \textbf{0.0750} & \textbf{0.1242}  & \textbf{0.1454}  \\
\textbf{Impr.}                   & -6.67\%         & -7.81\%          & -5.45\%          & 66.67\%         & 10.01\%          & 10.32\%          \\ \bottomrule
\end{tabular}
\label{tab:prompt-candidate-improvement-source}
\end{table}

\subsubsection{Observations and Discussion}

Findings of candidate items construction are as follows.

(1) \emph{Grounding forms of candidate items~(RQ1)}. Firstly, we focus on the grounding forms of candidate items. As shown in Table~\ref{tab:prompt-candidate-grounding}, we can see that whether in the movie dataset or the book dataset, the complete name of the output item is more convenient for extracting and grounding recommendation results than the specified identifier. In other words, description-based identifiers outperform Token-based identifiers when grounding candidate items for LLMs~\cite{zhao2023survey,hou2023large}. However, whether it is ID-based or description-based, the inference time of the grounding strategy based on generative output will increase with the increasing number of candidate items. A feasible solution is to use one prediction to obtain the probability score on all candidate items based on the output logits~\cite{yue2023llamarec}. Language models are more sensitive to the textual output rather than pure identifiers, but we can also use more advanced index strategies and grounding methods to improve the accuracy. 

(2) \emph{The evaluation for re-ranking abilities of LLMs~(RQ2)}. Secondly, we discuss the performance difference between traditional algorithms and re-ranking results after LLMs. As shown in Table~\ref{tab:prompt-candidate-improvement-source}, LLMs improve results of the traditional recommendation method BPR on two datasets by a large margin. 
As for the basic method Pop and SASRec, the re-ranking effect of LLMs varies between the two datasets. For the MovieLens-1M dataset in the movie domain, our initial prompts cannot instruct LLMs to achieve better results than traditional methods such as Pop and SASRec. The possible reason is that this movie dataset is very dense, and fully-trained collaborative filtering signals are more important than general knowledge of movies. While in the Amazon Books dataset, the re-ranking of LLMs can further improve results of Pop and SASRec, possibly because the general knowledge of LLMs for books can effectively adapt to the sparse recommendation data. The overall experimental results indicate that the re-ranking performance of LLMs for recommendation results varies depending on specific recommendation methods and datasets. However, it is acknowledged that the general knowledge of LLMs can complement the collaborative signals of traditional models, tapping the potential to improve traditional recommendation algorithms~\cite{li2023ctrl, zhang2023collm}.

\begin{tcolorbox} [title = {Observations of Candidate Items Construction}, left=0mm, right=8mm]
    \begin{itemize}
        \item \revision{LLMs may exhibit limitations in retaining critical information regarding the candidate list from prompt instructions, potentially leading to responses that deviate from the intended requirements.}
        \item \revision{Description-based identifiers outperform token-based identifiers when grounding candidate items.}
        \item Retrieving candidate items by traditional recommendation models first, and then re-ranking items by LLMs can further improve the results. 
    \end{itemize}
\end{tcolorbox}

%% file: sec-con.tex
\section{Conclusion}
\label{sec:con}


This paper aims to provide a comprehensive exploration of Large Language Models~(LLMs) to serve as recommender systems. It presents a systematic review of the advancements made in LLM-based recommendations, generalizing related work into multiple scenarios and tasks in terms of LLMs and prompts. We also conduct extensive experiments on two public datasets to investigate empirical findings for recommendation with LLMs. Our objective is to assist researchers in gaining a deeper understanding of the characteristics, strengths, and limitations of LLMs leveraged as recommender systems. Considering the significant progress in LLMs, the development of LLM-based recommendations holds the potential to better align the powerful capabilities of LLMs with the evolving needs of intended users in the field of recommender systems. By addressing current challenges, we hope that our work will contribute to the advancement of LLM-based recommendations and serve as an inspiration for future research efforts. Last but not least, we outline promising directions for future research in leveraging LLMs for recommendation.

\paratitle{Efficiency optimization of LLMs for recommendation}. The key limitation of leveraging LLMs in industrial recommender systems is efficiency\cite{li2023survey,wu2023survey,fan2023survey}, including considerations of both time and space. On the one hand, the fine-tuning and inference efficiency of LLMs cannot compare to traditional recommendation models~\cite{zheng2023LC-Rec,hou2023large}. While techniques such as parameter-efficient fine-tuning can aid in keeping LLMs updated in a computationally efficient manner, recommender systems need to iterate continuously over time, \ie incremental learning. Frequent updates of LLMs inevitably bring spatial and temporal burdens to recommender systems~\cite{shi2023LSAT}. On the other hand, billions of parameters in LLMs also pose challenges for the lightweight deployment of recommendation algorithms~\cite{touvron2023llama,touvron2023llama2}. Therefore, efficiency optimization of LLMs utilized as recommender systems is one of the prerequisites for large-scale applications, which has widespread application prospects and scientific research values~\cite{shi2023LSAT,wang2023key-value,huang2023recommender}.

\paratitle{Knowledge distillation of LLMs for recommendation}. Since LLMs as recommenders are limited by efficiency, another feasible approach is to distill~\cite{sun2024PRM-KD,li2023ctrl} the recommendation capabilities of LLMs into lightweight models, striking a balance between efficiency and effectiveness. Specifically, knowledge distillation is a classic compression method adopted in recommender systems~\cite{sun2024PRM-KD,tian2023directed}, with the core idea of guiding lightweight student models to ``imitate'' teacher models with better performance and more complex structures such as LLMs. In recommender systems, the collaborative optimization of LLMs and recommendation models can also be seen as the distillation process to inject knowledge from LLMs to traditional recommenders, enhancing representations of both users and items~\cite{li2023ctrl,qiu2023controlrec,xi2023towards-kar}. Due to the fact that knowledge distillation can improve efficiency while retaining the capabilities of LLMs for recommendation, more applications need to be fully explored. 

\paratitle{Multimodal recommendations with LLMs}. 
In addition to IDs and text, multimodal recommendations with LLMs hold considerable promise and warrant comprehensive exploration with the evolving landscape of media consumption~\cite{yang2023set-of-mark,zhang2023multimodal}. The essence of multimodal recommendations resides in the fusion of textual and visual information for enhanced user engagement~\cite{pan2022multimodal,harrison2023multimodal}, and the dual functionality of LLMs is pivotal in this context. LLMs possess the capability to function as multimodal LLMs, enabling the incorporation and encoding of visual information extracted from images. Furthermore, images can be transformed into textual representations by multimodal encoders first, and LLMs are mainly used for the subsequent integration of diverse modalities~\cite{harrison2023multimodal}. In addition, the rich multimodal attributes also provide a basis for diversified recommendation results~\cite{lin2023transrec}. As the field progresses, an emphasis on the reproducibility, benchmarking, and standardization of evaluation datasets and metrics will be essential to foster a cohesive and informed advancement in multimodal recommender systems leveraging LLMs~\cite{geng2023VIP5,wei2024UniMP}.

\paratitle{Fairness-aware recommendations of LLMs}. Future research in the domain of fairness-aware recommendations with LLMs presents a compelling avenue for scholarly inquiry~\cite{wang2023fair,zhang2023chatgptfair,li2023ChatGPT-news,dai2023llms-bias}. In line with existing work, researchers have explored that retrievers in information retrieval are biased towards contents generated by LLMs~\cite{dai2023llms-bias}, and LLMs utilized as recommender systems also output unfair recommendation results~\cite{wang2023fair,zhang2023chatgptfair,li2023ChatGPT-news}, necessitating a profound investigation into methodologies that ensure equitable and unbiased outcomes for LLM-based recommendations. It is imperative to scrutinize existing fairness-aware algorithms and develop novel approaches that cater to the intricacies of language models, particularly in understanding how fairness metrics align with user expectations. As LLMs continue to evolve, the research community must collaborate to establish standardized evaluation metrics and benchmarks for fairness-aware recommendations, ensuring the reproducibility and comparability of findings across diverse studies in the era of LLMs. In essence, fairness-aware recommendations with LLMs are poised to contribute substantially to the development of ethical and equitable recommender systems~\cite{wang2023fair,zhang2023chatgptfair}.

\paratitle{General-purpose LLMs in the vertical field of recommender systems}. Developing a comprehensive framework for LLMs to address multiple recommendation tasks represents a significant avenue for scholarly exploration~\cite{geng2022p5,chu2023RecSysLLM,zhang2023recommendation,wang2023recmind}. The endeavor involves formulating a unified and versatile structure that accommodates the intricacies of various recommendation tasks~\cite{dai2023uncovering}, encompassing diverse modalities and user preferences~\cite{lin2023transrec,harrison2023multimodal}. In addition, using agents of LLMs to simulate recommendation scenarios and make dynamic decisions is also a feasible application of general-purpose LLMs~\cite{wang2023recagent,zhang2023agentcf,wang2023survey}. Research efforts should focus on refining the architecture, training methodologies, and adaptability of such a general framework to ensure optimal performance across different recommendation domains. Investigating transfer learning techniques within this framework, enabling the transfer of knowledge between recommendation tasks, is crucial for enhancing efficiency and leveraging shared information~\cite{zhang2023recommendation,bao2023tallrec}. In general, general-purpose LLMs for recommendation have the potential to revolutionize recommender systems by providing a unified and scalable solution capable of addressing the multifaceted challenges. 

\paratitle{Privacy and ethical concerns}. Prior studies~\cite{shen2023chatgpt-reliability,weidinger2021ethical} have highlighted the potential issue of language models generating unreliable or personal contents based on certain prompts and insecure instructions. Recommendation systems involve massive amounts of the user data~\cite{ni2019amazon2018,guo2017deepfm}, and it is crucial to remove private and potentially harmful information stored in LLMs to enhance the privacy and security of LLM-based applications~\cite{carranza2023privacy,lei2023recexplainer}. Notably, researchers have observed that Reinforcement Learning from Human Feedback~(RLHF) and model editing techniques~\cite{geva2022transformer} have the potential to restrain the generation of poisonous or harmful contents from LLMs, thereby mitigating privacy and ethical concerns associated with privacy-preserving recommender systems~\cite{carranza2023privacy}. 
Nevertheless, it is imperative to acknowledge that the alignment technology of LLMs may be vulnerable to misuse. The apprehension exists that LLMs could be manipulated by malicious users to selectively influence agents and amplify specific viewpoints within recommender systems. Consequently, addressing the security and privacy implications of LLM-based recommendations is crucial, and there is a need to formulate public regulations to mitigate potential risks~\cite{weidinger2021ethical}.


\begin{acks}

The authors would like to thank Chuyuan Wang and Chenrui Zhang for participating in discussions of this paper. Lanling Xu is supported by Meituan Group during her research internship. Xin Zhao is the corresponding author.

\end{acks}